\documentclass[reprint,showkeys, amsmath,amssymb, aps, floatfix]{revtex4-2}
\usepackage{tgtermes}
\usepackage{graphicx,subfigure}
\graphicspath{{figures/}}
\usepackage{dcolumn}
\usepackage{bm}
\usepackage{amsmath}
\usepackage{amssymb}
\usepackage{physics}
\usepackage{xcolor}
\usepackage{slashed}
\usepackage{multirow}
\usepackage{float}
\usepackage{capt-of}
\usepackage{comment}
\usepackage{ mathrsfs }
\usepackage[colorlinks=true,linkcolor=blue,urlcolor=blue,citecolor=blue]{hyperref}
\DeclareUnicodeCharacter{2212}{\textendash}

\begin{document}
	
	\title{Study of anomalous $W^-W^+\gamma/Z$ couplings using polarizations and spin correlations in $e^-e^+\to W^-W^+$  with polarized beams}

	\author{Amir Subba}
	\email{as19rs008@iiserkol.ac.in}
	
	\author{Ritesh K. Singh}
	\email{ritesh.singh@iiserkol.ac.in}
	\affiliation{Department of Physical Sciences, Indian Institute of Science Education and Research Kolkata, Mohanpur, 741246, India\\
	}
	\date{\today}
	
\begin{abstract}
    We study the anomalous $W^-W^+\gamma/Z$ couplings in $e^-e^+\to W^-W^+$ followed by semileptonic decay using a complete set of polarization and spin correlation observables of $W$ boson with the longitudinally polarized beam. We consider a complete set of dimension-six operators affecting $W^-W^+\gamma/Z$ vertex, which are $SU(2)\times U(1)$ gauge invariant. Some of the polarization and spin correlation asymmetries average out if the daughter of $W^+$ is not tagged. We developed an artificial neural network and boosted decision trees to distinguish down-type jets from up-type jets. We obtain bounds on the anomalous couplings for center of mass energy $\sqrt{s} = 250$ GeV with integrated luminosities of~$\mathcal{L}\in\{100~\text{fb}^{-1}, 250~\text{fb}^{-1}, 1000~\text{fb}^{-1}, 3000~\text{fb}^{-1}\}$. We find that using spin-related observables and cross~section in the presence of initial beam polarization significantly improves the bounds on anomalous couplings compared to previous studies.
\end{abstract}
	
	\maketitle
	\section{Introduction}
 \label{sec:intro}
\begin{figure}
     \centering
     \includegraphics[scale=0.6]{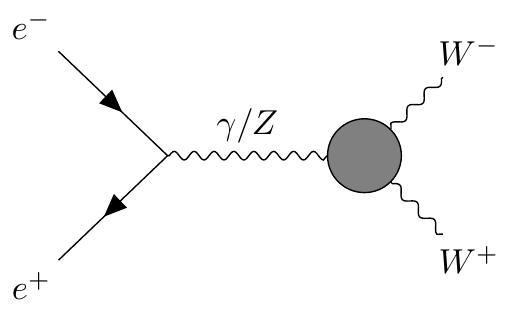}
     \includegraphics[scale=0.85]{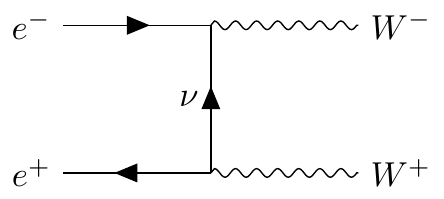}
     \caption{Leading order feynman diagrams for the $W^-W^+$ production process in $e^-e^+$ lepton collider. The blob in the diagram in the left panel represents anomalous vertex contribution.}
     \label{fig:feynman}
 \end{figure}

  The standard model is dimension-4 local quantum field theory, which has been successfully tested experimentally. With the discovery of a Higgs-like boson at LHC~\cite{CMS:2012qbp,ATLAS:2012yve}, the last missing piece of SM is finally in place. However, despite all these wonderful experimental justifications, many unexplained phenomena suggest SM's incompleteness. The mass of the Higgs boson~(125 GeV) cannot be explained within SM. The higher-order quantum corrections would shift the mass of the Higgs boson to the cut-off scale unless there are perfect cancellations of these corrections. These fine-tuning mechanisms are not present within the framework of SM. SM QCD still has an unexplained problem with the $\theta$ parameter known as the strong-CP problem. Dark matter constitutes approximately $80\%$ of all the matter in our universe~\cite{Planck:2013pxb}, yet the structure of dark matter is still a mystery. The recent report on the magnetic moment of muon~\cite{Muong-2:2021vma,Muong-2:2021ojo}, $W$ bosons mass~\cite{CDF:2022hxs} are some of the additional reasons to cast doubt that the SM is not a complete theory of fundamental particle. Though many theories beyond SM explain deviations with the inclusion of new particles, symmetry, or dimensions, experiments have not seen any signature in favor of any of these explicit models. In this article, we follow a more natural way of extending the SM called the effective field theory~(EFT), which provides a suitable framework to parameterize deviations from the SM predictions according to the decoupling theorem~\cite{Appelquist:1974tg}. This framework expands the Lagrangian of SM by adding higher dimensional terms. These higher dimensional terms are made out of SM fields, assuming that the new degrees of freedom are too heavy to be observed with currently available energy and are integrated out of the Lagrangian. The effects of new physics are encoded in the Wilson coefficient of these infinite series of higher dimensional operators. Considering the conservation of the lepton-baryon number, the effective Lagrangian is compactly written as~\cite{Buchmuller:1985jz}
    \begin{equation}
    	\mathscr{L}_{eft} = \mathscr{L}_{SM} + \frac{1}{\Lambda^2}\sum_i \mathcal{C}^{(6)}_i \mathscr{O}^{(6)}_i + \frac{1}{\Lambda^4}\sum_j\mathcal{C}^{(8)}_j\mathscr{O}^{(8)}_j + ..,
    	\label{efteqn}
    \end{equation}  
    where $c_i$ are the Wilson coefficient and $\mathscr{O}_i$ are higher order operators.
	Each higher dimensional term is suppressed by the power of $\Lambda^{(d-4)}$, where $\Lambda$ is a characteristic new physics scale usually taken to be several TeVs. This suggests that the contribution to SM decreases as one moves towards $d>6$. Thus, we can truncate the above Lagrangian at some lowest higher-order terms of one interest. This study focused on the dimension-6 operators which affect $WWV, V\in\{\gamma,Z\}$ vertex. Considering both CP-even and -odd operators, five relevant dimension-6 operator affects $WWV$ vertex, which in HISZ basis~\cite{Hagiwara:1986vm,Degrande:2012wf} are listed as:
	\begin{equation}
		\begin{aligned}
			&\mathscr{O}_{WWW} &=&\quad \text{Tr}[W_{\nu \rho}W^{\mu \nu} W_{\rho}^{\mu}],\\
			&\mathscr{O}_{W} &=&\quad (D_\mu \Phi)^\dagger W^{\mu \nu} (D_\nu \Phi), \\
			&\mathscr{O}_B &=&\quad (D_\mu \Phi)^\dagger B^{\mu \nu} (D_\nu \Phi), \\
			&\mathscr{O}_{\tilde{WWW}} &=&\quad \text{Tr}[\widetilde{W}_{\mu \nu} W^{\nu \rho} W^\mu_\rho], \\
			&\mathscr{O}_{\tilde{W}} &=&\quad (D_\mu \Phi)^\dagger \widetilde{W}^{\mu \nu} (D_\nu \Phi),
		\end{aligned}
	\label{eftop}
	\end{equation}
	where $\Phi = \begin{pmatrix}
		\phi^+ \\ \phi^0
	\end{pmatrix}$ is the Higgs doublet and $W^{\mu \nu}, B^{\mu \nu}$ represents the field strengths of $W$ and $B$ gauge Fields. The covariant derivative is defined as $D_\mu = \partial_\mu  +\frac{i}{2}g\tau^i W_\mu^i + \frac{i}{2}g'B_\mu$. The field tensor are given by $W_{\mu \nu} = \frac{i}{2}g\tau^i(\partial_\mu W_\nu^i-\partial_\nu W_\mu^i + g\epsilon_{ijk}W_\mu^i W_\nu^k)$ and $B_{\mu \nu} = \frac{i}{2}g'(\partial_\mu B_\nu - \partial_\nu B_\mu)$.
	The first three operators in Eq.~(\ref{eftop}) are $CP$-even, and the last two are $CP$-odd. After electroweak symmetry breaking~(EWSB), each operator of Eq.~(\ref{eftop}) generates anomalous $WWV$ vertex along with triple vertex containing Higgs and vector gauge boson. These operators~($c_{WWW}, c_W$ and their's dual) also generate quartic gauge boson couplings. The general couplings of two charged vector bosons with a neutral vector boson can be parameterized in an effective Lagrangian~\cite{Hagiwara:1986vm}:
	\begin{widetext}
	\begin{equation}
		\begin{split}
			\mathscr{L}^{eff}_{WWV} = ig_{WWV}[g_1^V(W^+_{\mu \nu}W^{-\mu} - W^{+\mu}W^-_{\mu \nu})V^\nu  
			+  k_V W^+_\mu W^-_\nu V^{\mu \nu} +\frac{\lambda_V}{m_W^2}W_\mu^{\nu+}W_\nu^{-\rho}V_{\rho}^{\mu}  \\
			+ ig_4^VW_\mu^+W_\nu^-(\partial^\mu V^\nu+\partial^\nu V^\mu)  
			- ig_5^V\epsilon^{\mu \nu \rho \sigma}(W_\mu^+ \partial_\rho W_\nu^- - \partial_\rho W_\mu^+W_\nu^-)V_\sigma \\ 
			+ \tilde{k}_VW_\mu^+W_\nu^-\tilde{V}^{\mu \nu} + \frac{\tilde{\lambda}_V}{m_W^2}W_\mu^{\nu+}W_\nu^{-\rho}\tilde{V}_\rho^{\mu}],
		\end{split} 
		\label{Lag:eff}	
	\end{equation}
	\end{widetext}
	where $W^{\pm}_{\mu\nu} = \partial_\mu W^{\pm}_\nu-\partial_\nu W^\pm_\mu$, $V_{\mu\nu} = \partial_\mu V_\nu-\partial_\nu V_\mu$, $g_{WW\gamma} = -e$ and $g_{WWZ} = -e\cot\theta_W$, where $e$ and $\theta_W$ are the proton charge and weak mixing angle respectively. The dual field is defined as $\tilde{V}^{\mu\nu} = 1/2\epsilon^{\mu\nu\rho\sigma}V_{\rho\sigma}$, with Levi-Civita tensor $\epsilon^{\mu\nu\rho\sigma}$ follows a standard convention, $\epsilon^{0123}=1$. As discussed in Ref.~\cite{Hagiwara:1986vm}, these seven operators exhaust all possible Lorentz structures to define the most general $WWV$ interaction. Within the SM, the couplings are given by $g_1^Z=g_1^\gamma = k_Z = k_\gamma = 1$ and all others are zero. The gauge invariance fixes the value couplings like $g_1^\gamma,g_4^V,g_5^V$, but in the presence of the effective operators given in Eq.~(\ref{eftop}), the values of other couplings changes, and are listed below:
	\begin{equation}
		\label{lep}
		\begin{aligned}
			&\Delta g_1^Z &=&\quad  c_W\frac{m^2_Z}{2\Lambda^2}, \quad \lambda_V = \quad c_{WWW}\frac{3m_W^2g^2}{2\Lambda^2}\\
			&\Delta k_Z &=&\quad  \left[c_W-\sin^2\theta_W\left(c_B+c_W\right)\right]\frac{m_Z^2}{2\Lambda^2} \\
			&\Delta k_\gamma &=&\quad \left(c_B+c_W\right)\frac{m_Z^2}{2\Lambda^2},\quad\tilde{\lambda}_V =c_{\widetilde{WWW}}\frac{3m_W^2g^2}{2\Lambda^2} \\
			&\tilde{k}_Z&=&\quad-c_{\tilde{W}}\sin^2\theta_W \frac{m_Z^2}{\Lambda^2},\quad\tilde{k}_\gamma =\quad c_{\tilde{W}}\frac{m_W^2}{2\Lambda^2}.
		\end{aligned}
	\end{equation}
	Here,
	\begin{equation}
       c_i = \{c_{WWW},c_W,c_B,c_{\widetilde{W}},c_{\widetilde{WWW}}\},
	\end{equation}
	  are the Wilson coefficient associated with the effective operators listed in Eq.~(\ref{eftop}). The presence of these anomalous couplings will bring change in the observable that could be measured in the current or future collider, given that the deviation is within the measurable reach of the collider. The BSM matrix element for a given process in the presence of effective operators of Eq.~(\ref{efteqn}) truncated at dimension-6 is given by
	\begin{equation}
		|\mathcal{M}_{BSM}|^2 = |\mathcal{M}_{SM}|^2 + 2\mathcal{R}e\left\{\mathcal{M}_{SM}\mathcal{M}^*_{d6}\right\}+|M_{d6}|^2.
		\label{bsmeq}
	\end{equation}
	The interference, i.e., the second term of Eq.~(\ref{bsmeq}) between the SM amplitude and the dim-6 amplitude, induces asymmetries in appropriately constructed $CP$-odd observable. The direct observation of the non-zero value of observable related to $CP$-odd operators would be a strong marker for the new physics as their values are predicted to be zero in SM at tree and loop level~\cite{Czyz:1988yt}. The last term, i.e., square term if the dimension-6 operator is of order $\Lambda^{-4}$, comparable to SM's interference with dimension-8 operators. However, we have assumed all dimension-8 couplings to be zero while performing the analyses upto order $\Lambda^{-4}$.\\ \\
	Many theoretical studies on $W^-W^+\gamma/Z$ couplings are done at $e^-e^+$ collider~\cite{Zhang:2016zsp,Rahaman:2019mnz,Bilchak:1984ur,Gaemers:1978hg,Hagiwara:1992eh,Hagiwara:1995se,Choudhury:1996ni,Choudhury:1999fz,Wells:2015eba,Buchalla:2013wpa,Berthier:2016tkq,Beyer:2020eas,Subba:2022czw,Bian:2015zha,Bian:2016umx} and hadron collider~\cite{Bian:2015zha,Bian:2016umx,Baglio:2019uty,Choudhury:2022iqz,RebelloTeles:2013kdy,Tizchang:2020tqs,Campanario:2020xaf,Ciulli:2020ygo,Yap:2020xjr,Hwang:2023wad,Deka:2022lmf,Falkowski:2016cxu,Butter:2016cvz,Azatov:2017kzw,Baglio:2017bfe,Li:2017esm,Bhatia:2018ndx,Chiesa:2018lcs,Rahaman:2019lab,Dixon:1999di,Baur:1987mt} and Large Hadron electron collider~(LHeC)~\cite{Biswal:2014oaa,Cakir:2014swa,Li:2017kfk,Koksal:2019oqt,Gutierrez-Rodriguez:2019hek}. On experimental side, the similar studies was reported by different collaborations like OPAL~\cite{OPAL:2000wbs,OPAL:2003xqq,OPAL:2000rnf,OPAL:1998ixj,OPAL:1997yzg,OPAL:1997xsu}, ALEPH~\cite{ALEPH:2004klc,ALEPH:2001ylz,ALEPH:1998nxw,ALEPH:1997agc,ALEPH:2013dgf}, DELPHI~\cite{DELPHI:2008uqu}, CDF~\cite{CDF:2012mnr,CDF:2007aqs}, D0~\cite{D0:2013rce,D0:2010jca,D0:2005djn}, ATLAS~\cite{ATLAS:2017pbb,ATLAS:2017luz,ATLAS:2016zwm,ATLAS:2016bkj,ATLAS:2014ofc,ATLAS:2013way,ATLAS:2012aid,ATLAS:2012bpb,ATLAS:2012upi,ATLAS:2011nle} and CMS~\cite{CMS:2021icx,CMS:2021foa,CMS:2016qth,CMS:2020ypo,CMS:2019ppl,CMS:2019efc,CMS:2017fhs,CMS:2017egm,CMS:2016qth,CMS:2014cdf,CMS:2013ryd,CMS:2013ant,CMS:2012wlr,CMS:2011egr} . We list the experimentally measured current tightest limits on anomalous couplings~($c_i$) at $95\%$ confidence level~(CL) in Table~\ref{tab:constraint}.
	\begin{table}[h!]
		\caption{The list of tightest constraints observed on the anomalous couplings in $SU(2) \times U(1)$ gauge at 95$\%$ confidence level~(CL) from various experiments.}
		\centering
		\begin{ruledtabular}
			\begin{tabular}{lll} 
				$c_i^{\mathscr{O}}$&Limits(TeV$^{-2}$)&Remarks \\ \hline 
				$c_{WWW}/{\Lambda^2}$&[-0.90,+0.91]&CMS~\cite{CMS:2021foa} \\
				$c_W/{\Lambda^2}$&[-2.50,+0.30]&CMS\cite{CMS:2021icx} \\ 
				$c_B/{\Lambda^2}$&[-8.78,+8.54]&CMS\cite{CMS:2019ppl} \\ 
				$c_{\widetilde{W}}/\Lambda^2$&[-20.0,+20.0]&CMS~\cite{CMS:2021foa}\\ 
				$c_{\widetilde{WWW}}/\Lambda^2$&[-0.45,+0.45]&CMS~\cite{CMS:2021foa} 
			\end{tabular}
		\end{ruledtabular}
		\label{tab:constraint}
	\end{table} \\
	The limits on the anomalous couplings listed in Table~\ref{tab:constraint} are obtained by varying one parameter at a time and keeping other couplings to zero. \\ \\
 Due to the $SU(2)_L\times U(1)_Y$ gauge structure of SM, the electro-weak interactions are susceptible to the chirality of fermions. Left-handed fermions interact fundamentally differently than their right-chiral counterpart with the weak boson. This fundamental difference in interaction can be used as a tool to probe a deviation from the Standard Model~(SM) prediction. Apart from increasing the luminosity, it has been shown~\cite{Swartz:1987xme} that the use of polarized beams can bring down the systematic error on the measurement of the left-right asymmetry~($A_{\text{LF}}$). Reducing systematic error of asymmetries becomes important when observables like polarization and spin correlation asymmetries are used as observables to probe beyond Standard Model~(BSM) physics. The polarized positron beams can be used to probe physics models that would allow some unconventional combinations of helicities like in Minimal Sypersymmetric SM~(MSSM) where $e^+_{R}e^-_R \to \tilde{e}^-_R\tilde{\chi}^+\bar{\nu}_e$ and $e^+_{L}e^-_L \to \tilde{e}^+_L\tilde{\chi}^-\nu_e$~\cite{Moortgat-Pick:2001vdh}. Also, the ability to adjust the polarization of both beams independently provides unique possibilities for directly probing the properties of the produced particles like quantum numbers, and chiral couplings. The important effects of initial beam polarization are~\cite{Fujii:2018mli}
	\begin{itemize}
		\item $e^-$ and $e^+$ polarization allow obtaining a subset of the sample with higher rates for interesting physics and lower background. It would lead to an overall increase in sensitivities,
		\item When both beams are polarized, one obtains four distinct data sets instead of the two available if only the $e^-$ beam can be polarized.
		\item The likely most important effect is the control of systematic error for precision studies in Lepton collider.
	\end{itemize}
The future Lepton collider viz. International Linear Collider~(ILC)~\cite{ILC:2013jhg,Adolphsen:2013kya}, Compact Linear Collider~(CLIC)~\cite{CLICdp:2018cto}, Future Circular Lepton Collider-ee~(FCC-ee)~\cite{FCC:2018evy} will be equipped with the potential of colliding polarized beams. With such specifications, these colliders will be the ideal place to undertake precision measurements.\\\\
In the current article, we use polarized $e^-$ and $e^+$ beams to probe a set of dimension-6 operators listed in Eq.~(\ref{eftop}) using various observable like cross~section, polarization asymmetries, and spin correlation asymmetries. The study is done at the center-of-mass energy~($\sqrt{s})$ of 250 GeV at machine parameters planned for ILC using longitudinal polarized beams. However, the analysis can be easily translated to other colliders. The process we probed is the di-boson production of $W^-W^+$ followed by their semi-leptonic decay,
	\begin{equation}
		\label{process}
		e^- + e^+ \to W^- + W^+ \to l^-jj\slashed{E},
	\end{equation}
	where $j$ are the light quarks and $l^- \in\{ e^-, \mu^-\}$. The amplitude of the process $e^-e^+ \to W^-W^+$ results from the $t$-channel neutrino and $s$-channel $\gamma$ and $Z$ exchange, see Fig.~\ref{fig:feynman}.
 The $s$-channel diagrams contain trilinear $\gamma W^-W^+$ and $ZW^-W^+$ gauge boson couplings whose deviations from SM due to the above given dim-6 EFT operators are being probed in this current article.\\\\
 Some of the asymmetries we use require knowledge about the $W$ boson decay products being \emph{up/down} fermions. It is straightforward in the leptonic channel but requires us to flavor tag the hadronic decay of $W^+$ into \emph{up}-type and \emph{down}-type jets. We use machine learning~(ML) models to develop such a tagger.
ML has been used to identify the jets originated from the gluon and quarks~\cite{Lonnblad:1990bi,Pumplin:1991kc,ATLAS:2017dfg,Komiske:2016rsd,Cheng:2017rdo,Kasieczka:2018lwf,Kasieczka:2020nyd}. The generic ML models have also been developed to classify the jets as originating from light-flavor or heavy-flavor quarks~\cite{Guest:2016iqz,CMS:2017wtu,Erdmann:2020ovh,Bedeschi:2022rnj,Nakai:2020kuu,Erdmann:2019blf}.
 In this article, we develop artificial neural network~(ANN) and boosted decision tree~(BDT) to assist in classifying the jets from the $W^+$ boson as originating from \emph{up}-type or \emph{down}-type light quarks.  
 \\ \\
We describe in section~\ref{spin related} the effect of beam polarization on the observable. We also describe spin-related observables like polarization asymmetries and spin correlation asymmetries and the method to calculate them. In section~\ref{flavor tagging}, we describe the machine learning~(ML) technique, especially artificial neural networks~(ANN) and boosted decision tree~(BDT) that will be used for flavor tagging or to reconstruct $W^+$ boson. We also discuss the performance of various models in classifying the jets. Section~\ref{parameter estimation} deals with the parameter estimation to find the limits on various anomalous couplings and discuss the results obtained. We conclude in section~\ref{conclude}.
\section{Beam Polarization and Observables}
\label{spin related}
\begin{figure}[!h]
	\centering
	\subfigure{\includegraphics[width=0.23\textwidth]{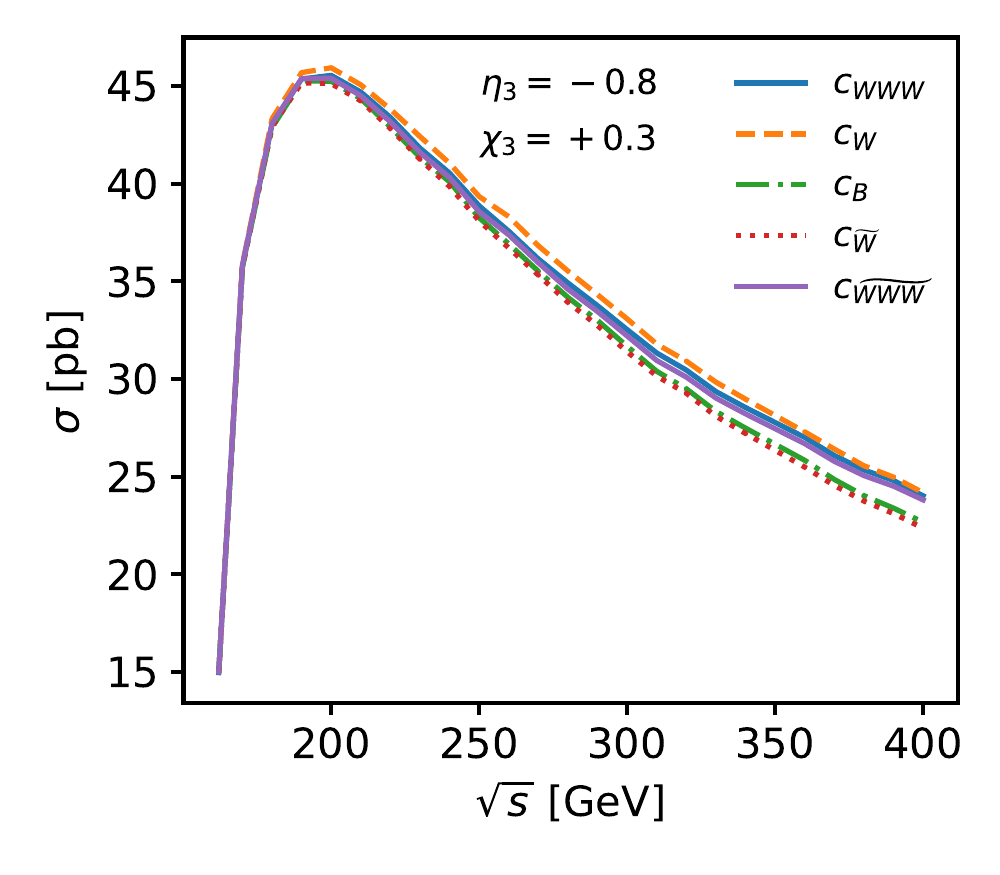}}
	\subfigure{\includegraphics[width=0.23\textwidth]{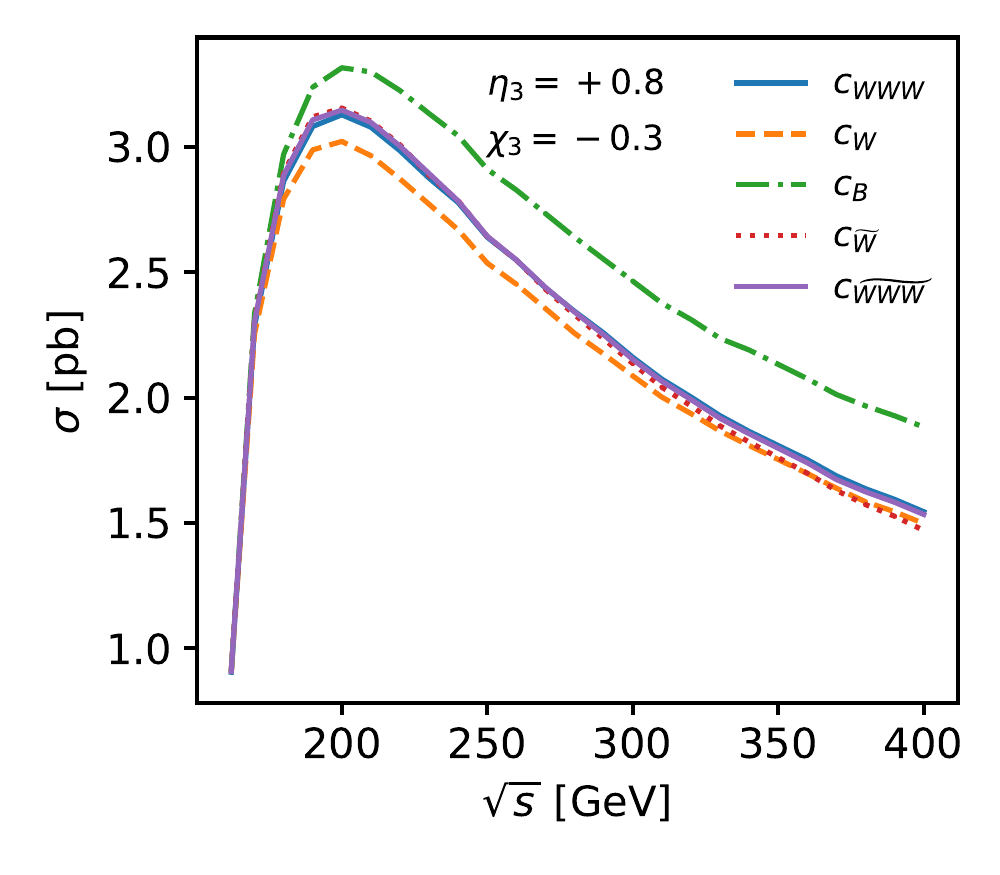}}
	\caption{\label{crosspol}Total production cross-section of $W^-W^+$ pair with respect to the center-of-mass energy $\sqrt{s}$ in the presence of initial beam polarization and anomalous triple gauge couplings. The left panel represents $(\eta_3,\xi_3) = (+0.8, -0.3), $ and right panel represents flip polarization.}
\end{figure}

The polarization of resonance $W$ bosons is reconstructed through the angular distribution of its decay products. The full hadronic channel suffers from the significant background contribution from the decay of $Z$ to hadrons. The full leptonic channel is cleaner to detect the polarization but comes with two missing neutrinos, which makes the reconstruction of 
the rest frame of the $W$ boson non-trivial. Thus, for a $W^-W^+$ final state, the optimal channel to study the polarization and spin correlations is the semi-leptonic channel. 
The normalized production density matrix of a $W$ boson which is a spin-1 boson can be written as~\cite{Boudjema:2009fz}:
\begin{equation}
	\label{normprod}
	\rho(\lambda,\lambda^\prime) = \frac{1}{3}\left[I+\frac{3}{2}\vec{P}\cdot\vec{S}+\sqrt{\frac{3}{2}}T_{ij}(S_iS_j+S_jS_i)\right],
\end{equation}  
where $\vec{P} = \{P_x,P_y,P_z\}$ is the vector polarization of $W$ boson, $\vec{S} = \{S_x,S_y,S_z\}$ is the spin basis and $T_{ij}(i,j=x,y,z)$ is the second-rank symmetric traceless tensor and $\lambda,\lambda^\prime$ are the helicities of the $W$ boson. Similarly, the normalized decay density matrix of $W$ boson decaying to two fermions with helicities $s_1$ and $s_2$ via an interaction vertex $Wf\bar{f}:$ $\gamma^\mu(L_fP_L+R_fP_R)$ is given as~\cite{Boudjema:2009fz}:
\begin{widetext}
	\begin{equation}
		\label{normdec}
		\Gamma_W(s_1,s_2)=
	\begin{bmatrix}
		\frac{1+\delta+(1-3\delta)\cos^2\theta_f+2\alpha\cos\theta_f}{4}&\frac{\sin\theta_f(\alpha+(1-3\delta)\cos\theta_f)}{2\sqrt{2}}e^{i\phi_f}&(1-3\delta)\frac{(1-\cos^2\theta_f)}{4}e^{i2\phi_f}\\
		\frac{\sin\theta_f(\alpha+(1-3\delta)\cos\theta_f)}{2\sqrt{2}}e^{-i\phi_f}&\delta+(1-3\delta)\frac{\sin^2\theta_f}{2}&\frac{\sin\theta_f(\alpha-(1-3\delta)\cos\theta_f)}{2\sqrt{2}}e^{i\phi_f} \\
		(1-3\delta)\frac{1-\cos^2\theta_f}{4}e^{-i2\phi_f}&\frac{\sin\theta_f(\alpha-(1-3\delta)\cos\theta_f)}{2\sqrt{2}}e^{-i\phi_f}&\frac{1+\delta+(1-3\delta)\cos^\theta_f-2\alpha\cos\theta_f}{4}
	\end{bmatrix}.
	\end{equation}
\end{widetext}
Assuming Narrow Width Approximation, the production and decay part can be factorized into different terms, and on combining Eq.~(\ref{normprod}) and Eq.~(\ref{normdec}), the differential cross-section would be:
\begin{widetext}
\begin{equation}
	\label{eqn:domega}
	\begin{aligned}
		\frac{1}{\sigma}\frac{d\sigma}{d\Omega_f} &= \frac{3}{8\pi}\left[\left(\frac{2}{3}-(1-3\delta)\frac{T_{zz}}{\sqrt{6}}\right)+\alpha P_z \text{cos}\theta_f + \sqrt{\frac{3}{2}}(1-3\delta)T_{zz}\text{cos}^2\theta_f +
		\left(\alpha P_x + 2\sqrt{\frac{2}{3}}(1-3\delta)T_{xz}\text{cos}\theta_f\right)\text{sin}\theta_f \text{cos}\theta_f \right.\\&+\left. \left(\alpha P_y + 2\sqrt{\frac{2}{3}}(1-3\delta)T_{yz}\text{cos}\theta_f\right)\text{sin}\theta_f \text{sin}\theta_f + (1-3\delta)\left(\frac{T_{xx}-T_{yy}}{\sqrt{6}}\right)\text{sin}^2\theta_f\text{cos}(2\phi_f) \right.\\&+\left. \sqrt{\frac{2}{3}}(1-3\delta)T_{xy}\text{sin}^2\theta_f\text{sin}(2\phi_f)\right]
	\end{aligned}
\end{equation} 
\end{widetext}
where $\theta_f,\phi_f$ are the polar and azimuth orientation of the fermion $f$, in the rest frame of $W$ boson with its would-be momentum along $z-$axis. The initial beam direction and the $W^-$ momentum in the lab frame define the $x$-$z$ plane, i.e., $\phi = 0$ plane, in the rest frame of $W^-$. In this case, along with assuming high energy limits, the mass of final state fermions can be zero, then $\alpha = -1$ and $\delta = 0$. The construction of various polarization at the rest frame of $W$ boson requires one to find the single missing neutrino. As the Lepton collider does not involve PDFs, reconstructing four momenta of a single neutrino is straightforward. One can construct several asymmetries to probe various polarization parameters using partial integration of the differential distribution given in Eq.~(\ref{eqn:domega}) of the final state $f$. For example, we can get $p_x$ from the left-right asymmetry $A_x$ as:
\begin{equation}
	P_x = \frac{4}{3\alpha\sigma}\left[\int_{\theta=0}^\pi\int_{\phi=-\frac{\pi}{2}}^{\frac{\pi}{2}}\frac{d\sigma}{d\Omega_f}d\Omega_f -\int_{\theta=0}^\pi\int_{\phi=\frac{\pi}{2}}^{\frac{3\pi}{2}}\frac{d\sigma}{d\Omega_f}d\Omega_f  \right]
\end{equation}
Finding $P_x$ would be a counting procedure, which can be written in terms of asymmetries as,
\begin{equation}
	A_x = \frac{3\alpha P_x}{4} \equiv \frac{\sigma(\cos\phi>0) - \sigma(\cos\phi<0)}{\sigma(\cos\phi>0) + \sigma(\cos\phi<0)}
\end{equation}
All other polarization parameters are obtained similarly using various angular functions listed in Table~\ref{tab:core}. 
\begin{table}
	\caption{\label{tab:core}List of correlators used to construct asymmetries.}
	\begin{ruledtabular}
		\begin{tabular}{@{}p{1.5cm}@{}p{2.5cm}@{}p{4.0cm}@{}} 
			$A_i$ & $c_j$ & functions \\ \hline
			$A_x$& $c_1 \equiv c_x$ &sin$\theta$cos$\phi$ \\ 
			$A_y$&  $c_2 \equiv c_y$ &sin$\theta$sin$\phi$ \\ 
			$A_z$&  $c_3 \equiv c_z$ &cos$\theta$ \\
			$A_{xy}$&  $c_4 \equiv c_{xy}$ &sin$^2\theta$sin$(2\phi)$ \\
			$A_{xz}$&  $c_5 \equiv c_{xz}$ &sin$\theta$cos$\theta$cos$\phi$ \\ 
			$A_{yz}$&  $c_6 \equiv c_{yz}$ &sin$\theta$cos$\theta$sin$\phi$ \\
			$A_{x^2-y^2}$&  $c_7 \equiv c_{x^2-y^2}$ &sin$^2\theta$cos$(2\phi)$ \\ 
			$A_{zz}$&  $c_8 \equiv c_{zz}$ &sin$(3\theta)$\\ 
		\end{tabular}
	\end{ruledtabular}
\end{table}\\ \\
When a pair of particles is produced as in the process probed in this article, i.e., $e^-e^+\to W^-W^+$, one can create various observables related to the spin correlations of these two $W$ bosons. Studies~\cite{Smillie:2006cd} show angular correlations between the decay products of two $W$ bosons. W-boson spin correlations are measured by tagging the helicity of the $W$ boson, which decays into hadrons, and measuring the helicity of the $W$ boson, which decays into leptons. Spin correlations of $W$ boson had been used to study the triple gauge boson couplings in Ref.~\cite{Subba:2022czw,ALEPH:2001ylz,OPAL:2003xqq}. We construct various spin correlation asymmetries using the angular distribution of the decay products of $W$ bosons using the correlators listed in Table~\ref{tab:core}. Similar to polarization, the spin correlation can be calculated in terms of asymmetries as,
\begin{equation}
	A_{ij}^{WW} = \frac{\sigma(c_i^ac_j^b > 0) - \sigma(c_i^ac_j^b < 0)}
	{\sigma(c_i^ac_j^b > 0) + \sigma(c_i^ac_j^b < 0)}, 
	\label{eq:spincorr}
\end{equation}   
where $a$ and $b$ are the final state leptons and jets coming from the decay of $W$ bosons, and $c's$ are the correlators listed in Table.~\ref{tab:core}. There will be 64 spin correlation asymmetries for a pair of spin-1 particles. Since the polarization and correlation depend strongly on the $\cos\theta$, where $\theta$ is the production angle of $W^-$ boson in the lab frame, we can divide the events in certain bins of $\cos\theta$ to increase the overall sensitivity.
\\\\ 
\textbf{Effect of initial beam polarization:} The future particle collider like ILC will use polarized beams for both electron and positron~\cite{Adolphsen:2013jya,Adolphsen:2013kya} to increase its sensitivity to new physics and to improve its measurement accuracy. The presumed design values of the beam polarization are 80$\%$ for the electron and 30$\%$ for the positron beam~\cite{ILC:2013jhg}. Here, we will discuss how various observable changes in the presence of initial beam polarization. The observables like cross section and asymmetries change in the presence of initial beam polarization, and we can show this with the change in the transition amplitude as defined in~\cite{Moortgat-Pick:2005jsx}
\begin{equation}
	|\mathcal{M}|^2 = \sum_{\lambda's}\rho_{\lambda_{e^-},\lambda_{e^-}'}\rho_{\lambda_{e^+},\lambda_{e^+}'}F_{\lambda_{e^-},\lambda_{e^+}}F^*_{\lambda_{e^-}'{\lambda_{e^+}'}}
	\label{polmat}
\end{equation}  
where $\rho$ are the spin density matrix of $e^-/e^+$ and $F$ are the helicity amplitude. \\
The cross-section with beam polarization $\eta_3$ and $\xi_3$ for electron and positron, respectively, can be written as:
\begin{equation}
	\begin{aligned}
		\sigma(\eta_3,\xi_3) &= \frac{(1+\eta_3)(1+\xi_3)}{4}\sigma_{\text{LR}} \\ &+ \frac{(1-\eta_3)(1-\xi_3)}{4}\sigma_{\text{RL}},
	\end{aligned}
\end{equation}
where $\sigma_{\text{LR}}(\sigma_{\text{RL}})$ is the cross-section with 100$\%$ left polarized~(right polarized) electron beam and 100$\%$ right polarized ~(left polarized) positron beam. The contribution like $\sigma_{\text{LL}}(\sigma_{\text{RR}})$ 
can be safely dropped as they are negligible. The SM cross section and its modifications due to the anomalous couplings are shown in Fig.~\ref{crosspol} for beam polarization ($\eta_3,\xi_3)=(-0.8,+0.3)$ in the left panel and for the flipped polarization in the right panel. The anomalous couplings are chosen to be 20 TeV$^{-2}$ one at a time, keeping others to zero. We see that the flipped polarization has smaller $\sigma$ for SM, and the fractional change due to anomalous couplings is large and increases with the increase in $\sqrt{s}$. We further observe that the contribution from the $CP$-odd couplings is much smaller than the $CP$-even couplings due to the interference effect, which will be discussed later.
\section{Flavor Tagging} 
\label{flavor tagging}
\begin{figure}[!t]
	\centering
	\includegraphics[scale=0.35]{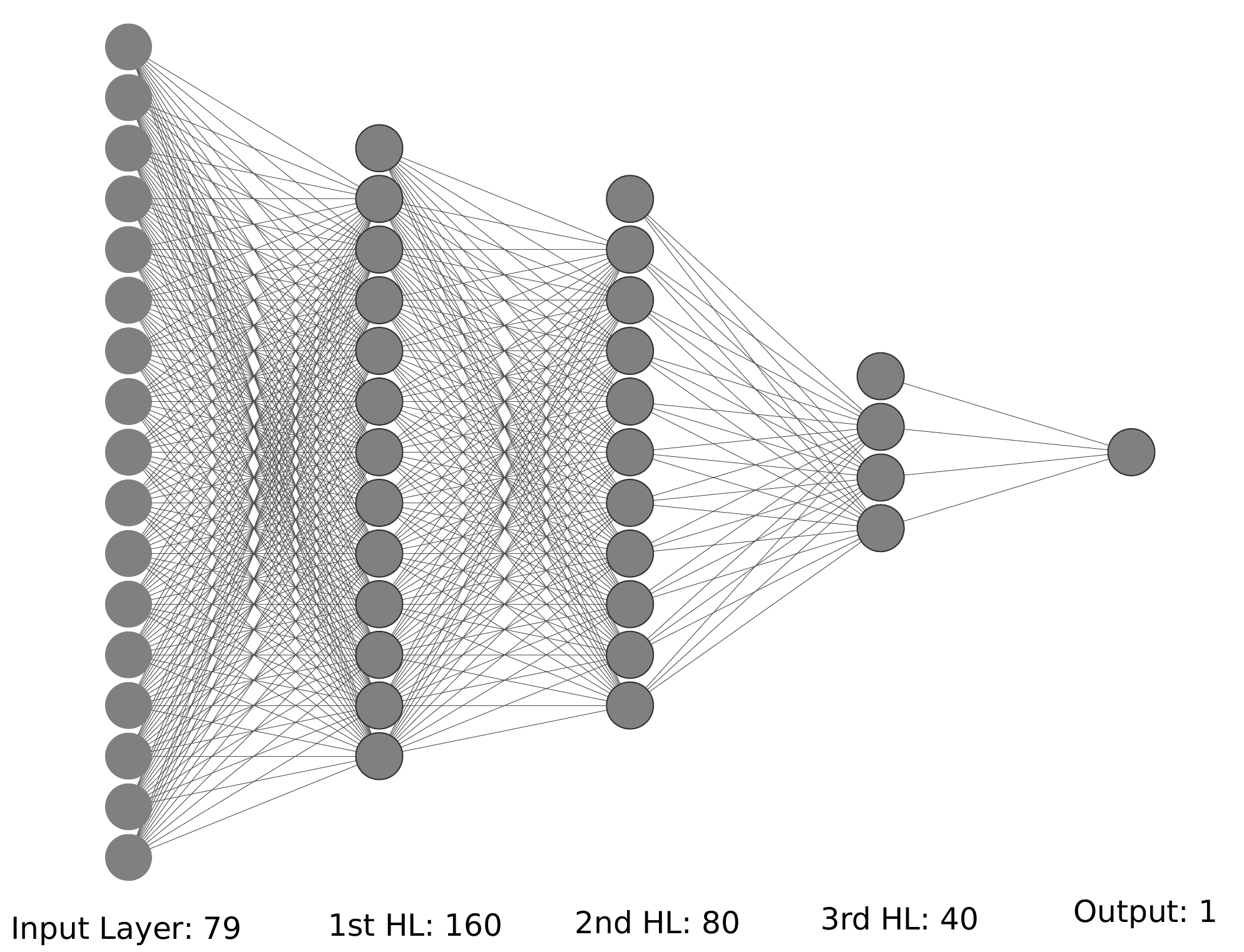}
	\caption{Architecture of ANN used for flavor tagging. It contains three hidden layers~(HL). The input layer contains 79 nodes, $1^{st}$ hidden layer with 160 nodes, $2^{nd}$ HL with 80, $3^{rd}$ HL contains 40 nodes, and the output layer has one node.}
	\label{fig:ANN-arch}
\end{figure}

As discussed in Sec.~\ref{sec:intro}, constructing some of the polarization asymmetries and spin correlation asymmetries requires identifying the daughter of $W$ boson. From Eq.~(\ref{eqn:domega}) and Table~\ref{tab:core}, we note that the correlators of vectorial polarization $\vec{P}\in \{P_x,P_y,P_z\}$ are parity odd and that of tensorial polarization, the correlators are parity even. It suggests that unless the daughter fermion is not tagged, the vectorial polarization and its related spin correlation parameters average out. Since the daughter of $W^-$ boson is a lepton, the identification becomes non-ambiguous in such a topology. Whereas, in the case of $W^+$ decaying hadronically, the identification of the jet initiator~(quark) remains fuzzy due to the close behavior of light quarks. In this article, we classify the final jets as initiated by \emph{up}-type or \emph{down}-type quarks. We used {\tt MadGraph5$\_$aMC$@$NLO}~\cite{Alwall:2014hca,Frederix:2018nkq} to generate event sets using a diagonal Cabibbo–Kobayashi–Maskawa~(CKM) matrix for training and testing purposes. The parton-level events were used as an input to the {\tt Pythia8}~\cite{Bierlich:2022pfr} for showering and hadronization of the colored partons. This process will lead to the formation of hadrons which can undergo further decay within a detector given their short lifetime. The final state, colorless particles, are clustered in a region called jets. The final state particles with $p_T\geq 0.3$ GeV and $|\eta| \le 3.0$ are selected for jet clustering using {\tt Fastjet}~\cite{Cacciari:2011ma}. The lepton from the decay of $W^-$ is excluded from jet analyses. The jet clustering is done using {\tt anti-}$k_T$~\cite{Cacciari:2008gp} algorithm with jet radius $R=0.7$, and the jets thus obtained are re-clustered using $k_T$~\cite{Catani:1993hr} algorithm with jet radius $R=1.0$. The two hardest jets account for more than $90\%$ of the parton level momentum of $W^+$ boson, and they are passed through ML models for tagging as \emph{up/down}-type. These two jets' truth labeling is done by using the distance $\Delta R_{jq}$ with the initiator quarks. In a case when both the jets become close to a single quark, the hardest jet is selected.\\\\ 
For tagging purposes, apart from the features listed in Ref.~\cite{Subba:2022czw}, additional features are obtained from the jets, which are used as input to ANN and BDT models. For a \emph{up/down} tagger, a strange tagging can enhance the overall efficiency of classification. In an event where a strange quark is present along with other light quarks, due to a comparatively longer lifetime of strange meson $K_S^0$ with $\tau = \mathcal{O}(10^{-10})$ s can decay within the detector range, one can obtain a secondary vertex resulting to displaced tracks. Similarly, the charmed mesons can provide displaced tracks provided such particles gets a significant $\beta\gamma$ factor. Such displaced tracks act as a good classifier for light quarks. A charged kaons $K^\pm$ with $\tau = \mathcal{O}(10^{-8})$ s are Collider-stable particles; hence the multiplicity and momenta of charged kaons are used as features. To use the charged kaons as a feature, it is necessary to distinguish them from charged pions, protons, and vice-versa. Many techniques have been developed to perform particle identification like using mean ionization energy loss, $dE/dx$~\cite{QUERTENMONT201195,LIPPMANN2012148,Boyarski:1989gw,Vavra:2000vag,Hauschild:1996np,Einhaus:2019jvg,Lehraus:1982em,Allison:1980vw,Adeva:1990kd}, time-of-flight~\cite{Sheaff:1999iv,DAgostini:1980jhh}.
For our analysis, we considered the ideal particle identification and used the available charged hadrons as an input to ML models. The higher multiplicity of kaons on strange jets can act as a good classifier of strange vs. other light quark jets. In general, we construct $x_i$ using the information of the jet and the objects contained within a jet. The variables we obtained to use as input to our network can be divided into two classes, discrete and continuous variables, and they are listed below.
\begin{itemize}
	\item Discrete Variables:
	\begin{itemize}
		\item Total number~({\tt nlep}), positive leptons~({\tt nl$+$}), negative leptons~({\tt nl$-$});
		\item Total number of visible particles~({\tt nvis});
		\item Total number of charged particles~({\tt nch}), positive charged particles~({\tt nch$+$}), negative charged particles~({\tt nch$-$});
		\item Total number of charged kaons~({\tt nK$+$, nK$-$})$^\star$;
		\item Total number of charged pions~({\tt npi$+$, npi$-$})$^\star$;
		\item Total number of hadrons~({\tt nhad});
		\item Total number of charged hadrons~({\tt nChad}), positively charged hadrons~({\tt nChad$+$}), negative hadrons~({\tt nChad$-$});
		\item Displaced tracks satisfying $p_T > 1.0$ GeV are used. They are binned with respect to the lifetime~($\tau$) in mm of their mother particles:
		\begin{itemize}
			\item {\tt c1}: $\tau < 3.0$ and $\tau > 0.3$,
			\item {\tt c2}: $\tau < 30.0$ and $\tau > 3.0$,
			\item {\tt c3}: $\tau < 300.0$ and $\tau > 30.0$,
			\item {\tt c4}: $\tau < 1200.0$ and $\tau > 300.0$,
			\item {\tt c5}: $\tau > 1200.0$.
		\end{itemize}
	\item Total number of $+$ve~({\tt pcl}) and $-$ve~({\tt ncl}) mother particles are also counted. The particle that decay and produces secondary displaced vertex are considered. 
		\end{itemize}
	\item Continuous Variables:
	\begin{itemize}
		\item Energy of photons~({\tt egamma})$^\star$;
		\item $P^{\mu\star}_i = \sum_{j \in i}p^\mu_j$, $ P_i^\star = \sum_{j \in i}|\vec{p}_j|$ , $|(P^\mu_i)_T|^\star$,\\ $i \in \{\text{Leptons}, K^+, K^-, \pi^+, \pi^-,K^0_L,\text{Hadrons}\}$;
        \item Energy of charged hadrons~($E^+_{\text{Had}},E^-_{\text{Had}}$). 
	\end{itemize}
\end{itemize}
The features with ($^\star$) represent the additional features on top of all the features used in Ref.~\cite{Subba:2022czw}. One can also obtain features related to the jet and use them as input to the classifying network. These variables, in general, provide a significant correlation to the jet class, which translates to increased classification accuracy. Some of the features related to the jet itself are:
\begin{itemize}
	\setlength\itemsep{0.1em}
	\item Transverse Momentum $p_T$, magnitude of momentum $|\vec{p}|$, transverse mass $m_T$ and mass of Jet;
	\item Pseudo rapidity $\eta$, azimuth angle $\phi$ of the jets;
	\item Momentum component~($p_x,p_y,p_z$).
	\end{itemize}
The jet variables like $p_T,\eta,E$, and the three-momenta have large correlation with the label of jets. These variables depend on the polarization of the mother particle, and since we are using the observable related to polarization to constrain the new physics parameters, using these variables to flavor tag will not be useful. Hence, these jet features are excluded from our inputs to train the ML models. We use the above-listed discrete and continuous variables in a concatenated form for our ANN and BDT models.\\\
\textbf{BDT}: The ML models using BDT are implemented in {\tt XGBoost}~\cite{Chen:2016:XST:2939672.2939785}. The parameter of our BDT models is as follows:
\begin{itemize}
	\setlength\itemsep{0.1em}
	\item Sub-sample ratio of columns for when constructing each tree: {\tt colsample$\_$bytree} = 0.6;
	\item Step size shrinkage used in the update to prevent overfitting: {\tt eta} = 0.3;
	\item Minimum loss reduction required to make a further partition on a leaf node of the tree: {\tt gamma} = 1.5;
	\item Maximum depth of a tree: {\tt max$\_$depth} = 6;
	\item Number of decision tree: {\tt n$\_$estimators} = 300;
	\item L2 regularization term on weights: {\tt lambda} = 1.0.
\end{itemize}
\textbf{ANN}: The ANN models are implemented using {\tt TensorFlow}. The architecture of ANN consists of three hidden layers, each consisting of 160, 80, and 40 nodes, respectively; see Fig.~\ref{fig:ANN-arch}. We used {\tt Relu} as an activation function for each hidden layer, and for the output layer, {\tt Sigmoid} function is used. The optimization is done using {\tt adam} algorithm.  
\\\\
constructing polarization parameters, which, otherwise, can increase the model's accuracy significantly.
The training datasets~($10^7$ events) and test datasets~($10^6$ events) were generated for three different beam polarization~($\eta_3,\xi_3$) $\in \{(0,0),(+0.8,-0.3),(-0.8,+0.3)\}$. We train ANN and BDT models separately for three polarization datasets and test them on all three test datasets. The final accuracy obtained using different sets of datasets~(with different polarization) are listed in Table.~\ref{tab:acc}.
We estimate the efficiency of our ML models as follows; a random subset with 60$\%$ of the test sample is taken, and we estimate the model accuracy on that subset. This process is repeated 1000 times, and the average accuracy is quoted in Table~\ref{tab:acc}.
We note that all the accuracy listed in Table~\ref{tab:acc} are comparable, i.e., one can use any of the models on any datasets with varying beam polarization~(and hence varying $W$ polarization) with comparable efficiency. For the rest of the paper, we choose BDT trained on an unpolarized beam.  
We want to highlight the fact that using the additional features~($^\star$) increases the accuracy from $\approx 70\%$ in Ref.~\cite{Subba:2022czw} to $\approx 80\%$, which should improve the constrain on anomalous couplings. In the later section, we describe the tagger's role in constraining the anomalous couplings.
\begin{table}
	\caption{\label{tab:acc}Accuracy on classifying two hard jets as up-type or down-type using different train and test models using artificial neural network~(ANN) and boosted decision tree~(BDT). The model differs in the initial degree of beam polarization.}
     \centering
     \begin{ruledtabular}
     	\begin{tabular}{rrlll}
     		ML&Train&\multicolumn{3}{c}{Test}\\
     		\cline{3-5}
     		&$(\eta_3,\xi_3)$&$(0.0,0.0)$&$(-0.8,+0.3)$&$(+0.8,-0.3)$\\
     		\hline
     		ANN&$(0.0,0.0)$&\textbf{82.0}&81.9&79.3\\
     		ANN&$(-0.8,+0.3)$&82.0&\textbf{82.3}&79.3\\
     		ANN&$(+0.8,-0.3)$&81.7&81.9&\textbf{80.0}\\   
     		BDT&$(0.0,0.0)$&\textbf{81.0}&80.5&77.5\\
     		BDT&$(-0.8,+0.3)$&80.1&\textbf{79.5}&78.6\\
     		BDT&$(+0.8,-0.3)$&80.1&79.5&\textbf{78.6}\\  		
     	\end{tabular}
     \end{ruledtabular}	
\end{table} 
\section{Parameter Estimation} 
\label{parameter estimation}
In order to constrain the new physics parameters effectively, it is advantageous to utilize a wide range of observables that are sensitive to such phenomena. This article employs cross~section, polarization, and spin correlation asymmetries as the observables of interest. With two spin-1 $W$ bosons, there exist 80 spin-related observables, of which 16 are polarization, and the rest 64 are spin correlation asymmetries. To further enhance the sensitivity to new physics, a division of these observables into intervals of $\cos\theta$ is proposed, where $\theta$ is the production angle of $W^-$ boson in the lab frame. It is motivated by the fact that due to the chiral couplings, the polarization and spin correlations depend on $\cos\theta$. By dividing the backward region~($\cos\theta \le 0.0$) into four equal intervals, which corresponds to the region where the new physics contributions are maximal, the statistical error dominance, which could be substantial due to the lower rate in this region, can be mitigated. In contrast, the forward region exhibits a significantly higher rate statistically and, therefore, can be divided into finer bins. However, to maintain uniformity, the forward region is also divided into four equal bins of $\cos\theta$. This binning approach allows for the identification of specific regions that possess greater sensitivity than the case without binning, leading to an overall increase in sensitivity when the contributions from all these bins are combined. A detailed demonstration of the enhanced sensitivity achieved through this binning scheme is provided in Appendix~\ref{app:binwise}. With this kind of binning, we would have 648 different observables. The value of these observables in each bin is obtained for a set of couplings. Then those are used for numerical fitting to obtain a semi-analytical expression of all the observables as a function of the couplings. For cross-section, which is a $CP$-even observable, the following parameterization is used to fit the data,
\begin{equation}
    \begin{split}		
        \sigma(\{c_i\}) = &\sigma_0 + \sum_{i=1}^3 c^i\sigma_i + \sum_{i=1}^5 c_i^2\sigma_{ii} \\&+ \frac{1}{2}\sum_{i,j(\neq i)=1}^3 c_ic_j\sigma_{ij} + c_4c_5\sigma_{45}.
    \end{split}
    \label{cpevenfit}
\end{equation}
For the asymmetries, the denominator is the cross-section, and the numerator $\Delta\sigma\{c_i\} = A\{c_i\}\sigma$ is parameterized separately. For the $CP$-even asymmetries the parameterization of $\Delta\sigma$ is same as in Eq.~(\ref{cpevenfit}) and for $CP$-odd asymmetries it is done as,
\begin{equation}
	\label{cpoddfit}
	\Delta\sigma(\{c_i\}) = \sum_{i=4}^5 c_i\sigma_i + \sum_{i=1}^3 c_ic_4\sigma_{i4} + \sum_{i=1}^3 c_ic_5\sigma_{i5}.
\end{equation}
Here, $c_i$ denotes the five couplings of the dimension-6 operators $c_i=\{c_{WWW},c_W,c_B,c_{\widetilde{W}},c_{\widetilde{WWW}}\}$. We define $\chi^2$ distance between the SM and SM plus anomalous point as,
\begin{equation}
	\label{eqn:chi}
	\chi^2(\{c_i\}) = \sum_k \sum_l \left(\frac{\mathscr{O}_k^l(\{c_i\}) - \mathscr{O}_k^l(0)}{\delta \mathscr{O}^l_k}\right)^2,
\end{equation}
where $k$ and $l$ corresponds to observable and bins respectively and $c_i$ denotes some non-zero anomalous couplings. The denominator $\delta \mathscr{O} = \sqrt{(\delta \mathscr{O}_{stat})^2 + (\delta \mathscr{O}_{sys})^2}$ is the estimated error in $\mathscr{O}$. If an observable is asymmetries $A = \frac{N^+-N^-}{N^++N^-}$, the error is given by \begin{equation}
	\delta A = \sqrt{\frac{1-A^2}{\mathcal{L}\sigma}+\epsilon_{A}^2}
\end{equation}
where $N^++N^- = \mathcal{L}\sigma,$ $\mathcal{L}$ being the integrated luminosity of the collider. The error in the cross-section $\sigma$ is given by 
\begin{equation}
	\delta \sigma = \sqrt{\frac{\sigma}{\mathcal{L}}+(\epsilon_\sigma \sigma)^2}.
\end{equation}
Here, $\epsilon_{A}$ and $\epsilon_{\sigma}$ are the fractional systematic error in asymmetries~(A) and cross-section~($\sigma$) respectively. The benchmark systematic errors chosen in our analysis are:
\begin{equation}
	(\epsilon_A, \epsilon_{\sigma}) \in \{(0,0),(0.25\%,0.5\%),(1\%,2\%)\}.
	\label{syst}
\end{equation}
We perform our analysis at $\sqrt{s}= 250$~GeV and different values of integrated luminosity,
\begin{equation}
	\mathcal{L} \in \{100\text{ fb}^{-1},250\text{ fb}^{-1},1000\text{ fb}^{-1},3000\text{ fb}^{-1}\}.
	\label{lumi}
\end{equation}
The SM cross~sections for the process given in Eq.~(\ref{process}) with initial beam polarizations of $(0,0),(+0.8,-0.3),(-0.8,+0.3)$ are 2.347~pb, 0.396~pb, and 5.678~pb, respectively. With these values of cross-sections, the relative statistical error for the chosen set of luminosities is:
\begin{equation}
\label{eqn:estat}
	\begin{split}
		\frac{\delta\sigma_{(0,0)}}{\sigma_{(0,0)}} = \{0.2\%,0.1\%,0.06\%,0.03\%\}, \\
		\frac{\delta\sigma_{(+,-)}}{\sigma_{(+,-)}} = \{0.5\%,0.3\%,0.015\%,0.09\%\},\\
		\frac{\delta\sigma_{(-,+)}}{\sigma_{(-,+)}} = \{0.1\%,0.08\%,0.04\%,0.02\%\}.
	\end{split}
\end{equation}
For the unpolarized case, the number corresponds to luminosity $\mathcal{L}$ listed in Eq.~(\ref{eqn:estat}), while the polarized cases stand for luminosity $\mathcal{L}$/2 each.

It has been shown in Ref.~\cite{Rahaman:2019mnz} that combining the two opposite beam polarization at the level of $\chi^2$ provides tighter constraints on anomalous couplings than combining the observable algebraically. The combined $\chi^2$ at different set of beam polarization~($\pm\eta_3,\mp\xi_3$), and using different observable $\mathcal{O}$ for a given value of anomalous couplings $c$ is defined as,
\begin{equation}
	\begin{aligned}
		&\chi^2(\mathcal{O},c_i,\pm\eta_3,\mp\xi_3) = \\ &\sum_{l,k} \left[\left(\frac{\mathscr{O}_k^l(c_i,+\eta_3,-\xi_3) - \mathscr{O}_k^l(0,+\eta_3,-\xi_3)}{\delta \mathscr{O}^l_k(0,+\eta_3,-\xi_3)}\right)^2 \right.\\ &\left.+  \left(\frac{\mathscr{O}_k^l(c_i,-\eta_3,+\xi_3) - \mathscr{O}_k^l(0,-\eta_3,+\xi_3)}{\delta \mathscr{O}^l_k(0,-\eta_3,+\xi_3)}\right)^2 \right],
	\end{aligned}
	\label{chisum}
\end{equation}
where $k$ and $l$ correspond to different observables and bins.\\ \\
We perform $\chi^2$ analysis by varying one parameter and two parameters at a time which will be described below. We also perform a set of Markov-Chain-Monte-Carlo~(MCMC) analyses with a different set of observable with polarized beams to obtain simultaneous limits on the anomalous couplings. 
\subsection{One Parameter Estimation}
\begin{figure}[!h]
	\centering
	\subfigure{\includegraphics[width=3.8cm,height=3.8cm]{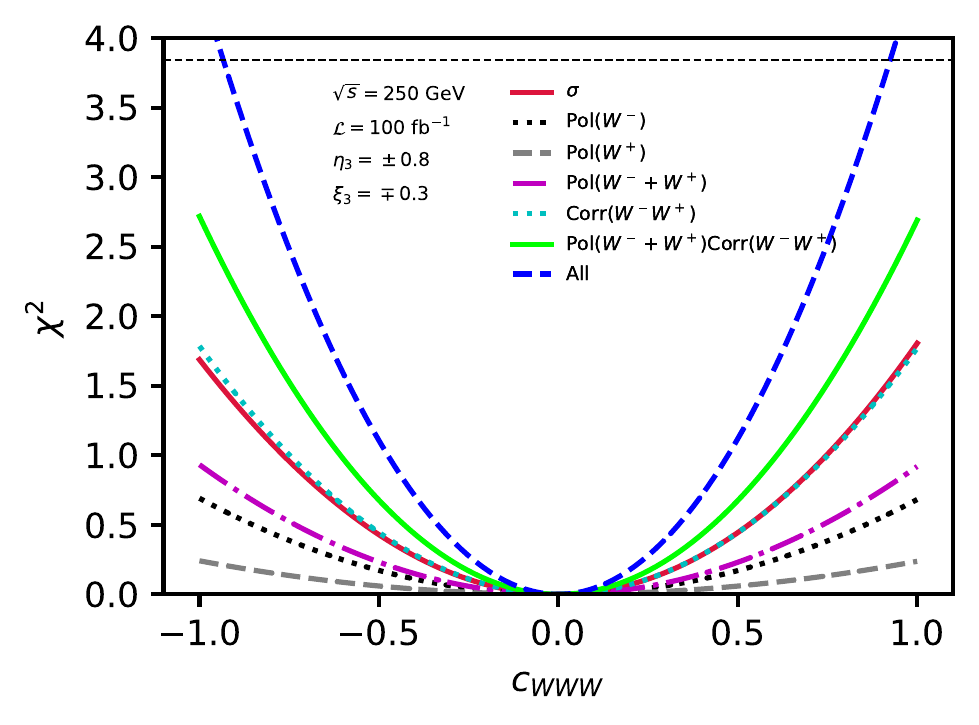}}
	\subfigure{\includegraphics[width=3.8cm,height=3.8cm]{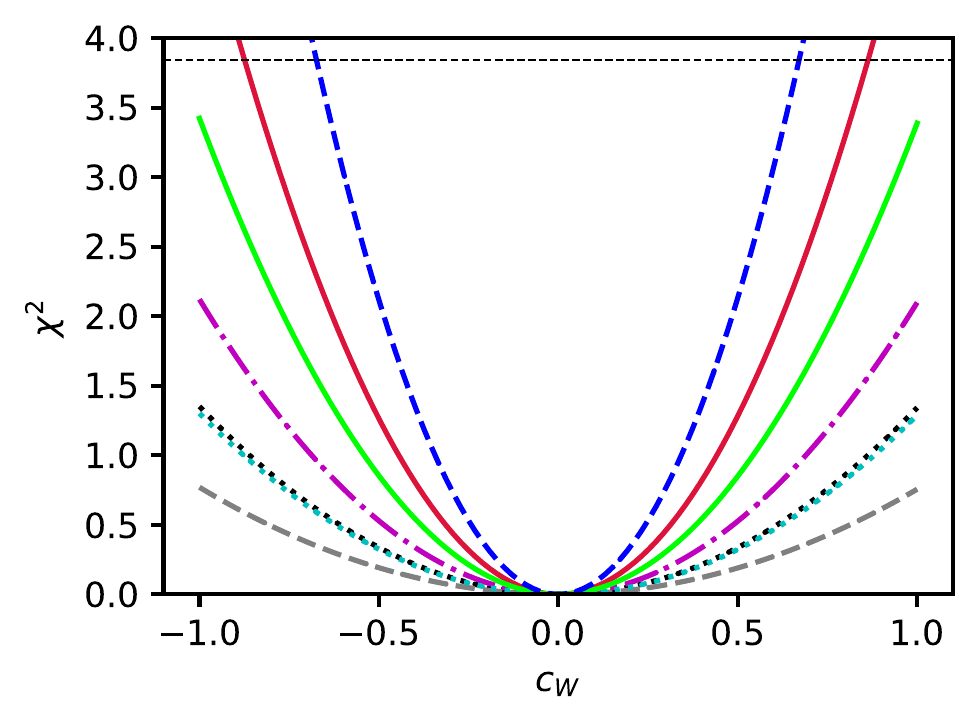}}
	\subfigure{\includegraphics[width=3.8cm,height=3.8cm]{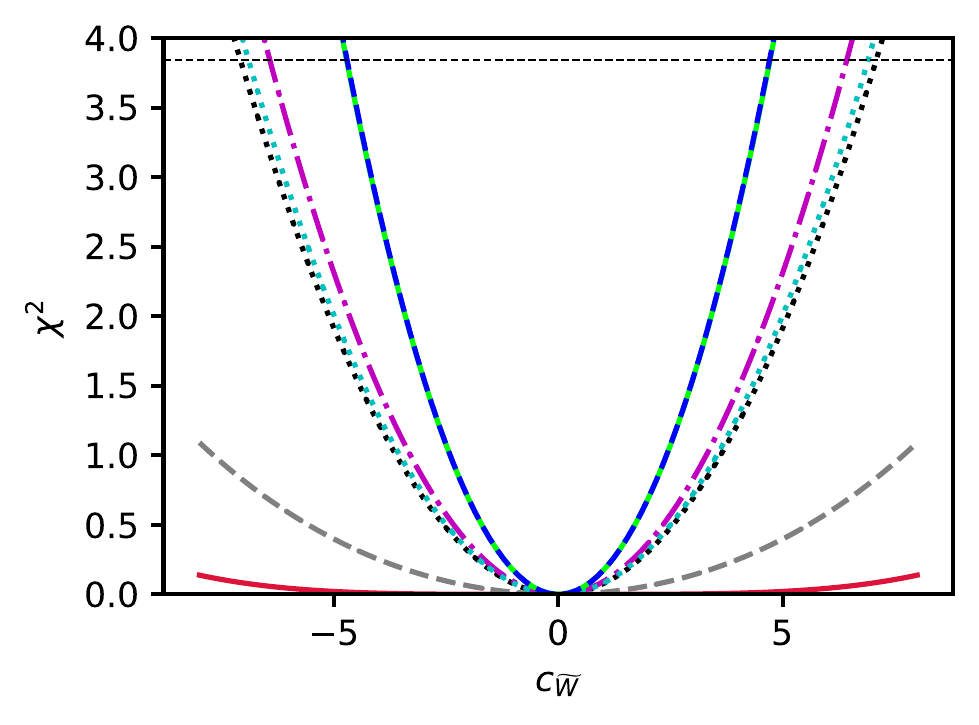}}
	\subfigure{\includegraphics[width=3.8cm,height=3.8cm]{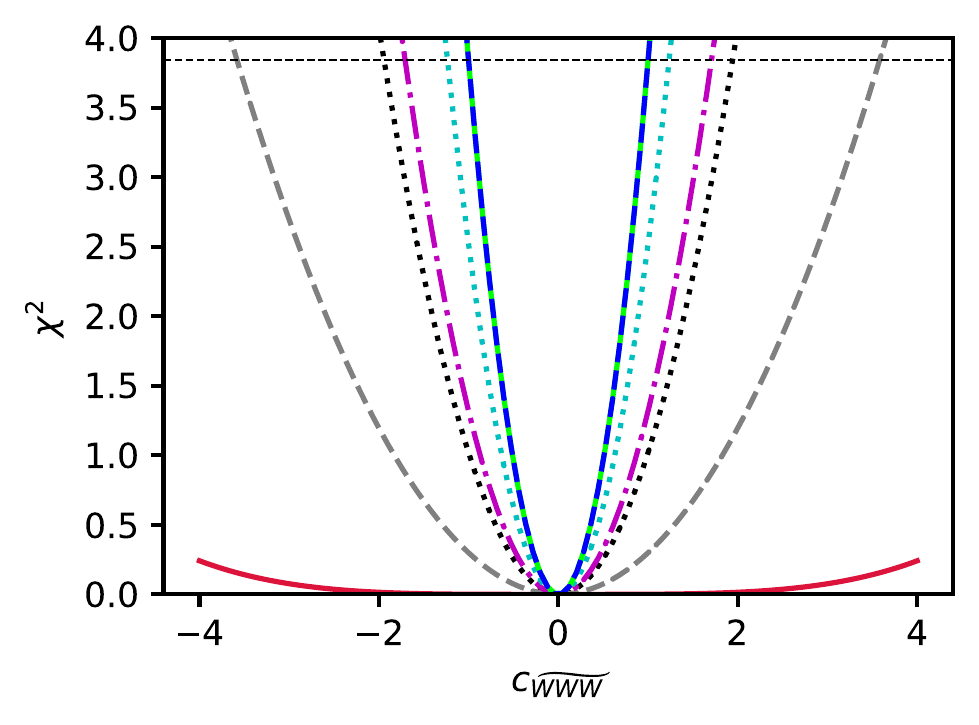}}
	\caption{\label{onedimplot}Variation of $\chi^2$ for a set of observables as a function of one anomalous coupling at a time. The systematic errors are chosen to be zero.}
\end{figure}

This section discusses the analysis done by varying one anomalous coupling $c_i$ at a time while others are kept at zero. The analysis are done at the center-of-mass energy, $\sqrt{s} = 250$ GeV, integrated luminosity, $\mathcal{L} = 100$ fb$^{-1}$ with zero systematic errors. The polarized beams are used with a degree of polarization $(+0.8,-0.3)$ and $ (-0.8,+0.3)$. 
Variation of $\chi^2$ for different sets of observables as a function of one anomalous coupling is shown in Fig.~\ref{onedimplot}. The different sets of observables are: cross~section~$\sigma$, set of $W^-$ asymmetries~Pol($W^-$), set of $W^+$ asymmetries~Pol($W^+$), union of $W^-$ and $W^+$ asymmetries~Pol($W^-+W^+$), spin correlations~Corr($W^-W^+$), union of all asymmetries~Pol($W^-+W^+$)~$\cup$~Corr($W^-W^+$) and set of all observables. 
The horizontal line in all panels represents $\chi^2$ at 95$\%$ confidence level~(CL).
From the bottom row of Fig.~\ref{onedimplot}, it is evident that the cross~section~(shown in the red curve) provides the least contribution to $\chi^2$ in the case of $CP$-odd couplings, i.e., $c_{\widetilde{W}}$ and $c_{\widetilde{WWW}}$. This behavior can be explained by the fact that in case of cross~section or any other $CP$-even observables, the contribution from $CP$-odd anomalous couplings comes only in a quadratic form or $1/\Lambda^4$ term, which naturally becomes tiny. Generally, cross~section, in the presence of one anomalous coupling $c_i$, can be parameterized as:
\begin{equation}
	\label{xsecPar}
	\sigma(c_i) = \sigma_0 + \sigma_i\times c_i + \sigma_{ii}\times c_i^2.
\end{equation}
Thus, the absence of a linear term in case of $CP$-odd couplings $ c_i~\in~\{c_{\widetilde{W}},c_{\widetilde{WWW}}\}$ suggests that the significant contribution will only arise when large values of anomalous couplings are taken. Whereas, in the case of $CP$-even couplings, the cross~section provides a reasonable contribution due to linear term, see top row of Fig.~\ref{onedimplot}.
The sensitivity of the polarization asymmetries constructed from the $W^-$ boson is comparatively more prominent than that of the $W^+$ boson because the accuracy with which we can tag the daughter lepton of $W^-$ boson is close to $100\%$. While, as discussed in Sec.~\ref{flavor tagging}, the machine learning~(ML) models achieved an approximate accuracy of $80\%$ in classifying $up$-type versus $down$-type jets from $W^+$ boson. In our specific scenario, these jets are initiated by light quarks, resulting in similar overall hadronization and jet contents. Consequently, distinguishing between them becomes challenging for the tagging process, and this would dilute the overall sensitivity of polarization asymmetries related to the $W^+$ boson. For the $CP$-even parameters, the contribution from the spin correlations is comparable to the $W^-$ polarization for $c_W$ and to the cross~section in the case of $c_{WWW}$. On the other hand, for the $CP$-odd parameters, the spin correlations perform better than polarizations of either $W$ bosons. 
We list the one parameter limits on all anomalous couplings $c_i$ obtained using all observables in Table~\ref{onepara95ci}. 
\begin{table}[!h]
	\centering
	\caption{\label{onepara95ci}The list of constraints on five anomalous couplings $c_i$~(TeV$^{-2}$) at $95\%$ confidence level obtained by varying one parameter at a time and keeping the other at zero. The limits are obtained for $\sqrt{s}=250$ GeV, integrated luminosity $\mathcal{L}=100$ fb$^{-1}$ and initial beam polarization of $\eta_3=\pm0.8$, $\xi_3=\pm0.3$. The systematic errors are kept to zero. }
	\begin{ruledtabular}
		\label{oneparaLim}
		\begin{tabular}{ll}
			Parameters~($c_i^{\mathscr{O}}$)&Limits~(TeV$^{-2}$) \\ \hline
			$c_{WWW}/{\Lambda^2}$&$\left[-0.92,+0.92\right]$\\
			$c_W/{\Lambda^2}$&$\left[-0.67,+0.67\right]$\\
			$c_B/{\Lambda^2}$&$\left[-1.46,+1.46\right]$\\
			$c_{\widetilde{W}}/{\Lambda^2}$&$\left[-4.62,+4.62\right]$\\
			$c_{\widetilde{WWW}}/{\Lambda^2}$&$\left[-1.00,+1.00\right]$\\
		\end{tabular}
	\end{ruledtabular}
\end{table}
The confidence interval obtained for various anomalous coupling $c_i \in \{c_{W},c_{B},c_{\widetilde{W}}\}$ are significantly tighter than that of the experimental confidence interval listed in Table~\ref{tab:constraint}. In contrast, for $c_{WWW}$ and $c_{\widetilde{WWW}}$, the limits remain comparable. Next, we discuss the significance of various observable on constraining the limits of two anomalous couplings at a time. 
\subsection{Two parameter estimation}  
\begin{figure}[!h]	
	\includegraphics[width=3.8cm,height=3.8cm]{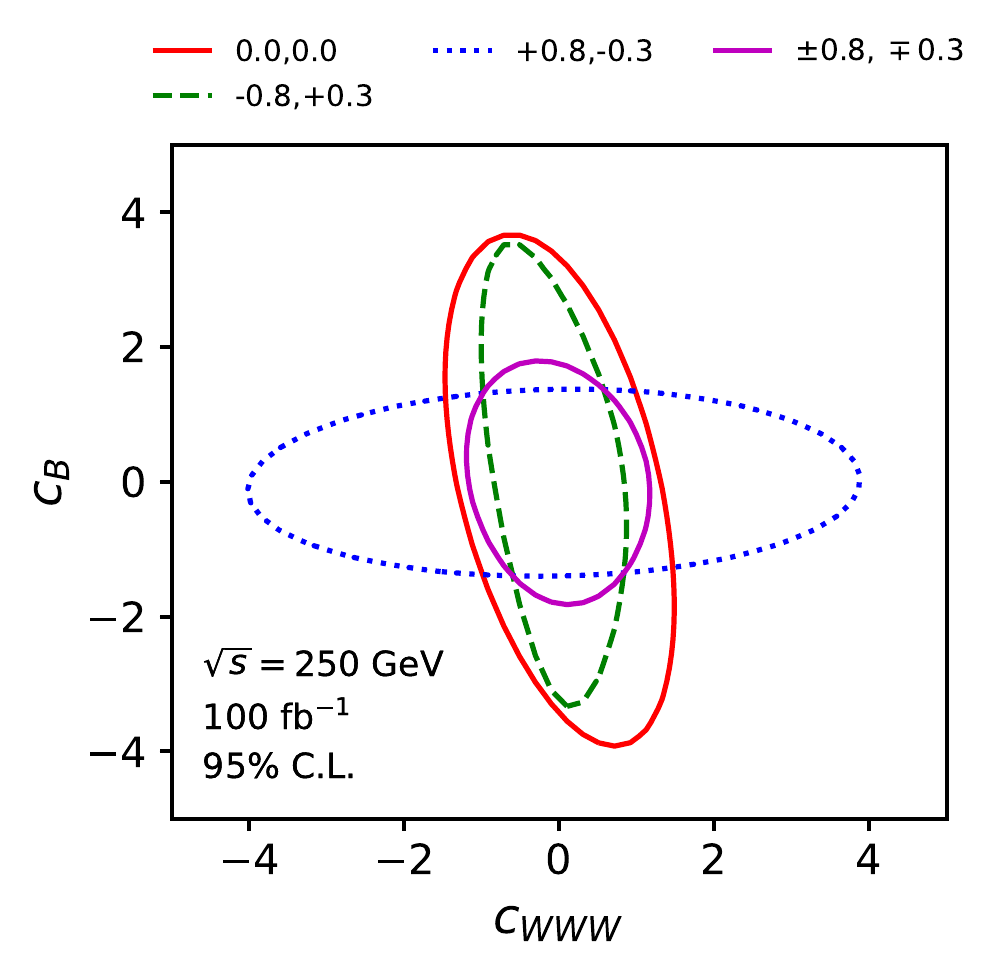}
	\includegraphics[width=3.8cm,height=3.8cm]{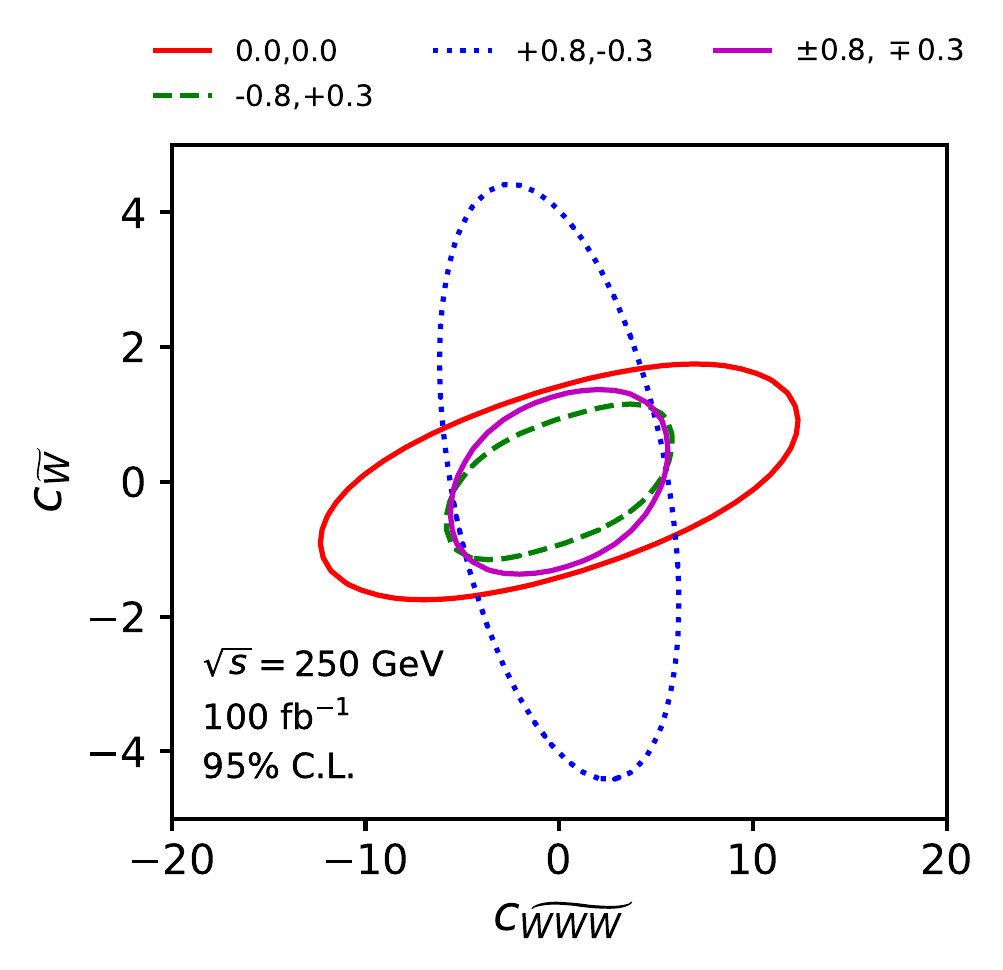}
        \centering
	\caption{\label{fig:polvsunpolchi}Two dimensional $95\%$ CL contours for combining all observables as a function of two anomalous couplings at a different set of beam polarization. The analysis is done with zero systematic errors.}
\end{figure}

\begin{figure}[!h]
	\centering
	\subfigure{\includegraphics[width=3.8cm,height=3.8cm]{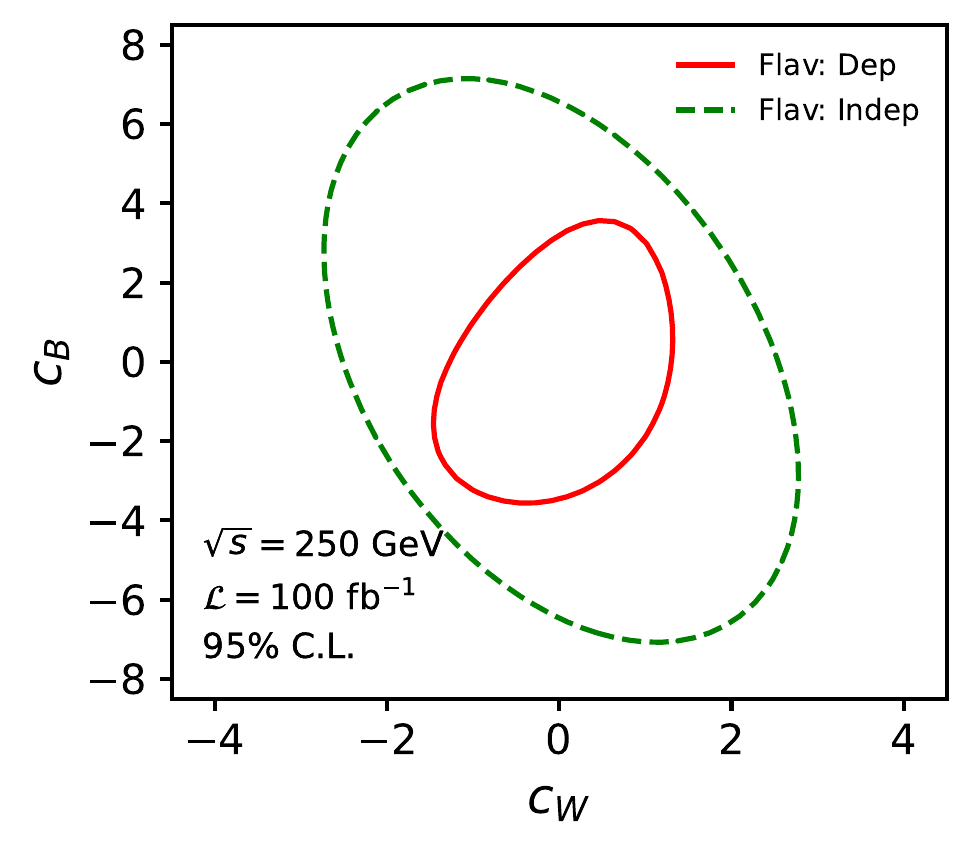}}
	\subfigure{\includegraphics[width=3.8cm,height=3.8cm]{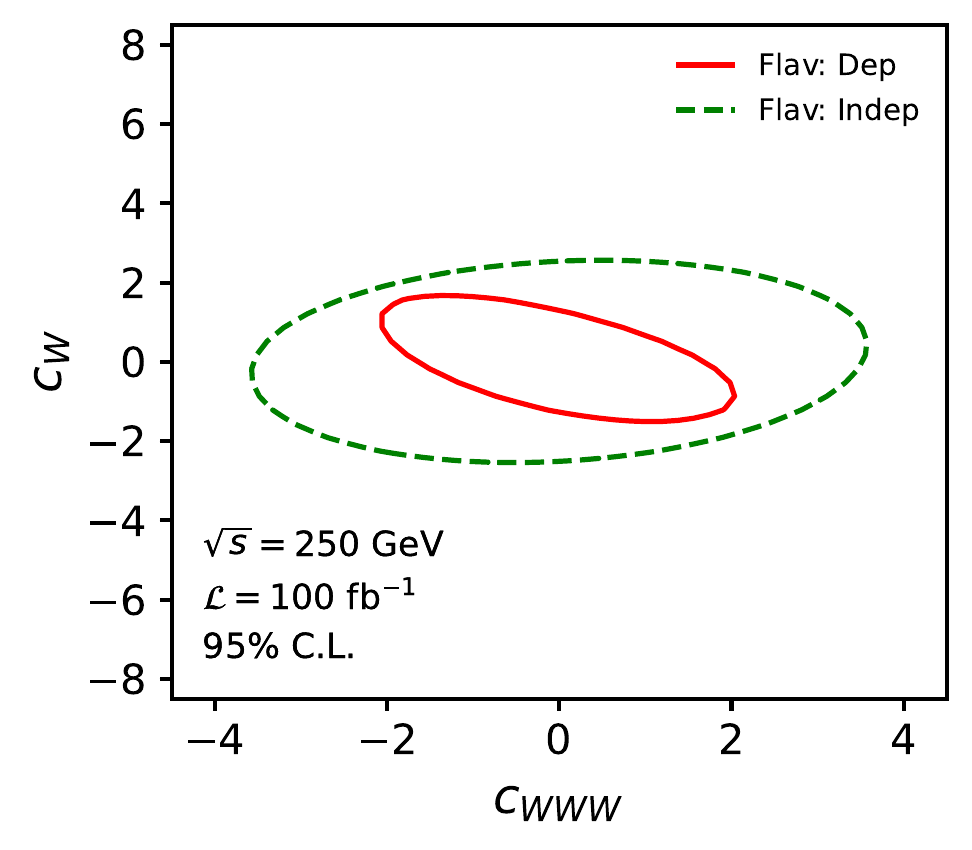}}
	\caption{\label{fig:flavdepvsindep}Two dimensional $95\%$ CL contour for combining all flavor-dependent and flavor-independent asymmetries as a function of two anomalous couplings. The systematic errors are kept at zero.}
\end{figure}

Here, we vary two anomalous couplings simultaneously while keeping all others to zero. The systematic error is taken to be zero, and luminosity is set to 100 fb$^{-1}$ with the center-of-mass energy, $\sqrt{s}$ of 250~GeV in the presence of polarized beams.
\begin{figure}[!h]
	\centering
        \subfigure{\includegraphics[width=4.2cm,height=3.9cm]{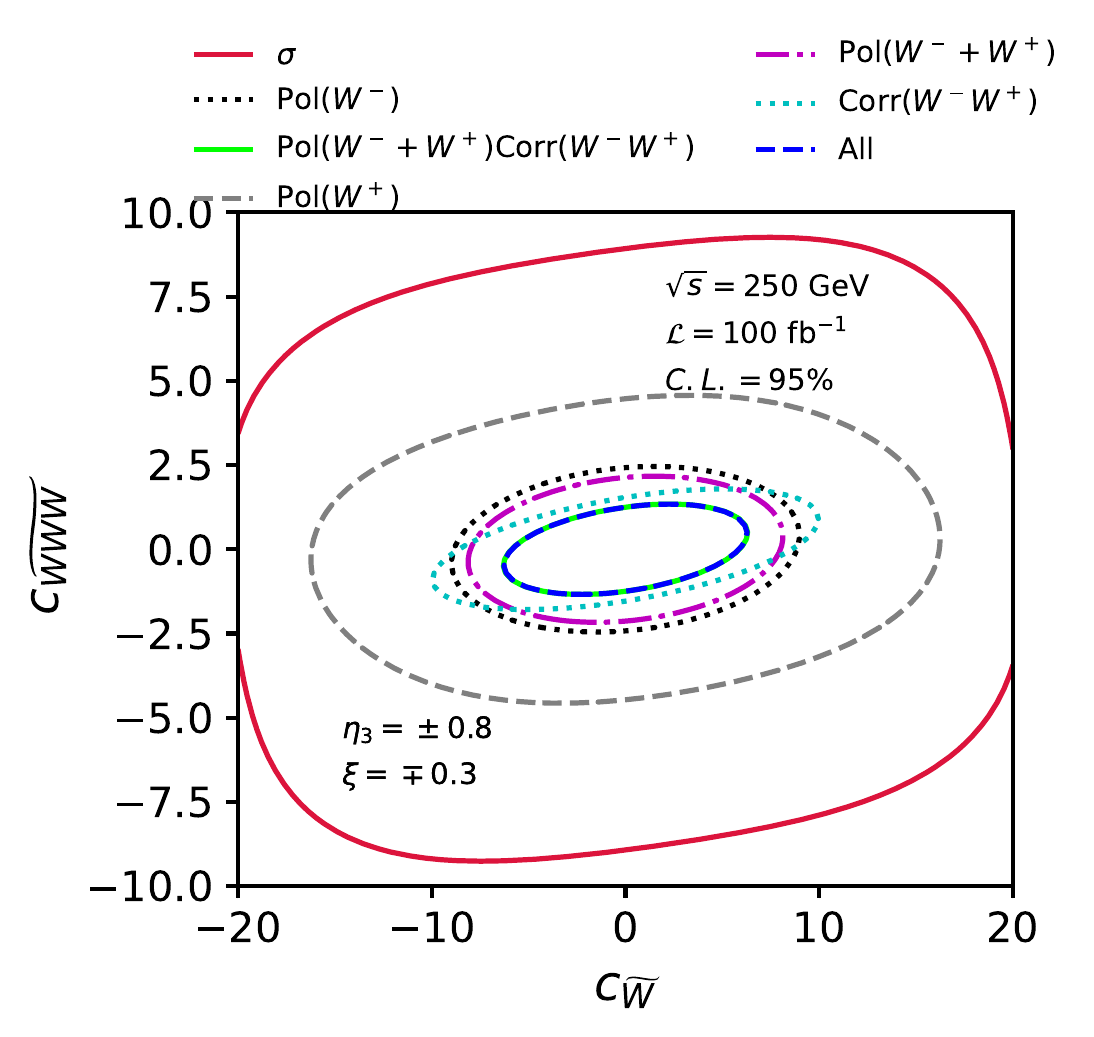}}
	\subfigure{\includegraphics[width=3.6cm,height=3.32cm]{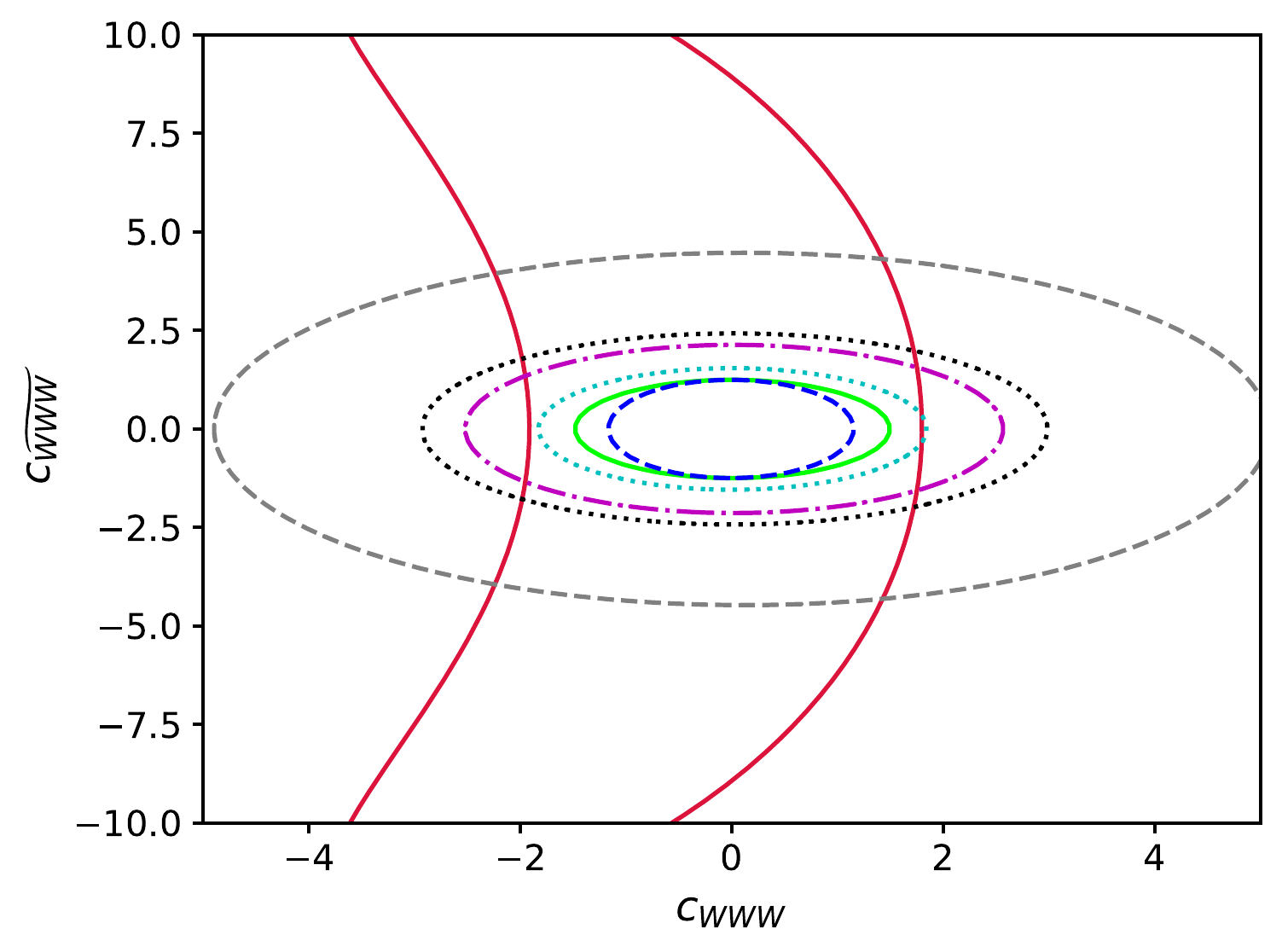}}
	\subfigure{\includegraphics[width=3.8cm,height=3.32cm]{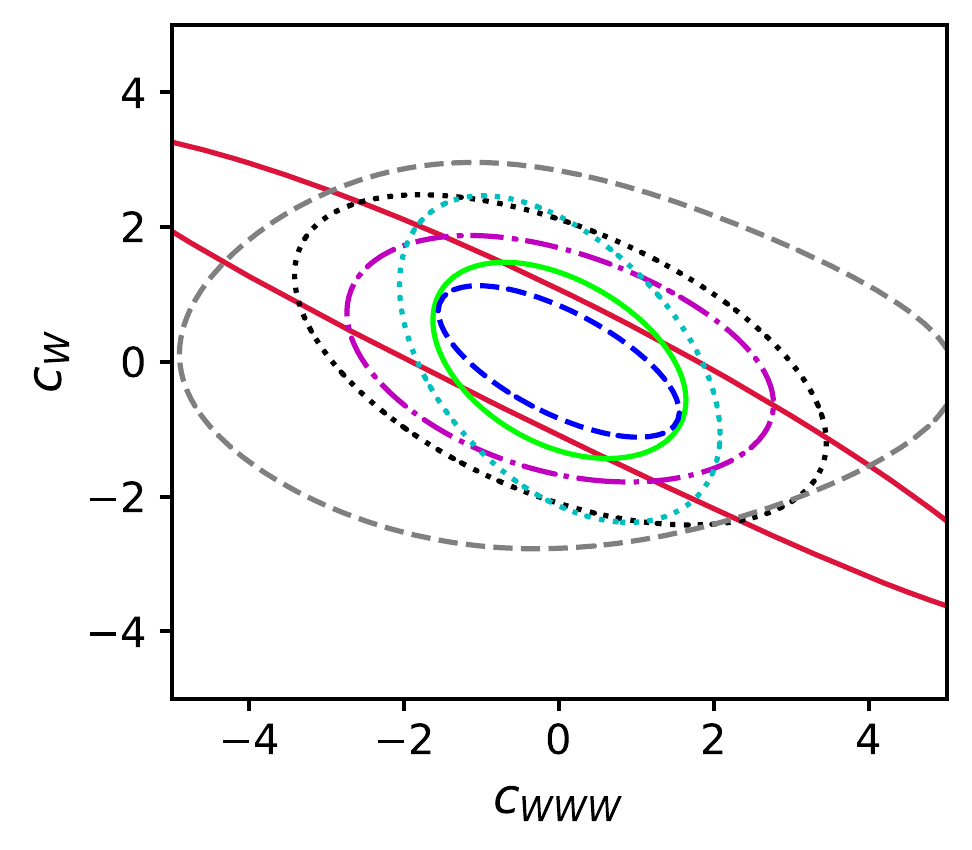}}
	\subfigure{\includegraphics[width=3.8cm,height=3.32cm]{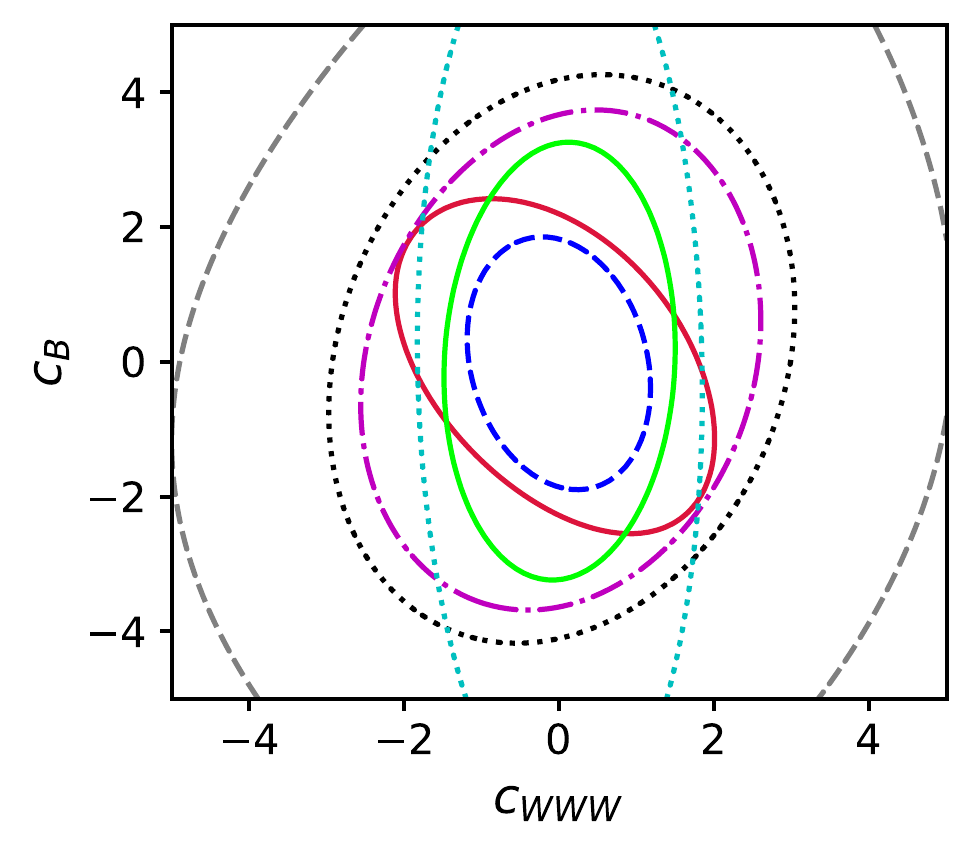}}
	\caption{\label{twoparachi}Two dimensional 95$\%$ CL contours for a set of observables as a function of two anomalous couplings at a time. The legend for all the panels follows the upper row first panel. The contours are shown for $\sqrt{s} = 250$ GeV, $\mathcal{L} = 100$ fb$^{-1}$ and zero systematic errors.}
\end{figure} \\
We study the role of various observables in setting simultaneous limits on different pairs of anomalous couplings. The observables are described in the previous section. To understand the role of these observables, we show two dimensional 95$\%$ CL contours for different pairs of anomalous couplings in Fig.~\ref{twoparachi}. For a case when both the parameters are $CP$-odd, cross~section provides the poorest bounds on both the axis~(see left panel top row) due to a tiny contribution from $1/\Lambda^4$ term. While for a case when both the parameters are $CP$-even, cross~section can be parameterized as,
\begin{equation}
	\begin{aligned}
		\sigma(c_i,c_j) &= \sigma_0 + \sigma_i \times c_i +\sigma_j\times c_j + \sigma_{ii}\times c_i^2 \\
		&+\sigma_{jj}\times c_j^2 + \sigma_{ij}\times c_ic_j.
	\end{aligned}
		\label{xsec}
\end{equation}
The cross~section provides tighter limits in the orthogonal direction and extended limits in the second and fourth quadrants. It is due to the cancellation of the linear terms in Eq.~(\ref{xsec}), i.e., $c_i\sigma_i+c_j\sigma_j = 0$, or $c_i\sim -c_j\sigma_j/\sigma_i$. The polarization of $W^-$ contributes more to $\chi^2$ than that of $W^+$ boson due to the reconstruction error of $W^+$ boson as seen in the 1-D case. The bounds obtained using spin correlations alone are tighter than that of a combination of polarization of $W$ boson in case of $(c_{WWW},c_{\widetilde{WWW}})$ pair. In the case of the first panel, spin correlation alone provides a maximal contribution to $\chi^2$ along the y-axis, i.e., $c_{\widetilde{WWW}}$ .\\\\
Below, we discuss the impact of beam polarization, flavor-dependent asymmetries, and tagger efficiency on determining the limits on anomalous couplings separately.\\
\textbf{Role of beam polarization:} We choose two sets of beam polarization for electron and positron beams, $(-0.8,+0.3)$ and $(+0.8,-0.3)$. Fig.~\ref{fig:polvsunpolchi} presents the 95$\%$ CL $\chi^2$ contours for a combination of all observables as a function of two anomalous couplings. The contours correspond to each beam polarization and unpolarized and combined cases. For individual set of beam polarization, $\chi^2$ is obtained at luminosity $\mathcal{L}=100$~fb$^{-1}$, while for a combined case each contribution is computed at 50~fb$^{-1}$.
 \\
The contour due to the two sets of opposite beam polarization provides directional limits on anomalous couplings. The intersection of these two contours corresponds to the bounds on the anomalous couplings obtained by combining the two sets of beam polarization~(see magenta curve). In case of ($c_{WWW},c_B$) pair, $(-0.8,+0.3)$ provides tighter limits on $c_{W}$ and loose bounds on $c_{WWW}$, while with $(+0.8,-0.3)$ limits on $c_W$ becomes loose and $c_{WWW}$ gets tighter. Moreover, on combining two beam polarization, the overall limits get tighter, which is also seen in $(c_{\widetilde{WWW}},c_{\widetilde{W}})$ pair. The contour without beam polarization is displayed in red and is wider than the one created by adding two sets of beam polarization. It emphasizes the importance of using polarized beams when investigating beyond the SM physics. Next, the contrast between the flavor-dependent and -independent asymmetries is emphasized.\\\\
\textbf{Comparison between Flavor dependent and independent observable:} As discussed in Sec.~\ref{flavor tagging}, some of the asymmetries are flavor-dependent, i.e., the value of those observables averages out unless the daughter of $W$ bosons is tagged. In each bin, there are 45 different flavor-dependent and 35 flavor-independent asymmetries. Here, we compare the role of these two different sets of asymmetries in constraining the anomalous couplings. For this, we show $95\%$ CL contours for all observables as a function of two anomalous couplings in Fig.~\ref{fig:flavdepvsindep}. We choose a pair ($c_W,c_B$) and ($c_{WWW},c_W$) for graphical demonstration. The contours are obtained by combining two sets of beam polarization at $\mathcal{L}=50$~fb$^{-1}$ each. In both panels, the limits set by flavor-dependent~(red curve) are tighter along both axes than those obtained by flavor-independent~(green curve). The 45 distinct flavor-dependent asymmetries thus make up most of the $\chi^2$ contribution in the case of spin-related observables. It strongly advises the development of taggers or machine learning models that are incredibly efficient. To better understand the direct impact of taggers on limiting anomalous couplings, we will compare two machine learning models with varying degrees of efficiency.
 \\\\  
\textbf{Role of Tagger Efficiency:} To understand the role of the efficiency of the tagger in constraining the anomalous couplings, we compare two different BDT models with different accuracy. One BDT model is trained using all the features listed in Sec.~\ref{flavor tagging}, and another model is trained using features apart from those listed as additional features.
\begin{figure}[!h]
	\centering
	\subfigure{\includegraphics[width=3.8cm,height=3.8cm]{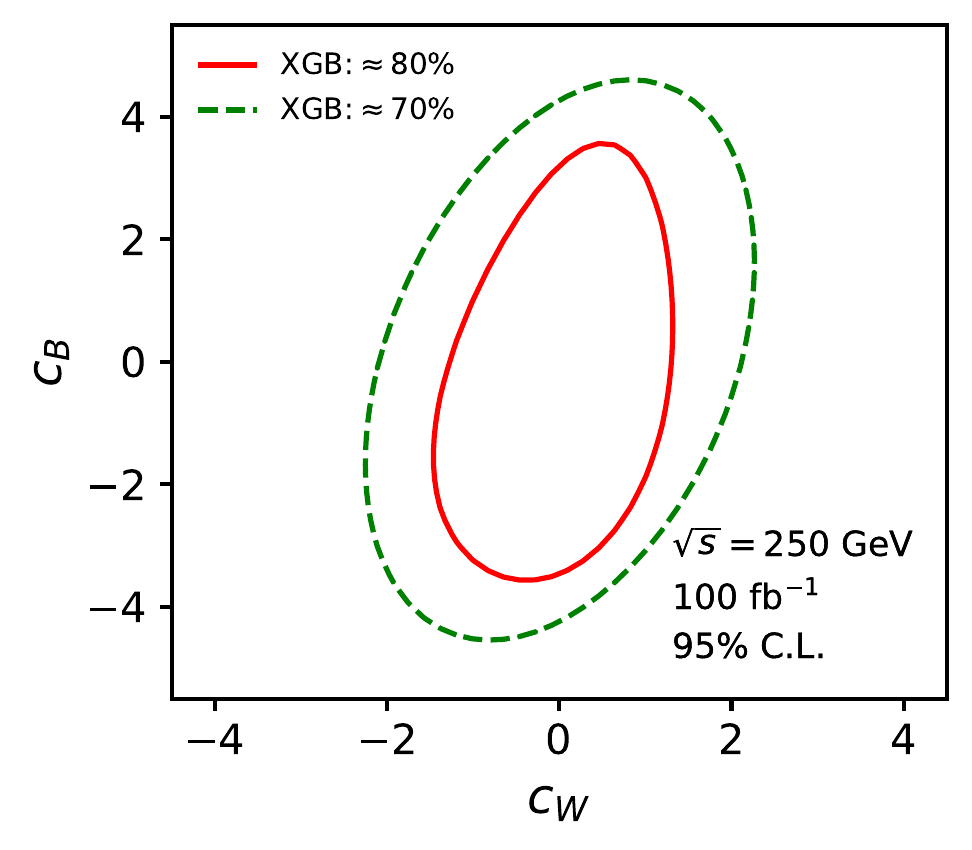}}
	\subfigure{\includegraphics[width=3.8cm,height=3.8cm]{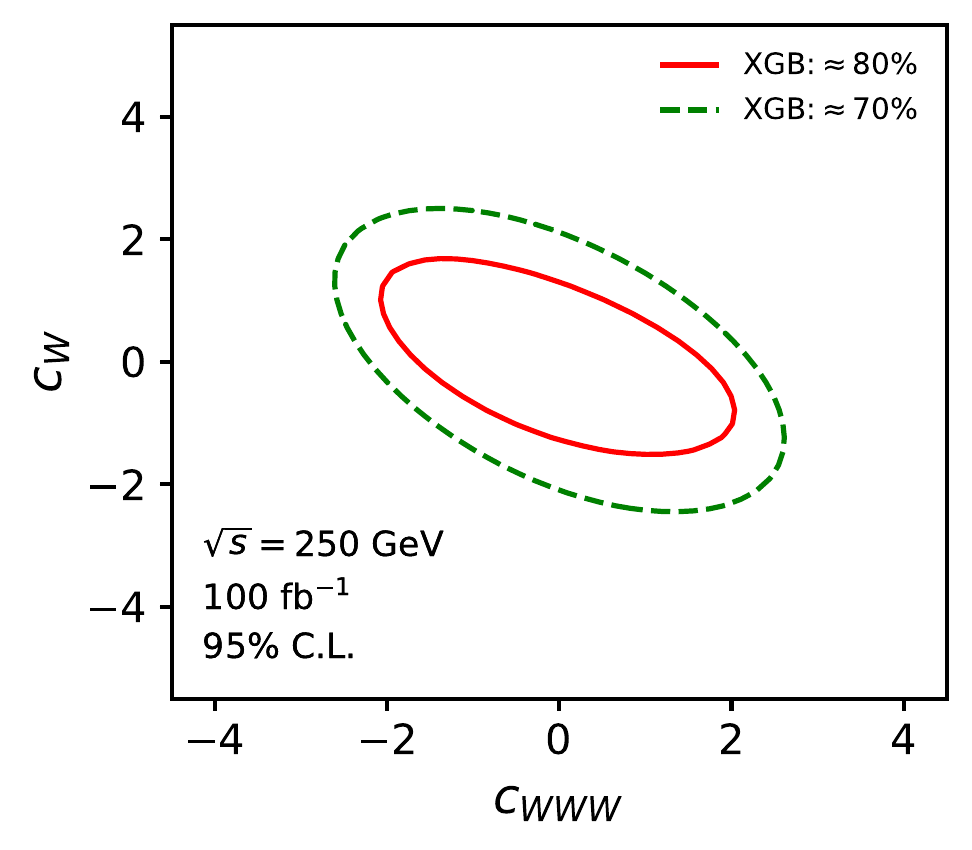}}
	\caption{\label{fig:newvsoldXGB}Two dimensional $\chi^2$ for a combination of all flavor-dependent observable as a function of two anomalous couplings. The systematic errors are kept at zero.}
\end{figure} 
\begin{figure*}[!htb]
	\centering
	\includegraphics[scale=0.48]{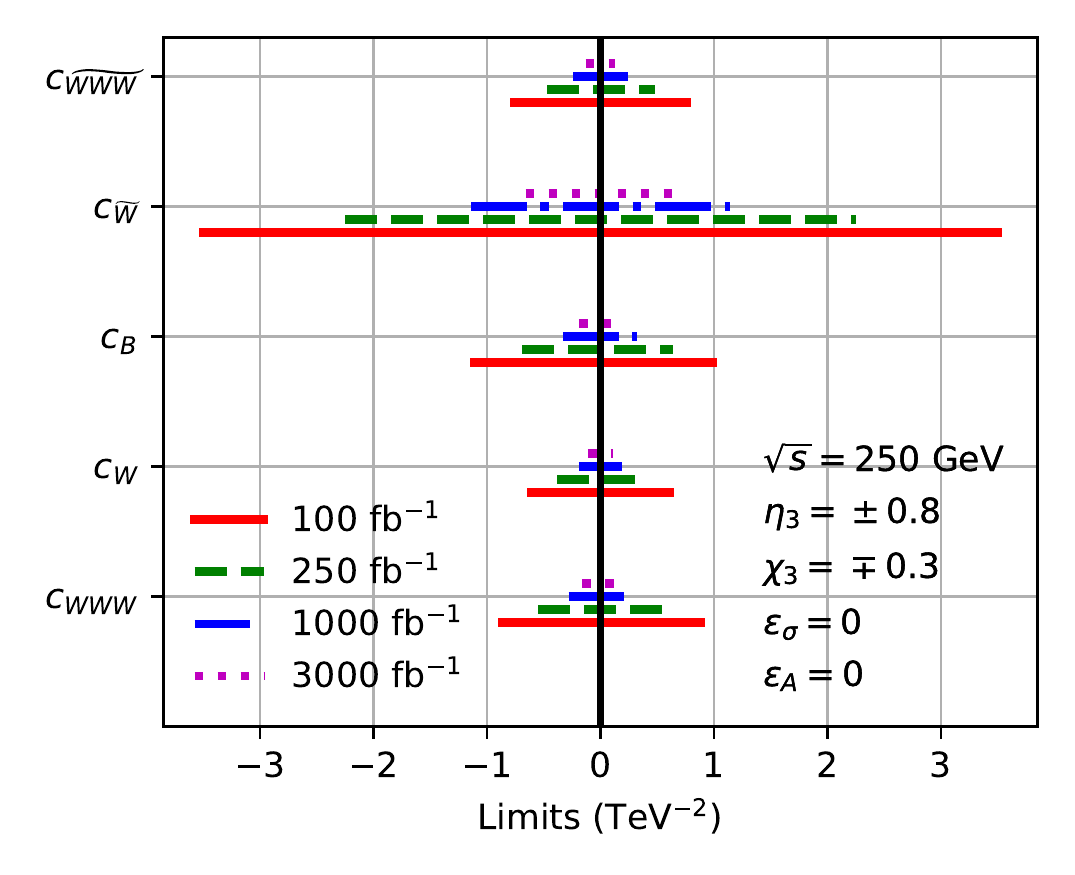}
	\includegraphics[scale=0.48]{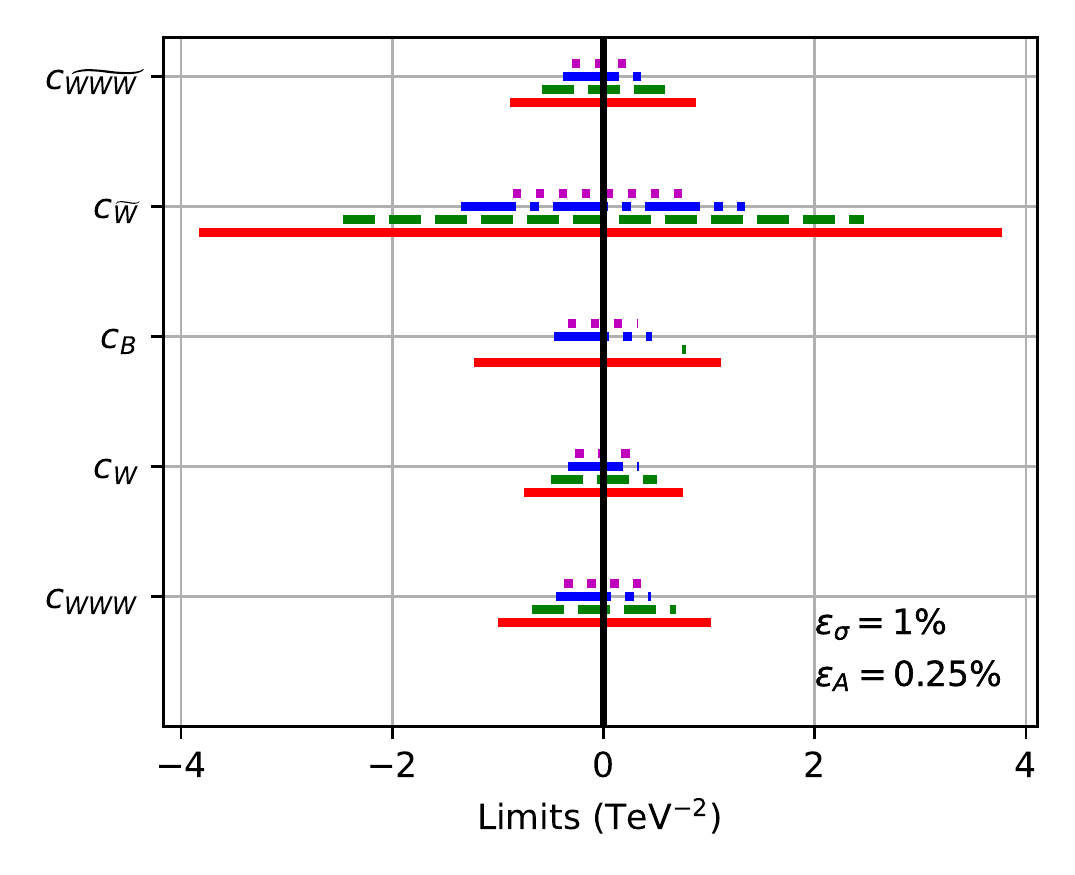}
	\includegraphics[scale=0.48]{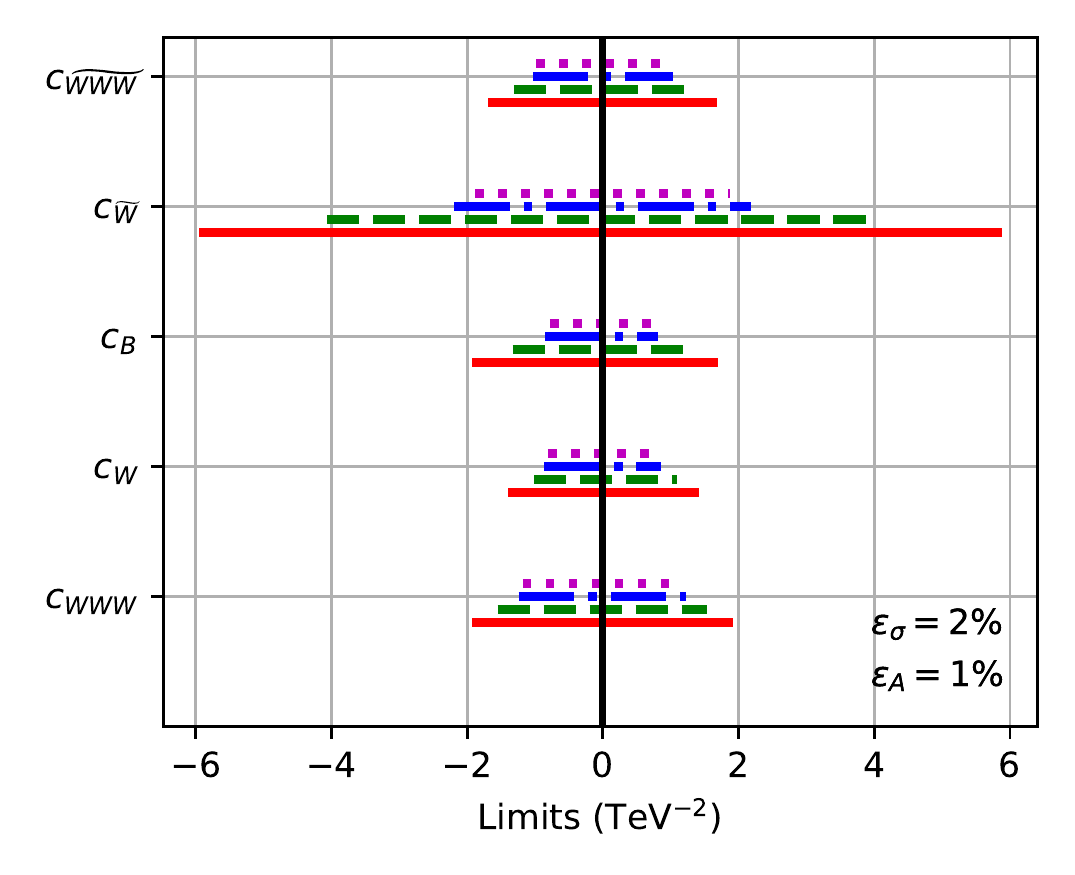}
	\caption{\label{fig:oneDlimit} The graphical visualizations of 95$\%$ BCI limits obtained from MCMC global fits on the anomalous couplings $c_i$ for a different set of systematic error, $(\epsilon_\sigma,\epsilon_{A})$ = (0,0) in the leftmost panel, ($1\%,0.25\%$) in the middle panel and ($2\%,1\%$) in the right-most panel. The limits are obtained at $\sqrt{s}=250$ GeV and luminosity given in Eq.~(\ref{lumi}).}
\end{figure*}
The second class of models was used for flavor tagging in Ref.~\cite{Subba:2022czw} with an accuracy of $\approx 70\%$, while the former model provides an accuracy of $\approx 80\%$. To demonstrate the role of the tagger, we compute $\chi^2$ for the combination of flavor-dependent asymmetries as a function of two anomalous couplings with two sets of beam polarization. For each set of beam polarization, $\chi^2$ is calculated at luminosity $\mathcal{L}$ = 50~fb$^{-1}$ and zero systematic errors. It is sufficient to demonstrate the involvement of the tagger with flavor-dependent asymmetries because the contribution from flavor-independent observables will be similar in both cases. In Fig.~\ref{fig:newvsoldXGB}, the resulting $95\%$ CL contours are displayed. According to the figures in both panels, the simultaneous bounds on a pair of anomalous couplings obtained using ML models with $\approx 80\%$ accuracy are tighter than those produced with $70\%$ accuracy. \\\\
In conclusion, we need to use two beam polarization combination with as good a flavor tagger as possible to include the flavor-dependent asymmetries in the analysis. With these choices, we do a full five coupling simultaneous analysis below.
\subsection{Five parameter analysis}
\label{sec:fivepara}
In this section, we comprehensively explain the methodology used to obtain simultaneous limits on anomalous couplings and the impact of systematic errors on those limits. The likelihood function used in this analysis is constructed using the chi-squared distance between the Standard Model~(SM) and the SM plus anomalous points, which is expressed in Eq.~(\ref{chisum}). The likelihood function is then defined as,
\begin{equation}
\lambda(\vec{x},c_i) \propto \text{Exp}\left[-\frac{\chi^2(\vec{x},c_i)}{2}\right],
\end{equation}
where $\vec{x} \in \{\mathcal{O},\eta_3,\xi_3\}$ represents observable and the degree of initial beam polarization. 
The Markov-Chain-Monte-Carlo~(MCMC) integration technique is used to obtain the posterior distribution of the anomalous couplings, $c_i$, based on the likelihood function. Specifically, the MCMC algorithm generates a set of samples from the posterior distribution by iteratively sampling the likelihood function and updating the current position in the parameter space. The resulting posterior distribution is then used as an input chain to the {\tt GetDist}~\cite{Lewis:2019xzd} package, which provides the marginalized limits on the couplings.\\\\
The analysis is performed for different values of luminosity values and systematic errors. The chi-squared values are calculated separately for each beam polarization and then combined to obtain the overall chi-squared value. Specifically, we combine the two beam polarizations at half the luminosity stated in Eq.~(\ref{lumi}). 
\\\\
To visualize the variation of limits on anomalous couplings with respect to luminosities, the $95\%$ Bayesian confidence intervals obtained from the MCMC global fits on the anomalous couplings for $\sqrt{s}=250$~GeV and one specific set of systematic errors are illustrated in Fig.~\ref{fig:oneDlimit}. The results show that the limits become tighter as the luminosity increases for zero systematic error. However, the saturation of the limits is observed for most of the couplings at particular luminosity values under conservative values of systematic error. Specifically, for the systematic error of ($2\%,1\%$) in the case of $c_{WWW}$, $c_W$, and $c_{\widetilde{WWW}}$, the boundaries of each anomalous coupling saturate at a luminosity value of 250~fb$^{-1}$. However, for $c_{\widetilde{W}}$ and $c_B$, the limits saturate at a luminosity value of 1000~fb$^{-1}$.\\\\
The significant influence of systematic errors on the constraints of anomalous $W^-W^+Z/\gamma$ couplings can be elucidated as follows: At the center-of-mass energy $\sqrt{s}=250$~GeV, the cross-section for $W^-W^+$ production approaches its maximum in the Standard Model scenario, thereby leading to lower statistical errors for large luminosities. Consequently, the estimated uncertainty in the measurement is primarily governed by systematic errors. It leads to the observed trend of saturation of limits for conservative levels of systematic errors. 
\begin{table}[!h]
	\caption{\label{95bci}The list of posterior $95\%$ BCI of anomalous
		couplings $c_i$ (TeV$^{-2}$) of effective operators for $\sqrt{s} = 250$ GeV with beam polarization $(\eta_3,\xi_3) = (\pm0.8,\pm0.3)$ at systematic error of $(\epsilon_{\sigma}, \epsilon_{A}) = (2\%,1\%)$ from MCMC global fits at the \emph{reconstruction level} at different values of $\mathcal{L}$. The reconstruction of $W^+$ is done using an artificial neural network.}
	\centering
	\begin{ruledtabular}
		\begin{tabular}{lllll}
			Parameters (TeV$^{-2}$)&100 fb$^{-1}$&250 fb$^{-1}$&1000 fb$^{-1}$&3000 fb$^{-1}$\\ \hline \\[-1em]
			$\frac{c_{WWW}}{\Lambda^2}$&$^{+1.8}_{-1.8}$&$^{+1.5}_{-1.5}$&$^{+1.2}_{-1.2}$&$^{+1.1}_{-1.1}$\\ \\[-0.8em]
			$\frac{c_W}{\Lambda^2}$&$^{+1.3}_{-1.3}$&$^{+1.1}_{-1.1}$&$^{+0.86}_{-0.86}$&$^{+0.82}_{-0.81}$ \\ \\[-0.8em]
			$\frac{c_B}{\Lambda^2}$&$^{+1.6}_{-1.8}$&$^{+1.2}_{-1.3}$&$^{+0.81}_{-0.85}$&$^{+0.78}_{-0.75}$\\ \\[-0.8em]
			$\frac{c_{\widetilde{W}}}{\Lambda^2}$&$^{+5.8}_{-5.8}$&$^{+4.0}_{-4.0}$&$^{+2.1}_{-2.1}$&$^{+1.8}_{-1.8}$ \\ \\[-0.8em]
			$\frac{c_{\widetilde{WWW}}}{\Lambda^2}$&$^{+1.6}_{-1.6}$&$^{+1.3}_{-1.3}$&$^{+1.0}_{-1.0}$&$^{+0.99}_{-0.98}$\\
		\end{tabular}
	\end{ruledtabular}	
\end{table}\\
Finally, we list down the 95$\%$ BCI of anomalous couplings $c_i$~(TeV$^{-2}$) of effective operators given in Eq.~(\ref{eftop}) at the center-of-mass energy of 250~GeV and systematic errors~($\epsilon_\sigma,\epsilon_{A}$) = (2$\%$,1$\%$) in Table~\ref{95bci}. The translation of these bounds to the LEP parameters can be found using Eq.~(\ref{lep}). In comparison to~\cite{Subba:2022czw}, the current limits on $c_{B},c_{\widetilde{W}}$ are $\approx$ a factor four and three tighter respectively, and all other couplings are reduced by a factor, $k$, of $1\le k \le 2$.
\section{Conclusion}
\label{conclude}
In this article, we examine the impact of dimension-6 operators on the charged triple gauge boson vertex in $e^-e^+$ Collider at $\sqrt{s}=250$~GeV. The notion is that when new physics is present at the top of the energy pyramid, sectors like electroweak are most likely to experience their indirect effects. We used a variety of observables, including asymmetries in cross~section, polarization, and spin correlation asymmetries, to restrict a set of anomalous couplings. Since some of these asymmetries requires the daughter of $W^+$ boson to be tagged, we developed machine learning models for flavor tagging. With the features listed in Sec.~\ref{flavor tagging}, we were able to classify the jets as \emph{up}-type and \emph{down}-type with an efficiency of around $80\%$. Several studies~\cite{Erdmann:2020ovh, Bedeschi:2022rnj, Nakai:2020kuu, Erdmann:2019blf} use features like jet energy, transverse momenta of jet for generic flavor tagging, while in our case, we excluded them because these are polarization dependent features. Using these features would increase the efficiency of our ML models above $90\%$ and thus tighten the anomalous couplings better.\\
The limits on each anomalous coupling are studied under different sets of luminosity, systematic error, and beam polarization. Initial beam polarization provides directional cuts, which results in tighter constraints. Hence a future collider like ILC would be a perfect machine to probe such weak effects of new physics in the electroweak sector.
Our five parameter simultaneous limits in Table~\ref{95bci} at 100~fb$^{-1}$ are tighter than the experimental one parameter limits listed in Table~\ref{tab:constraint} for $c_{W}, c_{B}$ and $c_{\widetilde{W}}$, while for $c_{WWW},c_{\widetilde{WWW}}$, the limits obtained by CMS using production rates alone remains better. It is due to the presence of $p^2$ term in case of $c_{WWW}$ and $c_{\widetilde{WWW}}$, which leads to enhanced contribution in machines like LHC running at 13~TeV. While in our case, the limits on these couplings are obtained using asymmetries at smaller momentum. Also, in the presence of beam polarization, the contribution of asymmetries increases significantly over the cross~section. There is a cancellation of the cross~section due to non-zero values of $CP$-even couplings in Fig.~\ref{twoparachi}. All these effects would add up, leading to a poorer limit on $c_{WWW}$ and $c_{\widetilde{WWW}}$ in Table~\ref{95bci}.\\ 
In our study, we found that systematic error is a significant challenge when it comes to constraining anomalous couplings. When we assume a conservative choice of systematic error $(\epsilon_{\sigma},\epsilon_{A}) = (2\%,1\%)$, we found that the limits on certain anomalous couplings like $c_{WWW},c_W$ and $c_{\widetilde{WWW}}$ only improved by a factor of approximately 1.5 when we increased the luminosity from 100~fb$^{-1}$ to 3000~fb$^{-1}$. The improvement on the limits of $c_B$ and $c_{\widetilde{W}}$ saturates to a factor of 2.3 and 3.1, respectively. This is not very encouraging, given the substantial increase in luminosity. Therefore, in significant systematic errors, it may be necessary to look for additional observables from various processes to constrain anomalous couplings more effectively. Our study suggests that a more efficient flavor tagging method could be implemented to reduce the dependence on certain polarization-dependent observables, which could help tighten the limits on anomalous couplings in the future.
\begin{figure}[H]
	\centering
	\subfigure{\includegraphics[width=3.9cm,height=3.3cm]{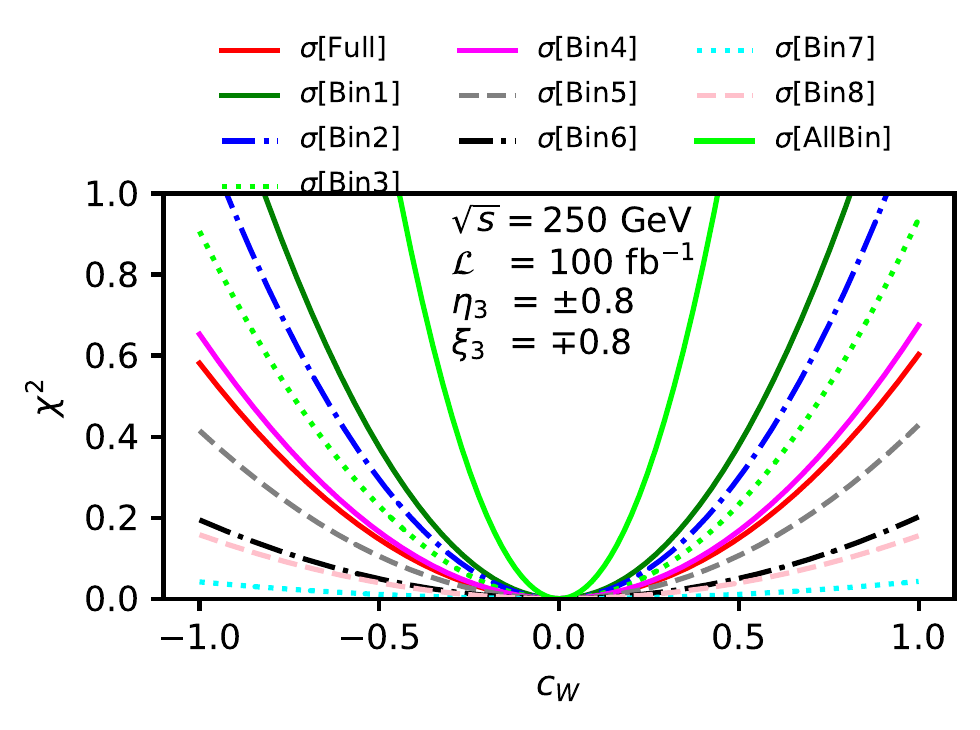}}
	\subfigure{\includegraphics[width=3.9cm,height=3.3cm]{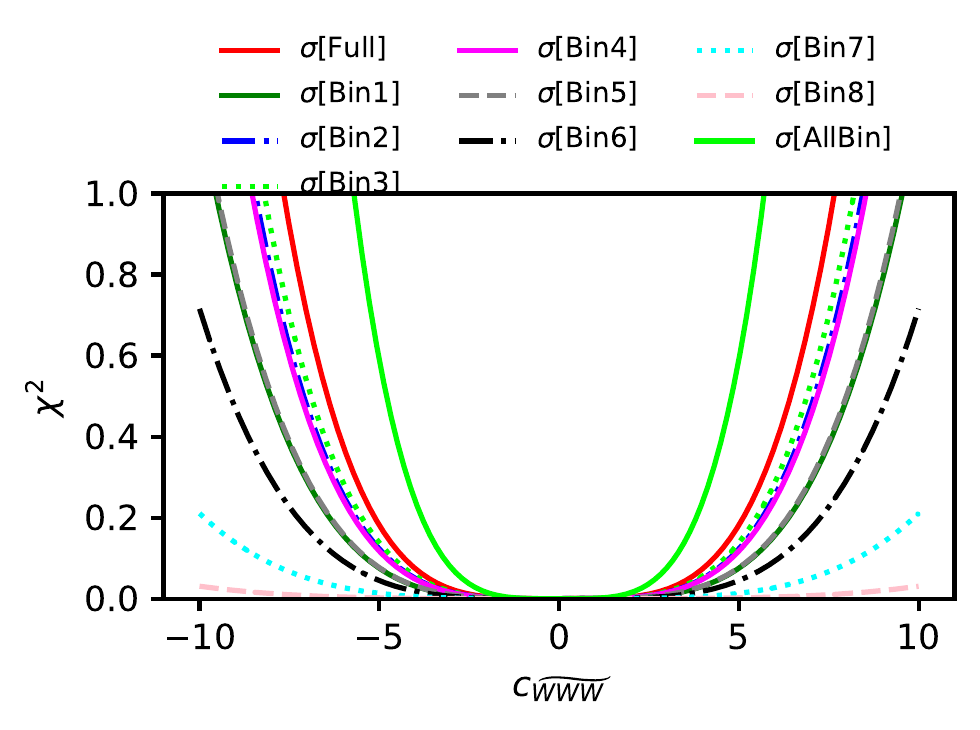}}
	\subfigure{\includegraphics[width=4cm,height=3.8cm]{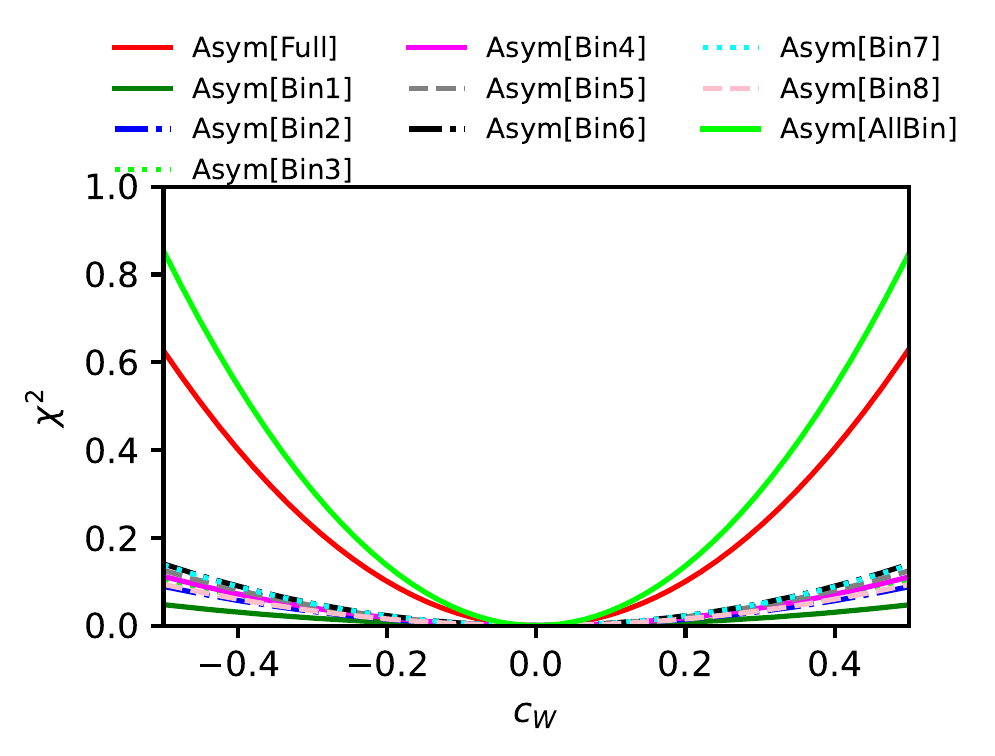}}
	\subfigure{\includegraphics[width=4cm,height=3.8cm]{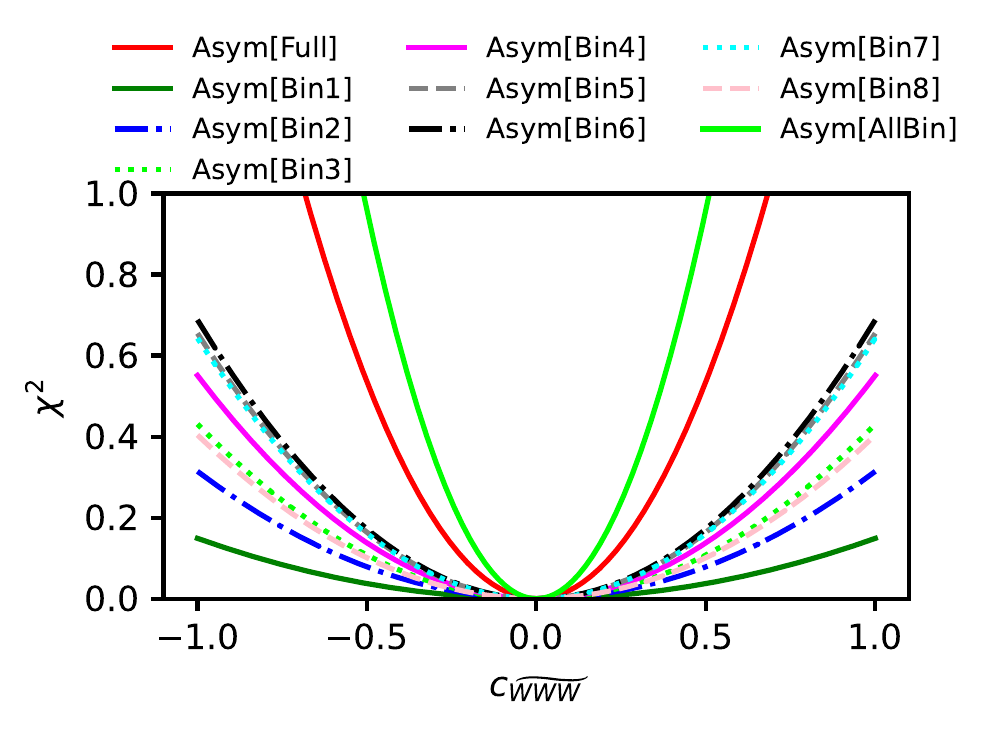}}
	\caption{\label{binwisefig} 1-D chi-squared plots for cross-section and asymmetries for each bin as a function of one anomalous coupling $c_i$ at a time. The $\chi^2$ is obtained at $\sqrt{s}=250$ GeV, $\mathcal{L}=100$ fb$^{-1}$ and zero systematic error.}
\end{figure}

\begin{acknowledgments}
A.S thanks University Grant Commission, Government of India, for financial support through UGC-NET Fellowship. 
\end{acknowledgments}

\appendix
\section{Significance of Binning:}
\label{app:binwise}
In the earlier section, we discussed that to obtain a more significant number of observables to scan the effect of new physics; each observable is divided into eight bins of $\cos\theta_{W^-}$. Moreover, in finding the final limits of anomalous couplings $c_i$, we combine all 648 observable of eight bins. Here, we tried to quantify the effect of binning, and for that, 1-D chi-squared plots for cross~section and asymmetries are shown for each bin along with a combination of all bins in Fig.~\ref{binwisefig}. In the left panel top row, the limit provided by cross~section without binning~(Full) is looser than some bins in the range $-1.0 \le \cos\theta_{W^-} \le 1.0$. Furthermore, in the case of $CP$-odd couplings $c_{\widetilde{WWW}}$, the unbinned limit is tighter but remains weaker than the limit provided by combined bins. In the bottom row of Fig.~\ref{binwisefig}, a forward-backward asymmetry exists only in $CP$-odd case~(right panel). For $c_{\widetilde{WWW}}$, the unbinned limit is tighter than that of individual bins obtained using asymmetries but remains poorer than the combined limit of all bins. The tightest limit is obtained in each panel by combining all the bins. It highlights the importance of binning technique to increase observables' sensitivity to new physics.

 \bibliography{biblio}

\begin{thebibliography}{133}%
\makeatletter
\providecommand \@ifxundefined [1]{%
 \@ifx{#1\undefined}
}%
\providecommand \@ifnum [1]{%
 \ifnum #1\expandafter \@firstoftwo
 \else \expandafter \@secondoftwo
 \fi
}%
\providecommand \@ifx [1]{%
 \ifx #1\expandafter \@firstoftwo
 \else \expandafter \@secondoftwo
 \fi
}%
\providecommand \natexlab [1]{#1}%
\providecommand \enquote  [1]{``#1''}%
\providecommand \bibnamefont  [1]{#1}%
\providecommand \bibfnamefont [1]{#1}%
\providecommand \citenamefont [1]{#1}%
\providecommand \href@noop [0]{\@secondoftwo}%
\providecommand \href [0]{\begingroup \@sanitize@url \@href}%
\providecommand \@href[1]{\@@startlink{#1}\@@href}%
\providecommand \@@href[1]{\endgroup#1\@@endlink}%
\providecommand \@sanitize@url [0]{\catcode `\\12\catcode `\$12\catcode
  `\&12\catcode `\#12\catcode `\^12\catcode `\_12\catcode `\%12\relax}%
\providecommand \@@startlink[1]{}%
\providecommand \@@endlink[0]{}%
\providecommand \url  [0]{\begingroup\@sanitize@url \@url }%
\providecommand \@url [1]{\endgroup\@href {#1}{\urlprefix }}%
\providecommand \urlprefix  [0]{URL }%
\providecommand \Eprint [0]{\href }%
\providecommand \doibase [0]{https://doi.org/}%
\providecommand \selectlanguage [0]{\@gobble}%
\providecommand \bibinfo  [0]{\@secondoftwo}%
\providecommand \bibfield  [0]{\@secondoftwo}%
\providecommand \translation [1]{[#1]}%
\providecommand \BibitemOpen [0]{}%
\providecommand \bibitemStop [0]{}%
\providecommand \bibitemNoStop [0]{.\EOS\space}%
\providecommand \EOS [0]{\spacefactor3000\relax}%
\providecommand \BibitemShut  [1]{\csname bibitem#1\endcsname}%
\let\auto@bib@innerbib\@empty
\bibitem [{\citenamefont {Chatrchyan}\ \emph {et~al.}(2012)\citenamefont
  {Chatrchyan} \emph {et~al.}}]{CMS:2012qbp}%
  \BibitemOpen
  \bibfield  {author} {\bibinfo {author} {\bibfnamefont {S.}~\bibnamefont
  {Chatrchyan}} \emph {et~al.} (\bibinfo {collaboration} {CMS}),\ }\bibfield
  {title} {\bibinfo {title} {{Observation of a New Boson at a Mass of 125 GeV
  with the CMS Experiment at the LHC}},\ }\href
  {https://doi.org/10.1016/j.physletb.2012.08.021} {\bibfield  {journal}
  {\bibinfo  {journal} {Phys. Lett. B}\ }\textbf {\bibinfo {volume} {716}},\
  \bibinfo {pages} {30} (\bibinfo {year} {2012})},\ \Eprint
  {https://arxiv.org/abs/1207.7235} {arXiv:1207.7235 [hep-ex]} \BibitemShut
  {NoStop}%
\bibitem [{\citenamefont {Aad}\ \emph {et~al.}(2012{\natexlab{a}})\citenamefont
  {Aad} \emph {et~al.}}]{ATLAS:2012yve}%
  \BibitemOpen
  \bibfield  {author} {\bibinfo {author} {\bibfnamefont {G.}~\bibnamefont
  {Aad}} \emph {et~al.} (\bibinfo {collaboration} {ATLAS}),\ }\bibfield
  {title} {\bibinfo {title} {{Observation of a new particle in the search for
  the Standard Model Higgs boson with the ATLAS detector at the LHC}},\ }\href
  {https://doi.org/10.1016/j.physletb.2012.08.020} {\bibfield  {journal}
  {\bibinfo  {journal} {Phys. Lett. B}\ }\textbf {\bibinfo {volume} {716}},\
  \bibinfo {pages} {1} (\bibinfo {year} {2012}{\natexlab{a}})},\ \Eprint
  {https://arxiv.org/abs/1207.7214} {arXiv:1207.7214 [hep-ex]} \BibitemShut
  {NoStop}%
\bibitem [{\citenamefont {Ade}\ \emph {et~al.}(2014)\citenamefont {Ade} \emph
  {et~al.}}]{Planck:2013pxb}%
  \BibitemOpen
  \bibfield  {author} {\bibinfo {author} {\bibfnamefont {P.~A.~R.}\
  \bibnamefont {Ade}} \emph {et~al.} (\bibinfo {collaboration} {Planck}),\
  }\bibfield  {title} {\bibinfo {title} {{Planck 2013 results. XVI.
  Cosmological parameters}},\ }\href
  {https://doi.org/10.1051/0004-6361/201321591} {\bibfield  {journal} {\bibinfo
   {journal} {Astron. Astrophys.}\ }\textbf {\bibinfo {volume} {571}},\
  \bibinfo {pages} {A16} (\bibinfo {year} {2014})},\ \Eprint
  {https://arxiv.org/abs/1303.5076} {arXiv:1303.5076 [astro-ph.CO]}
  \BibitemShut {NoStop}%
\bibitem [{\citenamefont {Albahri}\ \emph {et~al.}(2021)\citenamefont {Albahri}
  \emph {et~al.}}]{Muong-2:2021vma}%
  \BibitemOpen
  \bibfield  {author} {\bibinfo {author} {\bibfnamefont {T.}~\bibnamefont
  {Albahri}} \emph {et~al.} (\bibinfo {collaboration} {Muon g-2}),\ }\bibfield
  {title} {\bibinfo {title} {{Measurement of the anomalous precession frequency
  of the muon in the Fermilab Muon $g−2$ Experiment}},\ }\href
  {https://doi.org/10.1103/PhysRevD.103.072002} {\bibfield  {journal} {\bibinfo
   {journal} {Phys. Rev. D}\ }\textbf {\bibinfo {volume} {103}},\ \bibinfo
  {pages} {072002} (\bibinfo {year} {2021})},\ \Eprint
  {https://arxiv.org/abs/2104.03247} {arXiv:2104.03247 [hep-ex]} \BibitemShut
  {NoStop}%
\bibitem [{\citenamefont {Abi}\ \emph {et~al.}(2021)\citenamefont {Abi} \emph
  {et~al.}}]{Muong-2:2021ojo}%
  \BibitemOpen
  \bibfield  {author} {\bibinfo {author} {\bibfnamefont {B.}~\bibnamefont
  {Abi}} \emph {et~al.} (\bibinfo {collaboration} {Muon g-2}),\ }\bibfield
  {title} {\bibinfo {title} {{Measurement of the Positive Muon Anomalous
  Magnetic Moment to 0.46 ppm}},\ }\href
  {https://doi.org/10.1103/PhysRevLett.126.141801} {\bibfield  {journal}
  {\bibinfo  {journal} {Phys. Rev. Lett.}\ }\textbf {\bibinfo {volume} {126}},\
  \bibinfo {pages} {141801} (\bibinfo {year} {2021})},\ \Eprint
  {https://arxiv.org/abs/2104.03281} {arXiv:2104.03281 [hep-ex]} \BibitemShut
  {NoStop}%
\bibitem [{\citenamefont {Aaltonen}\ \emph {et~al.}(2022)\citenamefont
  {Aaltonen} \emph {et~al.}}]{CDF:2022hxs}%
  \BibitemOpen
  \bibfield  {author} {\bibinfo {author} {\bibfnamefont {T.}~\bibnamefont
  {Aaltonen}} \emph {et~al.} (\bibinfo {collaboration} {CDF}),\ }\bibfield
  {title} {\bibinfo {title} {{High-precision measurement of the $W$ boson mass
  with the CDF II detector}},\ }\href {https://doi.org/10.1126/science.abk1781}
  {\bibfield  {journal} {\bibinfo  {journal} {Science}\ }\textbf {\bibinfo
  {volume} {376}},\ \bibinfo {pages} {170} (\bibinfo {year}
  {2022})}\BibitemShut {NoStop}%
\bibitem [{\citenamefont {Appelquist}\ and\ \citenamefont
  {Carazzone}(1975)}]{Appelquist:1974tg}%
  \BibitemOpen
  \bibfield  {author} {\bibinfo {author} {\bibfnamefont {T.}~\bibnamefont
  {Appelquist}}\ and\ \bibinfo {author} {\bibfnamefont {J.}~\bibnamefont
  {Carazzone}},\ }\bibfield  {title} {\bibinfo {title} {{Infrared Singularities
  and Massive Fields}},\ }\href {https://doi.org/10.1103/PhysRevD.11.2856}
  {\bibfield  {journal} {\bibinfo  {journal} {Phys. Rev. D}\ }\textbf {\bibinfo
  {volume} {11}},\ \bibinfo {pages} {2856} (\bibinfo {year}
  {1975})}\BibitemShut {NoStop}%
\bibitem [{\citenamefont {Buchmuller}\ and\ \citenamefont
  {Wyler}(1986)}]{Buchmuller:1985jz}%
  \BibitemOpen
  \bibfield  {author} {\bibinfo {author} {\bibfnamefont {W.}~\bibnamefont
  {Buchmuller}}\ and\ \bibinfo {author} {\bibfnamefont {D.}~\bibnamefont
  {Wyler}},\ }\bibfield  {title} {\bibinfo {title} {{Effective Lagrangian
  Analysis of New Interactions and Flavor Conservation}},\ }\href
  {https://doi.org/10.1016/0550-3213(86)90262-2} {\bibfield  {journal}
  {\bibinfo  {journal} {Nucl. Phys. B}\ }\textbf {\bibinfo {volume} {268}},\
  \bibinfo {pages} {621} (\bibinfo {year} {1986})}\BibitemShut {NoStop}%
\bibitem [{\citenamefont {Hagiwara}\ \emph {et~al.}(1987)\citenamefont
  {Hagiwara}, \citenamefont {Peccei}, \citenamefont {Zeppenfeld},\ and\
  \citenamefont {Hikasa}}]{Hagiwara:1986vm}%
  \BibitemOpen
  \bibfield  {author} {\bibinfo {author} {\bibfnamefont {K.}~\bibnamefont
  {Hagiwara}}, \bibinfo {author} {\bibfnamefont {R.~D.}\ \bibnamefont
  {Peccei}}, \bibinfo {author} {\bibfnamefont {D.}~\bibnamefont {Zeppenfeld}},\
  and\ \bibinfo {author} {\bibfnamefont {K.}~\bibnamefont {Hikasa}},\
  }\bibfield  {title} {\bibinfo {title} {{Probing the Weak Boson Sector in e+
  e- ---\ensuremath{>} W+ W-}},\ }\href
  {https://doi.org/10.1016/0550-3213(87)90685-7} {\bibfield  {journal}
  {\bibinfo  {journal} {Nucl. Phys. B}\ }\textbf {\bibinfo {volume} {282}},\
  \bibinfo {pages} {253} (\bibinfo {year} {1987})}\BibitemShut {NoStop}%
\bibitem [{\citenamefont {Degrande}\ \emph {et~al.}(2013)\citenamefont
  {Degrande}, \citenamefont {Greiner}, \citenamefont {Kilian}, \citenamefont
  {Mattelaer}, \citenamefont {Mebane}, \citenamefont {Stelzer}, \citenamefont
  {Willenbrock},\ and\ \citenamefont {Zhang}}]{Degrande:2012wf}%
  \BibitemOpen
  \bibfield  {author} {\bibinfo {author} {\bibfnamefont {C.}~\bibnamefont
  {Degrande}}, \bibinfo {author} {\bibfnamefont {N.}~\bibnamefont {Greiner}},
  \bibinfo {author} {\bibfnamefont {W.}~\bibnamefont {Kilian}}, \bibinfo
  {author} {\bibfnamefont {O.}~\bibnamefont {Mattelaer}}, \bibinfo {author}
  {\bibfnamefont {H.}~\bibnamefont {Mebane}}, \bibinfo {author} {\bibfnamefont
  {T.}~\bibnamefont {Stelzer}}, \bibinfo {author} {\bibfnamefont
  {S.}~\bibnamefont {Willenbrock}},\ and\ \bibinfo {author} {\bibfnamefont
  {C.}~\bibnamefont {Zhang}},\ }\bibfield  {title} {\bibinfo {title}
  {{Effective Field Theory: A Modern Approach to Anomalous Couplings}},\ }\href
  {https://doi.org/10.1016/j.aop.2013.04.016} {\bibfield  {journal} {\bibinfo
  {journal} {Annals Phys.}\ }\textbf {\bibinfo {volume} {335}},\ \bibinfo
  {pages} {21} (\bibinfo {year} {2013})},\ \Eprint
  {https://arxiv.org/abs/1205.4231} {arXiv:1205.4231 [hep-ph]} \BibitemShut
  {NoStop}%
\bibitem [{\citenamefont {Czyz}\ \emph {et~al.}(1989)\citenamefont {Czyz},
  \citenamefont {Kolodziej},\ and\ \citenamefont {Zralek}}]{Czyz:1988yt}%
  \BibitemOpen
  \bibfield  {author} {\bibinfo {author} {\bibfnamefont {H.}~\bibnamefont
  {Czyz}}, \bibinfo {author} {\bibfnamefont {K.}~\bibnamefont {Kolodziej}},\
  and\ \bibinfo {author} {\bibfnamefont {M.}~\bibnamefont {Zralek}},\
  }\bibfield  {title} {\bibinfo {title} {{Composite $Z$ Boson and {CP}
  Violation in the Process $e^+ e^- \to Z \gamma$}},\ }\href
  {https://doi.org/10.1007/BF02430614} {\bibfield  {journal} {\bibinfo
  {journal} {Z. Phys. C}\ }\textbf {\bibinfo {volume} {43}},\ \bibinfo {pages}
  {97} (\bibinfo {year} {1989})}\BibitemShut {NoStop}%
\bibitem [{\citenamefont {Zhang}(2017)}]{Zhang:2016zsp}%
  \BibitemOpen
  \bibfield  {author} {\bibinfo {author} {\bibfnamefont {Z.}~\bibnamefont
  {Zhang}},\ }\bibfield  {title} {\bibinfo {title} {{Time to Go Beyond
  Triple-Gauge-Boson-Coupling Interpretation of $W$ Pair Production}},\ }\href
  {https://doi.org/10.1103/PhysRevLett.118.011803} {\bibfield  {journal}
  {\bibinfo  {journal} {Phys. Rev. Lett.}\ }\textbf {\bibinfo {volume} {118}},\
  \bibinfo {pages} {011803} (\bibinfo {year} {2017})},\ \Eprint
  {https://arxiv.org/abs/1610.01618} {arXiv:1610.01618 [hep-ph]} \BibitemShut
  {NoStop}%
\bibitem [{\citenamefont {Rahaman}\ and\ \citenamefont
  {Singh}(2020{\natexlab{a}})}]{Rahaman:2019mnz}%
  \BibitemOpen
  \bibfield  {author} {\bibinfo {author} {\bibfnamefont {R.}~\bibnamefont
  {Rahaman}}\ and\ \bibinfo {author} {\bibfnamefont {R.~K.}\ \bibnamefont
  {Singh}},\ }\bibfield  {title} {\bibinfo {title} {{Probing the anomalous
  triple gauge boson couplings in $e^+e^-\to W^+W^-$ using $W$ polarizations
  with polarized beams}},\ }\href {https://doi.org/10.1103/PhysRevD.101.075044}
  {\bibfield  {journal} {\bibinfo  {journal} {Phys. Rev. D}\ }\textbf {\bibinfo
  {volume} {101}},\ \bibinfo {pages} {075044} (\bibinfo {year}
  {2020}{\natexlab{a}})},\ \Eprint {https://arxiv.org/abs/1909.05496}
  {arXiv:1909.05496 [hep-ph]} \BibitemShut {NoStop}%
\bibitem [{\citenamefont {Bilchak}\ and\ \citenamefont
  {Stroughair}(1984)}]{Bilchak:1984ur}%
  \BibitemOpen
  \bibfield  {author} {\bibinfo {author} {\bibfnamefont {C.~L.}\ \bibnamefont
  {Bilchak}}\ and\ \bibinfo {author} {\bibfnamefont {J.~D.}\ \bibnamefont
  {Stroughair}},\ }\bibfield  {title} {\bibinfo {title} {{$W^+ W^-$ Pair
  Production in $e^+ e^-$ Colliders}},\ }\href
  {https://doi.org/10.1103/PhysRevD.30.1881} {\bibfield  {journal} {\bibinfo
  {journal} {Phys. Rev. D}\ }\textbf {\bibinfo {volume} {30}},\ \bibinfo
  {pages} {1881} (\bibinfo {year} {1984})}\BibitemShut {NoStop}%
\bibitem [{\citenamefont {Gaemers}\ and\ \citenamefont
  {Gounaris}(1979)}]{Gaemers:1978hg}%
  \BibitemOpen
  \bibfield  {author} {\bibinfo {author} {\bibfnamefont {K.~J.~F.}\
  \bibnamefont {Gaemers}}\ and\ \bibinfo {author} {\bibfnamefont {G.~J.}\
  \bibnamefont {Gounaris}},\ }\bibfield  {title} {\bibinfo {title}
  {{Polarization Amplitudes for e+ e- ---\ensuremath{>} W+ W- and e+ e-
  ---\ensuremath{>} Z Z}},\ }\href {https://doi.org/10.1007/BF01440226}
  {\bibfield  {journal} {\bibinfo  {journal} {Z. Phys. C}\ }\textbf {\bibinfo
  {volume} {1}},\ \bibinfo {pages} {259} (\bibinfo {year} {1979})}\BibitemShut
  {NoStop}%
\bibitem [{\citenamefont {Hagiwara}\ \emph {et~al.}(1992)\citenamefont
  {Hagiwara}, \citenamefont {Ishihara}, \citenamefont {Szalapski},\ and\
  \citenamefont {Zeppenfeld}}]{Hagiwara:1992eh}%
  \BibitemOpen
  \bibfield  {author} {\bibinfo {author} {\bibfnamefont {K.}~\bibnamefont
  {Hagiwara}}, \bibinfo {author} {\bibfnamefont {S.}~\bibnamefont {Ishihara}},
  \bibinfo {author} {\bibfnamefont {R.}~\bibnamefont {Szalapski}},\ and\
  \bibinfo {author} {\bibfnamefont {D.}~\bibnamefont {Zeppenfeld}},\ }\bibfield
   {title} {\bibinfo {title} {{Low-energy constraints on electroweak three
  gauge boson couplings}},\ }\href
  {https://doi.org/10.1016/0370-2693(92)90031-X} {\bibfield  {journal}
  {\bibinfo  {journal} {Phys. Lett. B}\ }\textbf {\bibinfo {volume} {283}},\
  \bibinfo {pages} {353} (\bibinfo {year} {1992})}\BibitemShut {NoStop}%
\bibitem [{\citenamefont {Hagiwara}(1995)}]{Hagiwara:1995se}%
  \BibitemOpen
  \bibfield  {author} {\bibinfo {author} {\bibfnamefont {K.}~\bibnamefont
  {Hagiwara}},\ }\bibfield  {title} {\bibinfo {title} {{Low energy constraints
  on electroweak vector boson self-interactions}},\ }\href
  {https://doi.org/10.1063/1.49303} {\bibfield  {journal} {\bibinfo  {journal}
  {AIP Conf. Proc.}\ }\textbf {\bibinfo {volume} {350}},\ \bibinfo {pages}
  {182} (\bibinfo {year} {1995})}\BibitemShut {NoStop}%
\bibitem [{\citenamefont {Choudhury}\ and\ \citenamefont
  {Kalinowski}(1997)}]{Choudhury:1996ni}%
  \BibitemOpen
  \bibfield  {author} {\bibinfo {author} {\bibfnamefont {D.}~\bibnamefont
  {Choudhury}}\ and\ \bibinfo {author} {\bibfnamefont {J.}~\bibnamefont
  {Kalinowski}},\ }\bibfield  {title} {\bibinfo {title} {{Unraveling the $W W
  \gamma$ and $W W Z$ vertices at the linear collider: Anti-neutrino neutrino
  $\gamma$ and anti-neutrino neutrino $\bar{q} q$ final states}},\ }\href
  {https://doi.org/10.1016/S0550-3213(97)00111-9} {\bibfield  {journal}
  {\bibinfo  {journal} {Nucl. Phys. B}\ }\textbf {\bibinfo {volume} {491}},\
  \bibinfo {pages} {129} (\bibinfo {year} {1997})},\ \Eprint
  {https://arxiv.org/abs/hep-ph/9608416} {arXiv:hep-ph/9608416} \BibitemShut
  {NoStop}%
\bibitem [{\citenamefont {Choudhury}\ \emph {et~al.}(1999)\citenamefont
  {Choudhury}, \citenamefont {Kalinowski},\ and\ \citenamefont
  {Kulesza}}]{Choudhury:1999fz}%
  \BibitemOpen
  \bibfield  {author} {\bibinfo {author} {\bibfnamefont {D.}~\bibnamefont
  {Choudhury}}, \bibinfo {author} {\bibfnamefont {J.}~\bibnamefont
  {Kalinowski}},\ and\ \bibinfo {author} {\bibfnamefont {A.}~\bibnamefont
  {Kulesza}},\ }\bibfield  {title} {\bibinfo {title} {{CP violating anomalous
  $W W \gamma$ couplings in $e^{+} e^{-}$ collisions}},\ }\href
  {https://doi.org/10.1016/S0370-2693(99)00527-4} {\bibfield  {journal}
  {\bibinfo  {journal} {Phys. Lett. B}\ }\textbf {\bibinfo {volume} {457}},\
  \bibinfo {pages} {193} (\bibinfo {year} {1999})},\ \Eprint
  {https://arxiv.org/abs/hep-ph/9904215} {arXiv:hep-ph/9904215} \BibitemShut
  {NoStop}%
\bibitem [{\citenamefont {Wells}\ and\ \citenamefont
  {Zhang}(2016)}]{Wells:2015eba}%
  \BibitemOpen
  \bibfield  {author} {\bibinfo {author} {\bibfnamefont {J.~D.}\ \bibnamefont
  {Wells}}\ and\ \bibinfo {author} {\bibfnamefont {Z.}~\bibnamefont {Zhang}},\
  }\bibfield  {title} {\bibinfo {title} {{Status and prospects of precision
  analyses with $e^+e^-\to W^+W^-$}},\ }\href
  {https://doi.org/10.1103/PhysRevD.93.034001} {\bibfield  {journal} {\bibinfo
  {journal} {Phys. Rev. D}\ }\textbf {\bibinfo {volume} {93}},\ \bibinfo
  {pages} {034001} (\bibinfo {year} {2016})},\ \Eprint
  {https://arxiv.org/abs/1507.01594} {arXiv:1507.01594 [hep-ph]} \BibitemShut
  {NoStop}%
\bibitem [{\citenamefont {Buchalla}\ \emph {et~al.}(2013)\citenamefont
  {Buchalla}, \citenamefont {Cata}, \citenamefont {Rahn},\ and\ \citenamefont
  {Schlaffer}}]{Buchalla:2013wpa}%
  \BibitemOpen
  \bibfield  {author} {\bibinfo {author} {\bibfnamefont {G.}~\bibnamefont
  {Buchalla}}, \bibinfo {author} {\bibfnamefont {O.}~\bibnamefont {Cata}},
  \bibinfo {author} {\bibfnamefont {R.}~\bibnamefont {Rahn}},\ and\ \bibinfo
  {author} {\bibfnamefont {M.}~\bibnamefont {Schlaffer}},\ }\bibfield  {title}
  {\bibinfo {title} {{Effective Field Theory Analysis of New Physics in e+e-
  -\ensuremath{>} W+W- at a Linear Collider}},\ }\href
  {https://doi.org/10.1140/epjc/s10052-013-2589-1} {\bibfield  {journal}
  {\bibinfo  {journal} {Eur. Phys. J. C}\ }\textbf {\bibinfo {volume} {73}},\
  \bibinfo {pages} {2589} (\bibinfo {year} {2013})},\ \Eprint
  {https://arxiv.org/abs/1302.6481} {arXiv:1302.6481 [hep-ph]} \BibitemShut
  {NoStop}%
\bibitem [{\citenamefont {Berthier}\ \emph {et~al.}(2016)\citenamefont
  {Berthier}, \citenamefont {Bj\o{}rn},\ and\ \citenamefont
  {Trott}}]{Berthier:2016tkq}%
  \BibitemOpen
  \bibfield  {author} {\bibinfo {author} {\bibfnamefont {L.}~\bibnamefont
  {Berthier}}, \bibinfo {author} {\bibfnamefont {M.}~\bibnamefont {Bj\o{}rn}},\
  and\ \bibinfo {author} {\bibfnamefont {M.}~\bibnamefont {Trott}},\ }\bibfield
   {title} {\bibinfo {title} {{Incorporating doubly resonant $W^\pm$ data in a
  global fit of SMEFT parameters to lift flat directions}},\ }\href
  {https://doi.org/10.1007/JHEP09(2016)157} {\bibfield  {journal} {\bibinfo
  {journal} {JHEP}\ }\textbf {\bibinfo {volume} {09}},\ \bibinfo {pages}
  {157}},\ \Eprint {https://arxiv.org/abs/1606.06693} {arXiv:1606.06693
  [hep-ph]} \BibitemShut {NoStop}%
\bibitem [{\citenamefont {Beyer}\ \emph {et~al.}(2020)\citenamefont {Beyer},
  \citenamefont {Karl},\ and\ \citenamefont {List}}]{Beyer:2020eas}%
  \BibitemOpen
  \bibfield  {author} {\bibinfo {author} {\bibfnamefont {J.}~\bibnamefont
  {Beyer}}, \bibinfo {author} {\bibfnamefont {R.}~\bibnamefont {Karl}},\ and\
  \bibinfo {author} {\bibfnamefont {J.}~\bibnamefont {List}},\ }\bibfield
  {title} {\bibinfo {title} {{Precision measurements of Triple Gauge Couplings
  at future electron-positron colliders}},\ }in\ \href@noop {} {\emph {\bibinfo
  {booktitle} {{International Workshop on Future Linear Colliders}}}}\
  (\bibinfo {year} {2020})\ \Eprint {https://arxiv.org/abs/2002.02777}
  {arXiv:2002.02777 [hep-ex]} \BibitemShut {NoStop}%
\bibitem [{\citenamefont {Subba}\ and\ \citenamefont
  {Singh}(2023)}]{Subba:2022czw}%
  \BibitemOpen
  \bibfield  {author} {\bibinfo {author} {\bibfnamefont {A.}~\bibnamefont
  {Subba}}\ and\ \bibinfo {author} {\bibfnamefont {R.~K.}\ \bibnamefont
  {Singh}},\ }\bibfield  {title} {\bibinfo {title} {{Role of polarizations and
  spin-spin correlations of W's in e-e+\textrightarrow{}W-W+ at s=250\,\,GeV to
  probe anomalous W-W+Z/\ensuremath{\gamma} couplings}},\ }\href
  {https://doi.org/10.1103/PhysRevD.107.073004} {\bibfield  {journal} {\bibinfo
   {journal} {Phys. Rev. D}\ }\textbf {\bibinfo {volume} {107}},\ \bibinfo
  {pages} {073004} (\bibinfo {year} {2023})},\ \Eprint
  {https://arxiv.org/abs/2212.12973} {arXiv:2212.12973 [hep-ph]} \BibitemShut
  {NoStop}%
\bibitem [{\citenamefont {Bian}\ \emph {et~al.}(2015)\citenamefont {Bian},
  \citenamefont {Shu},\ and\ \citenamefont {Zhang}}]{Bian:2015zha}%
  \BibitemOpen
  \bibfield  {author} {\bibinfo {author} {\bibfnamefont {L.}~\bibnamefont
  {Bian}}, \bibinfo {author} {\bibfnamefont {J.}~\bibnamefont {Shu}},\ and\
  \bibinfo {author} {\bibfnamefont {Y.}~\bibnamefont {Zhang}},\ }\bibfield
  {title} {\bibinfo {title} {{Prospects for Triple Gauge Coupling Measurements
  at Future Lepton Colliders and the 14 TeV LHC}},\ }\href
  {https://doi.org/10.1007/JHEP09(2015)206} {\bibfield  {journal} {\bibinfo
  {journal} {JHEP}\ }\textbf {\bibinfo {volume} {09}},\ \bibinfo {pages}
  {206}},\ \Eprint {https://arxiv.org/abs/1507.02238} {arXiv:1507.02238
  [hep-ph]} \BibitemShut {NoStop}%
\bibitem [{\citenamefont {Bian}\ \emph {et~al.}(2016)\citenamefont {Bian},
  \citenamefont {Shu},\ and\ \citenamefont {Zhang}}]{Bian:2016umx}%
  \BibitemOpen
  \bibfield  {author} {\bibinfo {author} {\bibfnamefont {L.}~\bibnamefont
  {Bian}}, \bibinfo {author} {\bibfnamefont {J.}~\bibnamefont {Shu}},\ and\
  \bibinfo {author} {\bibfnamefont {Y.}~\bibnamefont {Zhang}},\ }\bibfield
  {title} {\bibinfo {title} {{Triple gauge couplings at future hadron and
  lepton colliders}},\ }\href {https://doi.org/10.1142/S0217751X16440085}
  {\bibfield  {journal} {\bibinfo  {journal} {Int. J. Mod. Phys. A}\ }\textbf
  {\bibinfo {volume} {31}},\ \bibinfo {pages} {1644008} (\bibinfo {year}
  {2016})},\ \Eprint {https://arxiv.org/abs/1612.03888} {arXiv:1612.03888
  [hep-ph]} \BibitemShut {NoStop}%
\bibitem [{\citenamefont {Baglio}\ \emph {et~al.}(2019)\citenamefont {Baglio},
  \citenamefont {Dawson},\ and\ \citenamefont {Homiller}}]{Baglio:2019uty}%
  \BibitemOpen
  \bibfield  {author} {\bibinfo {author} {\bibfnamefont {J.}~\bibnamefont
  {Baglio}}, \bibinfo {author} {\bibfnamefont {S.}~\bibnamefont {Dawson}},\
  and\ \bibinfo {author} {\bibfnamefont {S.}~\bibnamefont {Homiller}},\
  }\bibfield  {title} {\bibinfo {title} {{QCD corrections in Standard Model EFT
  fits to $WZ$ and $WW$ production}},\ }\href
  {https://doi.org/10.1103/PhysRevD.100.113010} {\bibfield  {journal} {\bibinfo
   {journal} {Phys. Rev. D}\ }\textbf {\bibinfo {volume} {100}},\ \bibinfo
  {pages} {113010} (\bibinfo {year} {2019})},\ \Eprint
  {https://arxiv.org/abs/1909.11576} {arXiv:1909.11576 [hep-ph]} \BibitemShut
  {NoStop}%
\bibitem [{\citenamefont {Choudhury}\ \emph {et~al.}(2022)\citenamefont
  {Choudhury}, \citenamefont {Deka}, \citenamefont {Maharana},\ and\
  \citenamefont {Saini}}]{Choudhury:2022iqz}%
  \BibitemOpen
  \bibfield  {author} {\bibinfo {author} {\bibfnamefont {D.}~\bibnamefont
  {Choudhury}}, \bibinfo {author} {\bibfnamefont {K.}~\bibnamefont {Deka}},
  \bibinfo {author} {\bibfnamefont {S.}~\bibnamefont {Maharana}},\ and\
  \bibinfo {author} {\bibfnamefont {L.~K.}\ \bibnamefont {Saini}},\ }\bibfield
  {title} {\bibinfo {title} {{Anomalous gauge couplings vis-\`a-vis
  (g-2)\ensuremath{\mu} and flavor observables}},\ }\href
  {https://doi.org/10.1103/PhysRevD.106.115026} {\bibfield  {journal} {\bibinfo
   {journal} {Phys. Rev. D}\ }\textbf {\bibinfo {volume} {106}},\ \bibinfo
  {pages} {115026} (\bibinfo {year} {2022})},\ \Eprint
  {https://arxiv.org/abs/2203.04673} {arXiv:2203.04673 [hep-ph]} \BibitemShut
  {NoStop}%
\bibitem [{\citenamefont {Rebello~Teles}(2013)}]{RebelloTeles:2013kdy}%
  \BibitemOpen
  \bibfield  {author} {\bibinfo {author} {\bibfnamefont {P.}~\bibnamefont
  {Rebello~Teles}} (\bibinfo {collaboration} {CMS}),\ }\bibfield  {title}
  {\bibinfo {title} {{Search for anomalous gauge couplings in semi-leptonic
  decays of $WW\gamma$ and $WZ\gamma$ in pp collisions at $\sqrt{s} =$ 8
  TeV}},\ }in\ \href@noop {} {\emph {\bibinfo {booktitle} {{Meeting of the APS
  Division of Particles and Fields}}}}\ (\bibinfo {year} {2013})\ \Eprint
  {https://arxiv.org/abs/1310.0473} {arXiv:1310.0473 [hep-ex]} \BibitemShut
  {NoStop}%
\bibitem [{\citenamefont {Tizchang}\ and\ \citenamefont
  {Etesami}(2020)}]{Tizchang:2020tqs}%
  \BibitemOpen
  \bibfield  {author} {\bibinfo {author} {\bibfnamefont {S.}~\bibnamefont
  {Tizchang}}\ and\ \bibinfo {author} {\bibfnamefont {S.~M.}\ \bibnamefont
  {Etesami}},\ }\bibfield  {title} {\bibinfo {title} {{Pinning down the gauge
  boson couplings in WW\ensuremath{\gamma} production using forward proton
  tagging}},\ }\href {https://doi.org/10.1007/JHEP07(2020)191} {\bibfield
  {journal} {\bibinfo  {journal} {JHEP}\ }\textbf {\bibinfo {volume} {07}},\
  \bibinfo {pages} {191}},\ \Eprint {https://arxiv.org/abs/2004.12203}
  {arXiv:2004.12203 [hep-ph]} \BibitemShut {NoStop}%
\bibitem [{\citenamefont {Campanario}\ \emph {et~al.}(2020)\citenamefont
  {Campanario}, \citenamefont {Kerner}, \citenamefont {Le},\ and\ \citenamefont
  {Rosario}}]{Campanario:2020xaf}%
  \BibitemOpen
  \bibfield  {author} {\bibinfo {author} {\bibfnamefont {F.}~\bibnamefont
  {Campanario}}, \bibinfo {author} {\bibfnamefont {M.}~\bibnamefont {Kerner}},
  \bibinfo {author} {\bibfnamefont {N.~D.}\ \bibnamefont {Le}},\ and\ \bibinfo
  {author} {\bibfnamefont {I.}~\bibnamefont {Rosario}},\ }\bibfield  {title}
  {\bibinfo {title} {{Diphoton production in vector-boson scattering at the LHC
  at next-to-leading order QCD}},\ }\href
  {https://doi.org/10.1007/JHEP06(2020)072} {\bibfield  {journal} {\bibinfo
  {journal} {JHEP}\ }\textbf {\bibinfo {volume} {06}},\ \bibinfo {pages}
  {072}},\ \Eprint {https://arxiv.org/abs/2002.12109} {arXiv:2002.12109
  [hep-ph]} \BibitemShut {NoStop}%
\bibitem [{\citenamefont {Ciulli}(2020)}]{Ciulli:2020ygo}%
  \BibitemOpen
  \bibfield  {author} {\bibinfo {author} {\bibfnamefont {V.}~\bibnamefont
  {Ciulli}} (\bibinfo {collaboration} {CMS}),\ }\bibfield  {title} {\bibinfo
  {title} {{Electroweak Measurements with the CMS Detector}},\ }\href
  {https://doi.org/10.5506/APhysPolB.51.1315} {\bibfield  {journal} {\bibinfo
  {journal} {Acta Phys. Polon. B}\ }\textbf {\bibinfo {volume} {51}},\ \bibinfo
  {pages} {1315} (\bibinfo {year} {2020})}\BibitemShut {NoStop}%
\bibitem [{\citenamefont {Yap}(2020)}]{Yap:2020xjr}%
  \BibitemOpen
  \bibfield  {author} {\bibinfo {author} {\bibfnamefont {Y.~C.}\ \bibnamefont
  {Yap}},\ }\bibfield  {title} {\bibinfo {title} {{Recent observation and
  measurements of diboson processes from the ATLAS experiment}},\ }\href
  {https://doi.org/10.1142/S021773232030013X} {\bibfield  {journal} {\bibinfo
  {journal} {Mod. Phys. Lett. A}\ }\textbf {\bibinfo {volume} {35}},\ \bibinfo
  {pages} {2030013} (\bibinfo {year} {2020})},\ \Eprint
  {https://arxiv.org/abs/2006.08285} {arXiv:2006.08285 [hep-ex]} \BibitemShut
  {NoStop}%
\bibitem [{\citenamefont {Hwang}\ \emph {et~al.}(2023)\citenamefont {Hwang},
  \citenamefont {Min}, \citenamefont {Park}, \citenamefont {Son},\ and\
  \citenamefont {Yoo}}]{Hwang:2023wad}%
  \BibitemOpen
  \bibfield  {author} {\bibinfo {author} {\bibfnamefont {H.}~\bibnamefont
  {Hwang}}, \bibinfo {author} {\bibfnamefont {U.}~\bibnamefont {Min}}, \bibinfo
  {author} {\bibfnamefont {J.}~\bibnamefont {Park}}, \bibinfo {author}
  {\bibfnamefont {M.}~\bibnamefont {Son}},\ and\ \bibinfo {author}
  {\bibfnamefont {J.~H.}\ \bibnamefont {Yoo}},\ }\bibfield  {title} {\bibinfo
  {title} {{Anomalous triple gauge couplings in electroweak dilepton tails at
  the LHC and interference resurrection}},\ }\href@noop {} {\  (\bibinfo {year}
  {2023})},\ \Eprint {https://arxiv.org/abs/2301.13663} {arXiv:2301.13663
  [hep-ph]} \BibitemShut {NoStop}%
\bibitem [{\citenamefont {Deka}\ \emph {et~al.}(2022)\citenamefont {Deka},
  \citenamefont {Maharana},\ and\ \citenamefont {Saini}}]{Deka:2022lmf}%
  \BibitemOpen
  \bibfield  {author} {\bibinfo {author} {\bibfnamefont {K.}~\bibnamefont
  {Deka}}, \bibinfo {author} {\bibfnamefont {S.}~\bibnamefont {Maharana}},\
  and\ \bibinfo {author} {\bibfnamefont {L.~K.}\ \bibnamefont {Saini}},\
  }\bibfield  {title} {\bibinfo {title} {{Status of anomalous triple gauge
  couplings in the light of recent results from muon ($g-2$) and other flavor
  observables}},\ }\href {https://doi.org/10.22323/1.414.1211} {\bibfield
  {journal} {\bibinfo  {journal} {PoS}\ }\textbf {\bibinfo {volume}
  {ICHEP2022}},\ \bibinfo {pages} {1211} (\bibinfo {year} {2022})}\BibitemShut
  {NoStop}%
\bibitem [{\citenamefont {Falkowski}\ \emph {et~al.}(2017)\citenamefont
  {Falkowski}, \citenamefont {Gonzalez-Alonso}, \citenamefont {Greljo},
  \citenamefont {Marzocca},\ and\ \citenamefont {Son}}]{Falkowski:2016cxu}%
  \BibitemOpen
  \bibfield  {author} {\bibinfo {author} {\bibfnamefont {A.}~\bibnamefont
  {Falkowski}}, \bibinfo {author} {\bibfnamefont {M.}~\bibnamefont
  {Gonzalez-Alonso}}, \bibinfo {author} {\bibfnamefont {A.}~\bibnamefont
  {Greljo}}, \bibinfo {author} {\bibfnamefont {D.}~\bibnamefont {Marzocca}},\
  and\ \bibinfo {author} {\bibfnamefont {M.}~\bibnamefont {Son}},\ }\bibfield
  {title} {\bibinfo {title} {{Anomalous Triple Gauge Couplings in the Effective
  Field Theory Approach at the LHC}},\ }\href
  {https://doi.org/10.1007/JHEP02(2017)115} {\bibfield  {journal} {\bibinfo
  {journal} {JHEP}\ }\textbf {\bibinfo {volume} {02}},\ \bibinfo {pages}
  {115}},\ \Eprint {https://arxiv.org/abs/1609.06312} {arXiv:1609.06312
  [hep-ph]} \BibitemShut {NoStop}%
\bibitem [{\citenamefont {Butter}\ \emph {et~al.}(2016)\citenamefont {Butter},
  \citenamefont {\'Eboli}, \citenamefont {Gonzalez-Fraile}, \citenamefont
  {Gonzalez-Garcia}, \citenamefont {Plehn},\ and\ \citenamefont
  {Rauch}}]{Butter:2016cvz}%
  \BibitemOpen
  \bibfield  {author} {\bibinfo {author} {\bibfnamefont {A.}~\bibnamefont
  {Butter}}, \bibinfo {author} {\bibfnamefont {O.~J.~P.}\ \bibnamefont
  {\'Eboli}}, \bibinfo {author} {\bibfnamefont {J.}~\bibnamefont
  {Gonzalez-Fraile}}, \bibinfo {author} {\bibfnamefont {M.~C.}\ \bibnamefont
  {Gonzalez-Garcia}}, \bibinfo {author} {\bibfnamefont {T.}~\bibnamefont
  {Plehn}},\ and\ \bibinfo {author} {\bibfnamefont {M.}~\bibnamefont {Rauch}},\
  }\bibfield  {title} {\bibinfo {title} {{The Gauge-Higgs Legacy of the LHC Run
  I}},\ }\href {https://doi.org/10.1007/JHEP07(2016)152} {\bibfield  {journal}
  {\bibinfo  {journal} {JHEP}\ }\textbf {\bibinfo {volume} {07}},\ \bibinfo
  {pages} {152}},\ \Eprint {https://arxiv.org/abs/1604.03105} {arXiv:1604.03105
  [hep-ph]} \BibitemShut {NoStop}%
\bibitem [{\citenamefont {Azatov}\ \emph {et~al.}(2017)\citenamefont {Azatov},
  \citenamefont {Elias-Miro}, \citenamefont {Reyimuaji},\ and\ \citenamefont
  {Venturini}}]{Azatov:2017kzw}%
  \BibitemOpen
  \bibfield  {author} {\bibinfo {author} {\bibfnamefont {A.}~\bibnamefont
  {Azatov}}, \bibinfo {author} {\bibfnamefont {J.}~\bibnamefont {Elias-Miro}},
  \bibinfo {author} {\bibfnamefont {Y.}~\bibnamefont {Reyimuaji}},\ and\
  \bibinfo {author} {\bibfnamefont {E.}~\bibnamefont {Venturini}},\ }\bibfield
  {title} {\bibinfo {title} {{Novel measurements of anomalous triple gauge
  couplings for the LHC}},\ }\href {https://doi.org/10.1007/JHEP10(2017)027}
  {\bibfield  {journal} {\bibinfo  {journal} {JHEP}\ }\textbf {\bibinfo
  {volume} {10}},\ \bibinfo {pages} {027}},\ \Eprint
  {https://arxiv.org/abs/1707.08060} {arXiv:1707.08060 [hep-ph]} \BibitemShut
  {NoStop}%
\bibitem [{\citenamefont {Baglio}\ \emph {et~al.}(2017)\citenamefont {Baglio},
  \citenamefont {Dawson},\ and\ \citenamefont {Lewis}}]{Baglio:2017bfe}%
  \BibitemOpen
  \bibfield  {author} {\bibinfo {author} {\bibfnamefont {J.}~\bibnamefont
  {Baglio}}, \bibinfo {author} {\bibfnamefont {S.}~\bibnamefont {Dawson}},\
  and\ \bibinfo {author} {\bibfnamefont {I.~M.}\ \bibnamefont {Lewis}},\
  }\bibfield  {title} {\bibinfo {title} {{An NLO QCD effective field theory
  analysis of $W^+W^-$ production at the LHC including fermionic operators}},\
  }\href {https://doi.org/10.1103/PhysRevD.96.073003} {\bibfield  {journal}
  {\bibinfo  {journal} {Phys. Rev. D}\ }\textbf {\bibinfo {volume} {96}},\
  \bibinfo {pages} {073003} (\bibinfo {year} {2017})},\ \Eprint
  {https://arxiv.org/abs/1708.03332} {arXiv:1708.03332 [hep-ph]} \BibitemShut
  {NoStop}%
\bibitem [{\citenamefont {Li}\ and\ \citenamefont
  {Valencia}(2017)}]{Li:2017esm}%
  \BibitemOpen
  \bibfield  {author} {\bibinfo {author} {\bibfnamefont {H.~T.}\ \bibnamefont
  {Li}}\ and\ \bibinfo {author} {\bibfnamefont {G.}~\bibnamefont {Valencia}},\
  }\bibfield  {title} {\bibinfo {title} {{CP violating anomalous couplings in
  $W$ jet production at the LHC}},\ }\href
  {https://doi.org/10.1103/PhysRevD.96.075014} {\bibfield  {journal} {\bibinfo
  {journal} {Phys. Rev. D}\ }\textbf {\bibinfo {volume} {96}},\ \bibinfo
  {pages} {075014} (\bibinfo {year} {2017})},\ \Eprint
  {https://arxiv.org/abs/1708.04402} {arXiv:1708.04402 [hep-ph]} \BibitemShut
  {NoStop}%
\bibitem [{\citenamefont {Bhatia}\ \emph {et~al.}(2019)\citenamefont {Bhatia},
  \citenamefont {Maitra},\ and\ \citenamefont {Raychaudhuri}}]{Bhatia:2018ndx}%
  \BibitemOpen
  \bibfield  {author} {\bibinfo {author} {\bibfnamefont {D.}~\bibnamefont
  {Bhatia}}, \bibinfo {author} {\bibfnamefont {U.}~\bibnamefont {Maitra}},\
  and\ \bibinfo {author} {\bibfnamefont {S.}~\bibnamefont {Raychaudhuri}},\
  }\bibfield  {title} {\bibinfo {title} {{Pinning down anomalous $WW\gamma$
  couplings at the LHC}},\ }\href {https://doi.org/10.1103/PhysRevD.99.095017}
  {\bibfield  {journal} {\bibinfo  {journal} {Phys. Rev. D}\ }\textbf {\bibinfo
  {volume} {99}},\ \bibinfo {pages} {095017} (\bibinfo {year} {2019})},\
  \Eprint {https://arxiv.org/abs/1804.05357} {arXiv:1804.05357 [hep-ph]}
  \BibitemShut {NoStop}%
\bibitem [{\citenamefont {Chiesa}\ \emph {et~al.}(2018)\citenamefont {Chiesa},
  \citenamefont {Denner},\ and\ \citenamefont {Lang}}]{Chiesa:2018lcs}%
  \BibitemOpen
  \bibfield  {author} {\bibinfo {author} {\bibfnamefont {M.}~\bibnamefont
  {Chiesa}}, \bibinfo {author} {\bibfnamefont {A.}~\bibnamefont {Denner}},\
  and\ \bibinfo {author} {\bibfnamefont {J.-N.}\ \bibnamefont {Lang}},\
  }\bibfield  {title} {\bibinfo {title} {{Anomalous triple-gauge-boson
  interactions in vector-boson pair production with RECOLA2}},\ }\href
  {https://doi.org/10.1140/epjc/s10052-018-5949-z} {\bibfield  {journal}
  {\bibinfo  {journal} {Eur. Phys. J. C}\ }\textbf {\bibinfo {volume} {78}},\
  \bibinfo {pages} {467} (\bibinfo {year} {2018})},\ \Eprint
  {https://arxiv.org/abs/1804.01477} {arXiv:1804.01477 [hep-ph]} \BibitemShut
  {NoStop}%
\bibitem [{\citenamefont {Rahaman}\ and\ \citenamefont
  {Singh}(2020{\natexlab{b}})}]{Rahaman:2019lab}%
  \BibitemOpen
  \bibfield  {author} {\bibinfo {author} {\bibfnamefont {R.}~\bibnamefont
  {Rahaman}}\ and\ \bibinfo {author} {\bibfnamefont {R.~K.}\ \bibnamefont
  {Singh}},\ }\bibfield  {title} {\bibinfo {title} {{Unravelling the anomalous
  gauge boson couplings in $ZW^\pm$ production at the LHC and the role of
  spin-$1$ polarizations}},\ }\href {https://doi.org/10.1007/JHEP04(2020)075}
  {\bibfield  {journal} {\bibinfo  {journal} {JHEP}\ }\textbf {\bibinfo
  {volume} {04}},\ \bibinfo {pages} {075}},\ \Eprint
  {https://arxiv.org/abs/1911.03111} {arXiv:1911.03111 [hep-ph]} \BibitemShut
  {NoStop}%
\bibitem [{\citenamefont {Dixon}\ \emph {et~al.}(1999)\citenamefont {Dixon},
  \citenamefont {Kunszt},\ and\ \citenamefont {Signer}}]{Dixon:1999di}%
  \BibitemOpen
  \bibfield  {author} {\bibinfo {author} {\bibfnamefont {L.~J.}\ \bibnamefont
  {Dixon}}, \bibinfo {author} {\bibfnamefont {Z.}~\bibnamefont {Kunszt}},\ and\
  \bibinfo {author} {\bibfnamefont {A.}~\bibnamefont {Signer}},\ }\bibfield
  {title} {\bibinfo {title} {{Vector boson pair production in hadronic
  collisions at order $\alpha_s$ : Lepton correlations and anomalous
  couplings}},\ }\href {https://doi.org/10.1103/PhysRevD.60.114037} {\bibfield
  {journal} {\bibinfo  {journal} {Phys. Rev. D}\ }\textbf {\bibinfo {volume}
  {60}},\ \bibinfo {pages} {114037} (\bibinfo {year} {1999})},\ \Eprint
  {https://arxiv.org/abs/hep-ph/9907305} {arXiv:hep-ph/9907305} \BibitemShut
  {NoStop}%
\bibitem [{\citenamefont {Baur}\ and\ \citenamefont
  {Zeppenfeld}(1988)}]{Baur:1987mt}%
  \BibitemOpen
  \bibfield  {author} {\bibinfo {author} {\bibfnamefont {U.}~\bibnamefont
  {Baur}}\ and\ \bibinfo {author} {\bibfnamefont {D.}~\bibnamefont
  {Zeppenfeld}},\ }\bibfield  {title} {\bibinfo {title} {{Unitarity Constraints
  on the Electroweak Three Vector Boson Vertices}},\ }\href
  {https://doi.org/10.1016/0370-2693(88)91160-4} {\bibfield  {journal}
  {\bibinfo  {journal} {Phys. Lett. B}\ }\textbf {\bibinfo {volume} {201}},\
  \bibinfo {pages} {383} (\bibinfo {year} {1988})}\BibitemShut {NoStop}%
\bibitem [{\citenamefont {Biswal}\ \emph {et~al.}(2014)\citenamefont {Biswal},
  \citenamefont {Patra},\ and\ \citenamefont {Raychaudhuri}}]{Biswal:2014oaa}%
  \BibitemOpen
  \bibfield  {author} {\bibinfo {author} {\bibfnamefont {S.~S.}\ \bibnamefont
  {Biswal}}, \bibinfo {author} {\bibfnamefont {M.}~\bibnamefont {Patra}},\ and\
  \bibinfo {author} {\bibfnamefont {S.}~\bibnamefont {Raychaudhuri}},\
  }\bibfield  {title} {\bibinfo {title} {{Anomalous Triple Gauge Vertices at
  the Large Hadron-Electron Collider}},\ }\href@noop {} {\  (\bibinfo {year}
  {2014})},\ \Eprint {https://arxiv.org/abs/1405.6056} {arXiv:1405.6056
  [hep-ph]} \BibitemShut {NoStop}%
\bibitem [{\citenamefont {Cakir}\ \emph {et~al.}(2014)\citenamefont {Cakir},
  \citenamefont {Cakir}, \citenamefont {Senol},\ and\ \citenamefont
  {Tasci}}]{Cakir:2014swa}%
  \BibitemOpen
  \bibfield  {author} {\bibinfo {author} {\bibfnamefont {I.~T.}\ \bibnamefont
  {Cakir}}, \bibinfo {author} {\bibfnamefont {O.}~\bibnamefont {Cakir}},
  \bibinfo {author} {\bibfnamefont {A.}~\bibnamefont {Senol}},\ and\ \bibinfo
  {author} {\bibfnamefont {A.~T.}\ \bibnamefont {Tasci}},\ }\bibfield  {title}
  {\bibinfo {title} {{Search for anomalous $WW\gamma$ and $WWZ$ couplings with
  polarized $e$-beam at the LHeC}},\ }\href
  {https://doi.org/10.5506/APhysPolB.45.1947} {\bibfield  {journal} {\bibinfo
  {journal} {Acta Phys. Polon. B}\ }\textbf {\bibinfo {volume} {45}},\ \bibinfo
  {pages} {1947} (\bibinfo {year} {2014})},\ \Eprint
  {https://arxiv.org/abs/1406.7696} {arXiv:1406.7696 [hep-ph]} \BibitemShut
  {NoStop}%
\bibitem [{\citenamefont {Li}\ \emph {et~al.}(2018)\citenamefont {Li},
  \citenamefont {Shen}, \citenamefont {Wang}, \citenamefont {Xu}, \citenamefont
  {Zhang},\ and\ \citenamefont {Zhu}}]{Li:2017kfk}%
  \BibitemOpen
  \bibfield  {author} {\bibinfo {author} {\bibfnamefont {R.}~\bibnamefont
  {Li}}, \bibinfo {author} {\bibfnamefont {X.-M.}\ \bibnamefont {Shen}},
  \bibinfo {author} {\bibfnamefont {K.}~\bibnamefont {Wang}}, \bibinfo {author}
  {\bibfnamefont {T.}~\bibnamefont {Xu}}, \bibinfo {author} {\bibfnamefont
  {L.}~\bibnamefont {Zhang}},\ and\ \bibinfo {author} {\bibfnamefont
  {G.}~\bibnamefont {Zhu}},\ }\bibfield  {title} {\bibinfo {title} {{Probing
  anomalous $WW\gamma$ triple gauge bosons coupling at the LHeC}},\ }\href
  {https://doi.org/10.1103/PhysRevD.97.075043} {\bibfield  {journal} {\bibinfo
  {journal} {Phys. Rev. D}\ }\textbf {\bibinfo {volume} {97}},\ \bibinfo
  {pages} {075043} (\bibinfo {year} {2018})},\ \Eprint
  {https://arxiv.org/abs/1711.05607} {arXiv:1711.05607 [hep-ph]} \BibitemShut
  {NoStop}%
\bibitem [{\citenamefont {K\"oksal}\ \emph {et~al.}(2020)\citenamefont
  {K\"oksal}, \citenamefont {Billur}, \citenamefont
  {Guti\'errez-Rodr\'\i{}guez},\ and\ \citenamefont
  {Hern\'andez-Ru\'\i{}z}}]{Koksal:2019oqt}%
  \BibitemOpen
  \bibfield  {author} {\bibinfo {author} {\bibfnamefont {M.}~\bibnamefont
  {K\"oksal}}, \bibinfo {author} {\bibfnamefont {A.~A.}\ \bibnamefont
  {Billur}}, \bibinfo {author} {\bibfnamefont {A.}~\bibnamefont
  {Guti\'errez-Rodr\'\i{}guez}},\ and\ \bibinfo {author} {\bibfnamefont
  {M.~A.}\ \bibnamefont {Hern\'andez-Ru\'\i{}z}},\ }\bibfield  {title}
  {\bibinfo {title} {{Bounds on the non-standard $W^+W^-\gamma$ couplings at
  the LHeC and the FCC-he}},\ }\href
  {https://doi.org/10.1016/j.physletb.2020.135661} {\bibfield  {journal}
  {\bibinfo  {journal} {Phys. Lett. B}\ }\textbf {\bibinfo {volume} {808}},\
  \bibinfo {pages} {135661} (\bibinfo {year} {2020})},\ \Eprint
  {https://arxiv.org/abs/1910.06747} {arXiv:1910.06747 [hep-ph]} \BibitemShut
  {NoStop}%
\bibitem [{\citenamefont {Guti\'errez-Rodr\'\i{}guez}\ \emph
  {et~al.}(2020)\citenamefont {Guti\'errez-Rodr\'\i{}guez}, \citenamefont
  {K\"oksal}, \citenamefont {Billur},\ and\ \citenamefont
  {Hern\'andez-Ru\'\i{}z}}]{Gutierrez-Rodriguez:2019hek}%
  \BibitemOpen
  \bibfield  {author} {\bibinfo {author} {\bibfnamefont {A.}~\bibnamefont
  {Guti\'errez-Rodr\'\i{}guez}}, \bibinfo {author} {\bibfnamefont
  {M.}~\bibnamefont {K\"oksal}}, \bibinfo {author} {\bibfnamefont {A.~A.}\
  \bibnamefont {Billur}},\ and\ \bibinfo {author} {\bibfnamefont {M.~A.}\
  \bibnamefont {Hern\'andez-Ru\'\i{}z}},\ }\bibfield  {title} {\bibinfo {title}
  {{Probing model-independent limits on $W^+W^-\gamma$ triple gauge boson
  vertex at the LHeC and the FCC-he}},\ }\href
  {https://doi.org/10.1088/1361-6471/ab7ff9} {\bibfield  {journal} {\bibinfo
  {journal} {J. Phys. G}\ }\textbf {\bibinfo {volume} {47}},\ \bibinfo {pages}
  {055005} (\bibinfo {year} {2020})},\ \Eprint
  {https://arxiv.org/abs/1910.02307} {arXiv:1910.02307 [hep-ph]} \BibitemShut
  {NoStop}%
\bibitem [{\citenamefont {Abbiendi}\ \emph
  {et~al.}(2001{\natexlab{a}})\citenamefont {Abbiendi} \emph
  {et~al.}}]{OPAL:2000wbs}%
  \BibitemOpen
  \bibfield  {author} {\bibinfo {author} {\bibfnamefont {G.}~\bibnamefont
  {Abbiendi}} \emph {et~al.} (\bibinfo {collaboration} {OPAL}),\ }\bibfield
  {title} {\bibinfo {title} {{Measurement of $W$ boson polarizations and CP
  violating triple gauge couplings from $W^{+} W^{-}$ production at LEP}},\
  }\href {https://doi.org/10.1007/s100520100602} {\bibfield  {journal}
  {\bibinfo  {journal} {Eur. Phys. J. C}\ }\textbf {\bibinfo {volume} {19}},\
  \bibinfo {pages} {229} (\bibinfo {year} {2001}{\natexlab{a}})},\ \Eprint
  {https://arxiv.org/abs/hep-ex/0009021} {arXiv:hep-ex/0009021} \BibitemShut
  {NoStop}%
\bibitem [{\citenamefont {Abbiendi}\ \emph {et~al.}(2004)\citenamefont
  {Abbiendi} \emph {et~al.}}]{OPAL:2003xqq}%
  \BibitemOpen
  \bibfield  {author} {\bibinfo {author} {\bibfnamefont {G.}~\bibnamefont
  {Abbiendi}} \emph {et~al.} (\bibinfo {collaboration} {OPAL}),\ }\bibfield
  {title} {\bibinfo {title} {{Measurement of charged current triple gauge boson
  couplings using $W$ pairs at LEP}},\ }\href
  {https://doi.org/10.1140/epjc/s2003-01524-6} {\bibfield  {journal} {\bibinfo
  {journal} {Eur. Phys. J. C}\ }\textbf {\bibinfo {volume} {33}},\ \bibinfo
  {pages} {463} (\bibinfo {year} {2004})},\ \Eprint
  {https://arxiv.org/abs/hep-ex/0308067} {arXiv:hep-ex/0308067} \BibitemShut
  {NoStop}%
\bibitem [{\citenamefont {Abbiendi}\ \emph
  {et~al.}(2001{\natexlab{b}})\citenamefont {Abbiendi} \emph
  {et~al.}}]{OPAL:2000rnf}%
  \BibitemOpen
  \bibfield  {author} {\bibinfo {author} {\bibfnamefont {G.}~\bibnamefont
  {Abbiendi}} \emph {et~al.} (\bibinfo {collaboration} {OPAL}),\ }\bibfield
  {title} {\bibinfo {title} {{Measurement of triple gauge boson couplings from
  $W^{+} W^{-}$ production at LEP energies up to 189-GeV}},\ }\href
  {https://doi.org/10.1007/s100520100597} {\bibfield  {journal} {\bibinfo
  {journal} {Eur. Phys. J. C}\ }\textbf {\bibinfo {volume} {19}},\ \bibinfo
  {pages} {1} (\bibinfo {year} {2001}{\natexlab{b}})},\ \Eprint
  {https://arxiv.org/abs/hep-ex/0009022} {arXiv:hep-ex/0009022} \BibitemShut
  {NoStop}%
\bibitem [{\citenamefont {Abbiendi}\ \emph {et~al.}(1999)\citenamefont
  {Abbiendi} \emph {et~al.}}]{OPAL:1998ixj}%
  \BibitemOpen
  \bibfield  {author} {\bibinfo {author} {\bibfnamefont {G.}~\bibnamefont
  {Abbiendi}} \emph {et~al.} (\bibinfo {collaboration} {OPAL}),\ }\bibfield
  {title} {\bibinfo {title} {{$W^{+} W^{-}$ production and triple gauge boson
  couplings at LEP energies up to 183-GeV}},\ }\href
  {https://doi.org/10.1007/s100529901106} {\bibfield  {journal} {\bibinfo
  {journal} {Eur. Phys. J. C}\ }\textbf {\bibinfo {volume} {8}},\ \bibinfo
  {pages} {191} (\bibinfo {year} {1999})},\ \Eprint
  {https://arxiv.org/abs/hep-ex/9811028} {arXiv:hep-ex/9811028} \BibitemShut
  {NoStop}%
\bibitem [{\citenamefont {Ackerstaff}\ \emph {et~al.}(1998)\citenamefont
  {Ackerstaff} \emph {et~al.}}]{OPAL:1997yzg}%
  \BibitemOpen
  \bibfield  {author} {\bibinfo {author} {\bibfnamefont {K.}~\bibnamefont
  {Ackerstaff}} \emph {et~al.} (\bibinfo {collaboration} {OPAL}),\ }\bibfield
  {title} {\bibinfo {title} {{Measurement of triple gauge boson couplings from
  $W^{+} W^{-}$ production at $S^{(1/2)}$ = 172-GeV}},\ }\href
  {https://doi.org/10.1007/s100520050164} {\bibfield  {journal} {\bibinfo
  {journal} {Eur. Phys. J. C}\ }\textbf {\bibinfo {volume} {2}},\ \bibinfo
  {pages} {597} (\bibinfo {year} {1998})},\ \Eprint
  {https://arxiv.org/abs/hep-ex/9709023} {arXiv:hep-ex/9709023} \BibitemShut
  {NoStop}%
\bibitem [{\citenamefont {Ackerstaff}\ \emph {et~al.}(1997)\citenamefont
  {Ackerstaff} \emph {et~al.}}]{OPAL:1997xsu}%
  \BibitemOpen
  \bibfield  {author} {\bibinfo {author} {\bibfnamefont {K.}~\bibnamefont
  {Ackerstaff}} \emph {et~al.} (\bibinfo {collaboration} {OPAL}),\ }\bibfield
  {title} {\bibinfo {title} {{Measurement of the triple gauge boson coupling
  $\alpha$ (w $\phi^{)}$ from $W^{+} W^{-}$ production in $e^{+} e^{-}$
  collisions at $\sqrt{s}$ = 161-GeV}},\ }\href
  {https://doi.org/10.1016/S0370-2693(97)00162-7} {\bibfield  {journal}
  {\bibinfo  {journal} {Phys. Lett. B}\ }\textbf {\bibinfo {volume} {397}},\
  \bibinfo {pages} {147} (\bibinfo {year} {1997})}\BibitemShut {NoStop}%
\bibitem [{\citenamefont {Schael}\ \emph {et~al.}(2005)\citenamefont {Schael}
  \emph {et~al.}}]{ALEPH:2004klc}%
  \BibitemOpen
  \bibfield  {author} {\bibinfo {author} {\bibfnamefont {S.}~\bibnamefont
  {Schael}} \emph {et~al.} (\bibinfo {collaboration} {ALEPH}),\ }\bibfield
  {title} {\bibinfo {title} {{Improved measurement of the triple gauge-boson
  couplings gamma W W and Z W W in e+ e- collisions}},\ }\href
  {https://doi.org/10.1016/j.physletb.2005.03.058} {\bibfield  {journal}
  {\bibinfo  {journal} {Phys. Lett. B}\ }\textbf {\bibinfo {volume} {614}},\
  \bibinfo {pages} {7} (\bibinfo {year} {2005})}\BibitemShut {NoStop}%
\bibitem [{\citenamefont {Heister}\ \emph {et~al.}(2001)\citenamefont {Heister}
  \emph {et~al.}}]{ALEPH:2001ylz}%
  \BibitemOpen
  \bibfield  {author} {\bibinfo {author} {\bibfnamefont {A.}~\bibnamefont
  {Heister}} \emph {et~al.} (\bibinfo {collaboration} {ALEPH}),\ }\bibfield
  {title} {\bibinfo {title} {{Measurement of triple gauge boson couplings at
  LEP energies up to 189-GeV}},\ }\href {https://doi.org/10.1007/s100520100730}
  {\bibfield  {journal} {\bibinfo  {journal} {Eur. Phys. J. C}\ }\textbf
  {\bibinfo {volume} {21}},\ \bibinfo {pages} {423} (\bibinfo {year} {2001})},\
  \Eprint {https://arxiv.org/abs/hep-ex/0104034} {arXiv:hep-ex/0104034}
  \BibitemShut {NoStop}%
\bibitem [{\citenamefont {Barate}\ \emph
  {et~al.}(1998{\natexlab{a}})\citenamefont {Barate} \emph
  {et~al.}}]{ALEPH:1998nxw}%
  \BibitemOpen
  \bibfield  {author} {\bibinfo {author} {\bibfnamefont {R.}~\bibnamefont
  {Barate}} \emph {et~al.} (\bibinfo {collaboration} {ALEPH}),\ }\bibfield
  {title} {\bibinfo {title} {{Measurement of triple gauge W W gamma couplings
  at LEP-2 using photonic events}},\ }\href
  {https://doi.org/10.1016/S0370-2693(98)01475-0} {\bibfield  {journal}
  {\bibinfo  {journal} {Phys. Lett. B}\ }\textbf {\bibinfo {volume} {445}},\
  \bibinfo {pages} {239} (\bibinfo {year} {1998}{\natexlab{a}})},\ \Eprint
  {https://arxiv.org/abs/hep-ex/9901030} {arXiv:hep-ex/9901030} \BibitemShut
  {NoStop}%
\bibitem [{\citenamefont {Barate}\ \emph
  {et~al.}(1998{\natexlab{b}})\citenamefont {Barate} \emph
  {et~al.}}]{ALEPH:1997agc}%
  \BibitemOpen
  \bibfield  {author} {\bibinfo {author} {\bibfnamefont {R.}~\bibnamefont
  {Barate}} \emph {et~al.} (\bibinfo {collaboration} {ALEPH}),\ }\bibfield
  {title} {\bibinfo {title} {{Measurement of triple gauge boson couplings at
  172-GeV}},\ }\href {https://doi.org/10.1016/S0370-2693(98)00061-6} {\bibfield
   {journal} {\bibinfo  {journal} {Phys. Lett. B}\ }\textbf {\bibinfo {volume}
  {422}},\ \bibinfo {pages} {369} (\bibinfo {year}
  {1998}{\natexlab{b}})}\BibitemShut {NoStop}%
\bibitem [{\citenamefont {Schael}\ \emph {et~al.}(2013)\citenamefont {Schael}
  \emph {et~al.}}]{ALEPH:2013dgf}%
  \BibitemOpen
  \bibfield  {author} {\bibinfo {author} {\bibfnamefont {S.}~\bibnamefont
  {Schael}} \emph {et~al.} (\bibinfo {collaboration} {ALEPH, DELPHI, L3, OPAL,
  LEP Electroweak}),\ }\bibfield  {title} {\bibinfo {title} {{Electroweak
  Measurements in Electron-Positron Collisions at W-Boson-Pair Energies at
  LEP}},\ }\href {https://doi.org/10.1016/j.physrep.2013.07.004} {\bibfield
  {journal} {\bibinfo  {journal} {Phys. Rept.}\ }\textbf {\bibinfo {volume}
  {532}},\ \bibinfo {pages} {119} (\bibinfo {year} {2013})},\ \Eprint
  {https://arxiv.org/abs/1302.3415} {arXiv:1302.3415 [hep-ex]} \BibitemShut
  {NoStop}%
\bibitem [{\citenamefont {Abdallah}\ \emph {et~al.}(2008)\citenamefont
  {Abdallah} \emph {et~al.}}]{DELPHI:2008uqu}%
  \BibitemOpen
  \bibfield  {author} {\bibinfo {author} {\bibfnamefont {J.}~\bibnamefont
  {Abdallah}} \emph {et~al.} (\bibinfo {collaboration} {DELPHI}),\ }\bibfield
  {title} {\bibinfo {title} {{Study of W boson polarisations and Triple Gauge
  boson Couplings in the reaction e+e- ---\ensuremath{>} W+W- at LEP 2}},\
  }\href {https://doi.org/10.1140/epjc/s10052-008-0528-3} {\bibfield  {journal}
  {\bibinfo  {journal} {Eur. Phys. J. C}\ }\textbf {\bibinfo {volume} {54}},\
  \bibinfo {pages} {345} (\bibinfo {year} {2008})},\ \Eprint
  {https://arxiv.org/abs/0801.1235} {arXiv:0801.1235 [hep-ex]} \BibitemShut
  {NoStop}%
\bibitem [{\citenamefont {Aaltonen}\ \emph {et~al.}(2012)\citenamefont
  {Aaltonen} \emph {et~al.}}]{CDF:2012mnr}%
  \BibitemOpen
  \bibfield  {author} {\bibinfo {author} {\bibfnamefont {T.}~\bibnamefont
  {Aaltonen}} \emph {et~al.} (\bibinfo {collaboration} {CDF}),\ }\bibfield
  {title} {\bibinfo {title} {{Measurement of the $WZ$ Cross Section and Triple
  Gauge Couplings in $p \bar p$ Collisions at $\sqrt{s} = 1.96$ TeV}},\ }\href
  {https://doi.org/10.1103/PhysRevD.86.031104} {\bibfield  {journal} {\bibinfo
  {journal} {Phys. Rev. D}\ }\textbf {\bibinfo {volume} {86}},\ \bibinfo
  {pages} {031104} (\bibinfo {year} {2012})},\ \Eprint
  {https://arxiv.org/abs/1202.6629} {arXiv:1202.6629 [hep-ex]} \BibitemShut
  {NoStop}%
\bibitem [{\citenamefont {Aaltonen}\ \emph {et~al.}(2007)\citenamefont
  {Aaltonen} \emph {et~al.}}]{CDF:2007aqs}%
  \BibitemOpen
  \bibfield  {author} {\bibinfo {author} {\bibfnamefont {T.}~\bibnamefont
  {Aaltonen}} \emph {et~al.} (\bibinfo {collaboration} {CDF}),\ }\bibfield
  {title} {\bibinfo {title} {{Limits on Anomalous Triple Gauge Couplings in $p
  \bar{p}$ Collisions at $\sqrt{s}$ = 1.96-TeV}},\ }\href
  {https://doi.org/10.1103/PhysRevD.76.111103} {\bibfield  {journal} {\bibinfo
  {journal} {Phys. Rev. D}\ }\textbf {\bibinfo {volume} {76}},\ \bibinfo
  {pages} {111103} (\bibinfo {year} {2007})},\ \Eprint
  {https://arxiv.org/abs/0705.2247} {arXiv:0705.2247 [hep-ex]} \BibitemShut
  {NoStop}%
\bibitem [{\citenamefont {Abazov}\ \emph {et~al.}(2013)\citenamefont {Abazov}
  \emph {et~al.}}]{D0:2013rce}%
  \BibitemOpen
  \bibfield  {author} {\bibinfo {author} {\bibfnamefont {V.~M.}\ \bibnamefont
  {Abazov}} \emph {et~al.} (\bibinfo {collaboration} {D0}),\ }\bibfield
  {title} {\bibinfo {title} {{Search for anomalous quartic $WW{\gamma}{\gamma}$
  couplings in dielectron and missing energy final states in $p\bar{p}$
  collisions at $\sqrt{s}$ = 1.96 TeV}},\ }\href
  {https://doi.org/10.1103/PhysRevD.88.012005} {\bibfield  {journal} {\bibinfo
  {journal} {Phys. Rev. D}\ }\textbf {\bibinfo {volume} {88}},\ \bibinfo
  {pages} {012005} (\bibinfo {year} {2013})},\ \Eprint
  {https://arxiv.org/abs/1305.1258} {arXiv:1305.1258 [hep-ex]} \BibitemShut
  {NoStop}%
\bibitem [{\citenamefont {Abazov}\ \emph {et~al.}(2011)\citenamefont {Abazov}
  \emph {et~al.}}]{D0:2010jca}%
  \BibitemOpen
  \bibfield  {author} {\bibinfo {author} {\bibfnamefont {V.~M.}\ \bibnamefont
  {Abazov}} \emph {et~al.} (\bibinfo {collaboration} {D0}),\ }\bibfield
  {title} {\bibinfo {title} {{Measurement of the $WZ\rightarrow
  \ell\nu\ell\ell$ Cross Section and Limits on Anomalous Triple Gauge Couplings
  in $p\bar{p}$ Collisions at $\sqrt{s}$ = 1.96 TeV}},\ }\href
  {https://doi.org/10.1016/j.physletb.2010.10.047} {\bibfield  {journal}
  {\bibinfo  {journal} {Phys. Lett. B}\ }\textbf {\bibinfo {volume} {695}},\
  \bibinfo {pages} {67} (\bibinfo {year} {2011})},\ \Eprint
  {https://arxiv.org/abs/1006.0761} {arXiv:1006.0761 [hep-ex]} \BibitemShut
  {NoStop}%
\bibitem [{\citenamefont {Abazov}\ \emph {et~al.}(2005)\citenamefont {Abazov}
  \emph {et~al.}}]{D0:2005djn}%
  \BibitemOpen
  \bibfield  {author} {\bibinfo {author} {\bibfnamefont {V.~M.}\ \bibnamefont
  {Abazov}} \emph {et~al.} (\bibinfo {collaboration} {D0}),\ }\bibfield
  {title} {\bibinfo {title} {{Measurement of the $p$ - $\bar{p} \to W \gamma$ +
  $X$ cross section at $\sqrt{s}$ = 1.96-TeV and $W W \gamma$ anomalous
  coupling limits}},\ }\href {https://doi.org/10.1103/PhysRevD.71.091108}
  {\bibfield  {journal} {\bibinfo  {journal} {Phys. Rev. D}\ }\textbf {\bibinfo
  {volume} {71}},\ \bibinfo {pages} {091108} (\bibinfo {year} {2005})},\
  \Eprint {https://arxiv.org/abs/hep-ex/0503048} {arXiv:hep-ex/0503048}
  \BibitemShut {NoStop}%
\bibitem [{\citenamefont {Aaboud}\ \emph
  {et~al.}(2017{\natexlab{a}})\citenamefont {Aaboud} \emph
  {et~al.}}]{ATLAS:2017pbb}%
  \BibitemOpen
  \bibfield  {author} {\bibinfo {author} {\bibfnamefont {M.}~\bibnamefont
  {Aaboud}} \emph {et~al.} (\bibinfo {collaboration} {ATLAS}),\ }\bibfield
  {title} {\bibinfo {title} {{Measurement of $WW/WZ \to \ell \nu q q^{\prime}$
  production with the hadronically decaying boson reconstructed as one or two
  jets in $pp$ collisions at $\sqrt{s}=8$ TeV with ATLAS, and constraints on
  anomalous gauge couplings}},\ }\href
  {https://doi.org/10.1140/epjc/s10052-017-5084-2} {\bibfield  {journal}
  {\bibinfo  {journal} {Eur. Phys. J. C}\ }\textbf {\bibinfo {volume} {77}},\
  \bibinfo {pages} {563} (\bibinfo {year} {2017}{\natexlab{a}})},\ \Eprint
  {https://arxiv.org/abs/1706.01702} {arXiv:1706.01702 [hep-ex]} \BibitemShut
  {NoStop}%
\bibitem [{\citenamefont {Aaboud}\ \emph
  {et~al.}(2017{\natexlab{b}})\citenamefont {Aaboud} \emph
  {et~al.}}]{ATLAS:2017luz}%
  \BibitemOpen
  \bibfield  {author} {\bibinfo {author} {\bibfnamefont {M.}~\bibnamefont
  {Aaboud}} \emph {et~al.} (\bibinfo {collaboration} {ATLAS}),\ }\bibfield
  {title} {\bibinfo {title} {{Measurements of electroweak $Wjj$ production and
  constraints on anomalous gauge couplings with the ATLAS detector}},\ }\href
  {https://doi.org/10.1140/epjc/s10052-017-5007-2} {\bibfield  {journal}
  {\bibinfo  {journal} {Eur. Phys. J. C}\ }\textbf {\bibinfo {volume} {77}},\
  \bibinfo {pages} {474} (\bibinfo {year} {2017}{\natexlab{b}})},\ \Eprint
  {https://arxiv.org/abs/1703.04362} {arXiv:1703.04362 [hep-ex]} \BibitemShut
  {NoStop}%
\bibitem [{\citenamefont {Aad}\ \emph {et~al.}(2016{\natexlab{a}})\citenamefont
  {Aad} \emph {et~al.}}]{ATLAS:2016zwm}%
  \BibitemOpen
  \bibfield  {author} {\bibinfo {author} {\bibfnamefont {G.}~\bibnamefont
  {Aad}} \emph {et~al.} (\bibinfo {collaboration} {ATLAS}),\ }\bibfield
  {title} {\bibinfo {title} {{Measurement of total and differential $W^+W^-$
  production cross sections in proton-proton collisions at $\sqrt{s}=$ 8 TeV
  with the ATLAS detector and limits on anomalous triple-gauge-boson
  couplings}},\ }\href {https://doi.org/10.1007/JHEP09(2016)029} {\bibfield
  {journal} {\bibinfo  {journal} {JHEP}\ }\textbf {\bibinfo {volume} {09}},\
  \bibinfo {pages} {029}},\ \Eprint {https://arxiv.org/abs/1603.01702}
  {arXiv:1603.01702 [hep-ex]} \BibitemShut {NoStop}%
\bibitem [{\citenamefont {Aad}\ \emph {et~al.}(2016{\natexlab{b}})\citenamefont
  {Aad} \emph {et~al.}}]{ATLAS:2016bkj}%
  \BibitemOpen
  \bibfield  {author} {\bibinfo {author} {\bibfnamefont {G.}~\bibnamefont
  {Aad}} \emph {et~al.} (\bibinfo {collaboration} {ATLAS}),\ }\bibfield
  {title} {\bibinfo {title} {{Measurements of $W^\pm Z$ production cross
  sections in $pp$ collisions at $\sqrt{s} = 8$ TeV with the ATLAS detector and
  limits on anomalous gauge boson self-couplings}},\ }\href
  {https://doi.org/10.1103/PhysRevD.93.092004} {\bibfield  {journal} {\bibinfo
  {journal} {Phys. Rev. D}\ }\textbf {\bibinfo {volume} {93}},\ \bibinfo
  {pages} {092004} (\bibinfo {year} {2016}{\natexlab{b}})},\ \Eprint
  {https://arxiv.org/abs/1603.02151} {arXiv:1603.02151 [hep-ex]} \BibitemShut
  {NoStop}%
\bibitem [{\citenamefont {Aad}\ \emph {et~al.}(2015)\citenamefont {Aad} \emph
  {et~al.}}]{ATLAS:2014ofc}%
  \BibitemOpen
  \bibfield  {author} {\bibinfo {author} {\bibfnamefont {G.}~\bibnamefont
  {Aad}} \emph {et~al.} (\bibinfo {collaboration} {ATLAS}),\ }\bibfield
  {title} {\bibinfo {title} {{Measurement of the $WW+WZ$ cross section and
  limits on anomalous triple gauge couplings using final states with one
  lepton, missing transverse momentum, and two jets with the ATLAS detector at
  $\sqrt{\rm{s}} = 7$ TeV}},\ }\href {https://doi.org/10.1007/JHEP01(2015)049}
  {\bibfield  {journal} {\bibinfo  {journal} {JHEP}\ }\textbf {\bibinfo
  {volume} {01}},\ \bibinfo {pages} {049}},\ \Eprint
  {https://arxiv.org/abs/1410.7238} {arXiv:1410.7238 [hep-ex]} \BibitemShut
  {NoStop}%
\bibitem [{\citenamefont {Aad}\ \emph {et~al.}(2013)\citenamefont {Aad} \emph
  {et~al.}}]{ATLAS:2013way}%
  \BibitemOpen
  \bibfield  {author} {\bibinfo {author} {\bibfnamefont {G.}~\bibnamefont
  {Aad}} \emph {et~al.} (\bibinfo {collaboration} {ATLAS}),\ }\bibfield
  {title} {\bibinfo {title} {{Measurements of $W \gamma$ and $Z \gamma$
  production in $pp$ collisions at $\sqrt{s}$=7 TeV with the ATLAS detector at
  the LHC}},\ }\href {https://doi.org/10.1103/PhysRevD.87.112003} {\bibfield
  {journal} {\bibinfo  {journal} {Phys. Rev. D}\ }\textbf {\bibinfo {volume}
  {87}},\ \bibinfo {pages} {112003} (\bibinfo {year} {2013})},\ \bibinfo {note}
  {[Erratum: Phys.Rev.D 91, 119901 (2015)]},\ \Eprint
  {https://arxiv.org/abs/1302.1283} {arXiv:1302.1283 [hep-ex]} \BibitemShut
  {NoStop}%
\bibitem [{\citenamefont {Aad}\ \emph {et~al.}(2012{\natexlab{b}})\citenamefont
  {Aad} \emph {et~al.}}]{ATLAS:2012aid}%
  \BibitemOpen
  \bibfield  {author} {\bibinfo {author} {\bibfnamefont {G.}~\bibnamefont
  {Aad}} \emph {et~al.} (\bibinfo {collaboration} {ATLAS}),\ }\bibfield
  {title} {\bibinfo {title} {{Measurement of $WZ$ production in proton-proton
  collisions at $\sqrt{s}=7$ TeV with the ATLAS detector}},\ }\href
  {https://doi.org/10.1140/epjc/s10052-012-2173-0} {\bibfield  {journal}
  {\bibinfo  {journal} {Eur. Phys. J. C}\ }\textbf {\bibinfo {volume} {72}},\
  \bibinfo {pages} {2173} (\bibinfo {year} {2012}{\natexlab{b}})},\ \Eprint
  {https://arxiv.org/abs/1208.1390} {arXiv:1208.1390 [hep-ex]} \BibitemShut
  {NoStop}%
\bibitem [{\citenamefont {Aad}\ \emph {et~al.}(2012{\natexlab{c}})\citenamefont
  {Aad} \emph {et~al.}}]{ATLAS:2012bpb}%
  \BibitemOpen
  \bibfield  {author} {\bibinfo {author} {\bibfnamefont {G.}~\bibnamefont
  {Aad}} \emph {et~al.} (\bibinfo {collaboration} {ATLAS}),\ }\bibfield
  {title} {\bibinfo {title} {{Measurement of $W \gamma$ and $Z \gamma$
  production cross sections in $pp$ collisions at $\sqrt{s}=7$ TeV and limits
  on anomalous triple gauge couplings with the ATLAS detector}},\ }\href
  {https://doi.org/10.1016/j.physletb.2012.09.017} {\bibfield  {journal}
  {\bibinfo  {journal} {Phys. Lett. B}\ }\textbf {\bibinfo {volume} {717}},\
  \bibinfo {pages} {49} (\bibinfo {year} {2012}{\natexlab{c}})},\ \Eprint
  {https://arxiv.org/abs/1205.2531} {arXiv:1205.2531 [hep-ex]} \BibitemShut
  {NoStop}%
\bibitem [{\citenamefont {Aad}\ \emph {et~al.}(2012{\natexlab{d}})\citenamefont
  {Aad} \emph {et~al.}}]{ATLAS:2012upi}%
  \BibitemOpen
  \bibfield  {author} {\bibinfo {author} {\bibfnamefont {G.}~\bibnamefont
  {Aad}} \emph {et~al.} (\bibinfo {collaboration} {ATLAS}),\ }\bibfield
  {title} {\bibinfo {title} {{Measurement of the $W W$ cross section in
  $\sqrt{s}=7$ TeV $pp$ collisions with the ATLAS detector and limits on
  anomalous gauge couplings}},\ }\href
  {https://doi.org/10.1016/j.physletb.2012.05.003} {\bibfield  {journal}
  {\bibinfo  {journal} {Phys. Lett. B}\ }\textbf {\bibinfo {volume} {712}},\
  \bibinfo {pages} {289} (\bibinfo {year} {2012}{\natexlab{d}})},\ \Eprint
  {https://arxiv.org/abs/1203.6232} {arXiv:1203.6232 [hep-ex]} \BibitemShut
  {NoStop}%
\bibitem [{\citenamefont {Aad}\ \emph {et~al.}(2012{\natexlab{e}})\citenamefont
  {Aad} \emph {et~al.}}]{ATLAS:2011nle}%
  \BibitemOpen
  \bibfield  {author} {\bibinfo {author} {\bibfnamefont {G.}~\bibnamefont
  {Aad}} \emph {et~al.} (\bibinfo {collaboration} {ATLAS}),\ }\bibfield
  {title} {\bibinfo {title} {{Measurement of the $W^\pm Z$ production cross
  section and limits on anomalous triple gauge couplings in proton-proton
  collisions at $\sqrt{s}=7$ TeV with the ATLAS detector}},\ }\href
  {https://doi.org/10.1016/j.physletb.2012.02.053} {\bibfield  {journal}
  {\bibinfo  {journal} {Phys. Lett. B}\ }\textbf {\bibinfo {volume} {709}},\
  \bibinfo {pages} {341} (\bibinfo {year} {2012}{\natexlab{e}})},\ \Eprint
  {https://arxiv.org/abs/1111.5570} {arXiv:1111.5570 [hep-ex]} \BibitemShut
  {NoStop}%
\bibitem [{\citenamefont {Tumasyan}\ \emph {et~al.}(2022)\citenamefont
  {Tumasyan} \emph {et~al.}}]{CMS:2021icx}%
  \BibitemOpen
  \bibfield  {author} {\bibinfo {author} {\bibfnamefont {A.}~\bibnamefont
  {Tumasyan}} \emph {et~al.} (\bibinfo {collaboration} {CMS}),\ }\bibfield
  {title} {\bibinfo {title} {{Measurement of the inclusive and differential WZ
  production cross sections, polarization angles, and triple gauge couplings in
  pp collisions at $ \sqrt{s} $ = 13 TeV}},\ }\href
  {https://doi.org/10.1007/JHEP07(2022)032} {\bibfield  {journal} {\bibinfo
  {journal} {JHEP}\ }\textbf {\bibinfo {volume} {07}},\ \bibinfo {pages}
  {032}},\ \Eprint {https://arxiv.org/abs/2110.11231} {arXiv:2110.11231
  [hep-ex]} \BibitemShut {NoStop}%
\bibitem [{\citenamefont {Sirunyan}\ \emph {et~al.}(2021)\citenamefont
  {Sirunyan} \emph {et~al.}}]{CMS:2021foa}%
  \BibitemOpen
  \bibfield  {author} {\bibinfo {author} {\bibfnamefont {A.~M.}\ \bibnamefont
  {Sirunyan}} \emph {et~al.} (\bibinfo {collaboration} {CMS}),\ }\bibfield
  {title} {\bibinfo {title} {{Measurement of the W$\gamma$ Production Cross
  Section in Proton-Proton Collisions at $\sqrt {s}$=13\,\,TeV and Constraints
  on Effective Field Theory Coefficients}},\ }\href
  {https://doi.org/10.1103/PhysRevLett.126.252002} {\bibfield  {journal}
  {\bibinfo  {journal} {Phys. Rev. Lett.}\ }\textbf {\bibinfo {volume} {126}},\
  \bibinfo {pages} {252002} (\bibinfo {year} {2021})},\ \Eprint
  {https://arxiv.org/abs/2102.02283} {arXiv:2102.02283 [hep-ex]} \BibitemShut
  {NoStop}%
\bibitem [{\citenamefont {Khachatryan}\ \emph {et~al.}(2017)\citenamefont
  {Khachatryan} \emph {et~al.}}]{CMS:2016qth}%
  \BibitemOpen
  \bibfield  {author} {\bibinfo {author} {\bibfnamefont {V.}~\bibnamefont
  {Khachatryan}} \emph {et~al.} (\bibinfo {collaboration} {CMS}),\ }\bibfield
  {title} {\bibinfo {title} {{Measurement of the WZ production cross section in
  pp collisions at $\sqrt{s} = 7$ and 8 $\,\text{TeV}$ and search for anomalous
  triple gauge couplings at $\sqrt{s} = 8\,\text{TeV} $}},\ }\href
  {https://doi.org/10.1140/epjc/s10052-017-4730-z} {\bibfield  {journal}
  {\bibinfo  {journal} {Eur. Phys. J. C}\ }\textbf {\bibinfo {volume} {77}},\
  \bibinfo {pages} {236} (\bibinfo {year} {2017})},\ \Eprint
  {https://arxiv.org/abs/1609.05721} {arXiv:1609.05721 [hep-ex]} \BibitemShut
  {NoStop}%
\bibitem [{\citenamefont {Sirunyan}\ \emph {et~al.}(2020)\citenamefont
  {Sirunyan} \emph {et~al.}}]{CMS:2020ypo}%
  \BibitemOpen
  \bibfield  {author} {\bibinfo {author} {\bibfnamefont {A.~M.}\ \bibnamefont
  {Sirunyan}} \emph {et~al.} (\bibinfo {collaboration} {CMS}),\ }\bibfield
  {title} {\bibinfo {title} {{Observation of electroweak production of
  W$\gamma$ with two jets in proton-proton collisions at $\sqrt {s}$ = 13
  TeV}},\ }\href {https://doi.org/10.1016/j.physletb.2020.135988} {\bibfield
  {journal} {\bibinfo  {journal} {Phys. Lett. B}\ }\textbf {\bibinfo {volume}
  {811}},\ \bibinfo {pages} {135988} (\bibinfo {year} {2020})},\ \Eprint
  {https://arxiv.org/abs/2008.10521} {arXiv:2008.10521 [hep-ex]} \BibitemShut
  {NoStop}%
\bibitem [{\citenamefont {Sirunyan}\ \emph
  {et~al.}(2019{\natexlab{a}})\citenamefont {Sirunyan} \emph
  {et~al.}}]{CMS:2019ppl}%
  \BibitemOpen
  \bibfield  {author} {\bibinfo {author} {\bibfnamefont {A.~M.}\ \bibnamefont
  {Sirunyan}} \emph {et~al.} (\bibinfo {collaboration} {CMS}),\ }\bibfield
  {title} {\bibinfo {title} {{Search for anomalous triple gauge couplings in WW
  and WZ production in lepton + jet events in proton-proton collisions at
  $\sqrt{s} =$ 13 TeV}},\ }\href {https://doi.org/10.1007/JHEP12(2019)062}
  {\bibfield  {journal} {\bibinfo  {journal} {JHEP}\ }\textbf {\bibinfo
  {volume} {12}},\ \bibinfo {pages} {062}},\ \Eprint
  {https://arxiv.org/abs/1907.08354} {arXiv:1907.08354 [hep-ex]} \BibitemShut
  {NoStop}%
\bibitem [{\citenamefont {Sirunyan}\ \emph
  {et~al.}(2019{\natexlab{b}})\citenamefont {Sirunyan} \emph
  {et~al.}}]{CMS:2019efc}%
  \BibitemOpen
  \bibfield  {author} {\bibinfo {author} {\bibfnamefont {A.~M.}\ \bibnamefont
  {Sirunyan}} \emph {et~al.} (\bibinfo {collaboration} {CMS}),\ }\bibfield
  {title} {\bibinfo {title} {{Measurements of the pp $\to$ WZ inclusive and
  differential production cross section and constraints on charged anomalous
  triple gauge couplings at $\sqrt{s} =$ 13 TeV}},\ }\href
  {https://doi.org/10.1007/JHEP04(2019)122} {\bibfield  {journal} {\bibinfo
  {journal} {JHEP}\ }\textbf {\bibinfo {volume} {04}},\ \bibinfo {pages}
  {122}},\ \Eprint {https://arxiv.org/abs/1901.03428} {arXiv:1901.03428
  [hep-ex]} \BibitemShut {NoStop}%
\bibitem [{\citenamefont {Sirunyan}\ \emph
  {et~al.}(2018{\natexlab{a}})\citenamefont {Sirunyan} \emph
  {et~al.}}]{CMS:2017fhs}%
  \BibitemOpen
  \bibfield  {author} {\bibinfo {author} {\bibfnamefont {A.~M.}\ \bibnamefont
  {Sirunyan}} \emph {et~al.} (\bibinfo {collaboration} {CMS}),\ }\bibfield
  {title} {\bibinfo {title} {{Observation of electroweak production of
  same-sign W boson pairs in the two jet and two same-sign lepton final state
  in proton-proton collisions at $\sqrt{s} = $ 13 TeV}},\ }\href
  {https://doi.org/10.1103/PhysRevLett.120.081801} {\bibfield  {journal}
  {\bibinfo  {journal} {Phys. Rev. Lett.}\ }\textbf {\bibinfo {volume} {120}},\
  \bibinfo {pages} {081801} (\bibinfo {year} {2018}{\natexlab{a}})},\ \Eprint
  {https://arxiv.org/abs/1709.05822} {arXiv:1709.05822 [hep-ex]} \BibitemShut
  {NoStop}%
\bibitem [{\citenamefont {Sirunyan}\ \emph {et~al.}(2017)\citenamefont
  {Sirunyan} \emph {et~al.}}]{CMS:2017egm}%
  \BibitemOpen
  \bibfield  {author} {\bibinfo {author} {\bibfnamefont {A.~M.}\ \bibnamefont
  {Sirunyan}} \emph {et~al.} (\bibinfo {collaboration} {CMS}),\ }\bibfield
  {title} {\bibinfo {title} {{Search for anomalous couplings in boosted
  $\mathrm{ WW/WZ }\to\ell\nu\mathrm{ q \bar{q} }$ production in proton-proton
  collisions at $\sqrt{s} =$ 8 TeV}},\ }\href
  {https://doi.org/10.1016/j.physletb.2017.06.009} {\bibfield  {journal}
  {\bibinfo  {journal} {Phys. Lett. B}\ }\textbf {\bibinfo {volume} {772}},\
  \bibinfo {pages} {21} (\bibinfo {year} {2017})},\ \Eprint
  {https://arxiv.org/abs/1703.06095} {arXiv:1703.06095 [hep-ex]} \BibitemShut
  {NoStop}%
\bibitem [{\citenamefont {Chatrchyan}\ \emph
  {et~al.}(2014{\natexlab{a}})\citenamefont {Chatrchyan} \emph
  {et~al.}}]{CMS:2014cdf}%
  \BibitemOpen
  \bibfield  {author} {\bibinfo {author} {\bibfnamefont {S.}~\bibnamefont
  {Chatrchyan}} \emph {et~al.} (\bibinfo {collaboration} {CMS}),\ }\bibfield
  {title} {\bibinfo {title} {{Search for $WW \gamma$ and $WZ \gamma$ production
  and constraints on anomalous quartic gauge couplings in $pp$ collisions at
  $\sqrt s =$ 8 TeV}},\ }\href {https://doi.org/10.1103/PhysRevD.90.032008}
  {\bibfield  {journal} {\bibinfo  {journal} {Phys. Rev. D}\ }\textbf {\bibinfo
  {volume} {90}},\ \bibinfo {pages} {032008} (\bibinfo {year}
  {2014}{\natexlab{a}})},\ \Eprint {https://arxiv.org/abs/1404.4619}
  {arXiv:1404.4619 [hep-ex]} \BibitemShut {NoStop}%
\bibitem [{\citenamefont {Chatrchyan}\ \emph
  {et~al.}(2014{\natexlab{b}})\citenamefont {Chatrchyan} \emph
  {et~al.}}]{CMS:2013ryd}%
  \BibitemOpen
  \bibfield  {author} {\bibinfo {author} {\bibfnamefont {S.}~\bibnamefont
  {Chatrchyan}} \emph {et~al.} (\bibinfo {collaboration} {CMS}),\ }\bibfield
  {title} {\bibinfo {title} {{Measurement of the $W\gamma$ and $Z\gamma$
  Inclusive Cross Sections in $pp$ Collisions at $\sqrt s=7$ TeV and Limits on
  Anomalous Triple Gauge Boson Couplings}},\ }\href
  {https://doi.org/10.1103/PhysRevD.89.092005} {\bibfield  {journal} {\bibinfo
  {journal} {Phys. Rev. D}\ }\textbf {\bibinfo {volume} {89}},\ \bibinfo
  {pages} {092005} (\bibinfo {year} {2014}{\natexlab{b}})},\ \Eprint
  {https://arxiv.org/abs/1308.6832} {arXiv:1308.6832 [hep-ex]} \BibitemShut
  {NoStop}%
\bibitem [{\citenamefont {Chatrchyan}\ \emph
  {et~al.}(2013{\natexlab{a}})\citenamefont {Chatrchyan} \emph
  {et~al.}}]{CMS:2013ant}%
  \BibitemOpen
  \bibfield  {author} {\bibinfo {author} {\bibfnamefont {S.}~\bibnamefont
  {Chatrchyan}} \emph {et~al.} (\bibinfo {collaboration} {CMS}),\ }\bibfield
  {title} {\bibinfo {title} {{Measurement of the $W^+W^-$ Cross Section in $pp$
  Collisions at $\sqrt{s} = 7$ TeV and Limits on Anomalous $WW\gamma$ and $WWZ$
  Couplings}},\ }\href {https://doi.org/10.1140/epjc/s10052-013-2610-8}
  {\bibfield  {journal} {\bibinfo  {journal} {Eur. Phys. J. C}\ }\textbf
  {\bibinfo {volume} {73}},\ \bibinfo {pages} {2610} (\bibinfo {year}
  {2013}{\natexlab{a}})},\ \Eprint {https://arxiv.org/abs/1306.1126}
  {arXiv:1306.1126 [hep-ex]} \BibitemShut {NoStop}%
\bibitem [{\citenamefont {Chatrchyan}\ \emph
  {et~al.}(2013{\natexlab{b}})\citenamefont {Chatrchyan} \emph
  {et~al.}}]{CMS:2012wlr}%
  \BibitemOpen
  \bibfield  {author} {\bibinfo {author} {\bibfnamefont {S.}~\bibnamefont
  {Chatrchyan}} \emph {et~al.} (\bibinfo {collaboration} {CMS}),\ }\bibfield
  {title} {\bibinfo {title} {{Measurement of the Sum of $W W$ and $WZ$
  Production with $W+$Dijet Events in $pp$ Collisions at $\sqrt{s}=7$ TeV}},\
  }\href {https://doi.org/10.1140/epjc/s10052-013-2283-3} {\bibfield  {journal}
  {\bibinfo  {journal} {Eur. Phys. J. C}\ }\textbf {\bibinfo {volume} {73}},\
  \bibinfo {pages} {2283} (\bibinfo {year} {2013}{\natexlab{b}})},\ \Eprint
  {https://arxiv.org/abs/1210.7544} {arXiv:1210.7544 [hep-ex]} \BibitemShut
  {NoStop}%
\bibitem [{\citenamefont {Chatrchyan}\ \emph {et~al.}(2011)\citenamefont
  {Chatrchyan} \emph {et~al.}}]{CMS:2011egr}%
  \BibitemOpen
  \bibfield  {author} {\bibinfo {author} {\bibfnamefont {S.}~\bibnamefont
  {Chatrchyan}} \emph {et~al.} (\bibinfo {collaboration} {CMS}),\ }\bibfield
  {title} {\bibinfo {title} {{Measurement of $W^+ W^-$ production and search
  for the Higgs boson in pp collisions at $\sqrt s=7$ TeV}},\ }\href
  {https://doi.org/10.1016/j.physletb.2011.03.056} {\bibfield  {journal}
  {\bibinfo  {journal} {Phys. Lett. B}\ }\textbf {\bibinfo {volume} {699}},\
  \bibinfo {pages} {25} (\bibinfo {year} {2011})},\ \Eprint
  {https://arxiv.org/abs/1102.5429} {arXiv:1102.5429 [hep-ex]} \BibitemShut
  {NoStop}%
\bibitem [{\citenamefont {Swartz}(1987)}]{Swartz:1987xme}%
  \BibitemOpen
  \bibfield  {author} {\bibinfo {author} {\bibfnamefont {M.~L.}\ \bibnamefont
  {Swartz}},\ }\bibfield  {title} {\bibinfo {title} {{PHYSICS WITH POLARIZED
  ELECTRON BEAMS}},\ }\href@noop {} {\bibfield  {journal} {\bibinfo  {journal}
  {Conf. Proc. C}\ }\textbf {\bibinfo {volume} {8708101}},\ \bibinfo {pages}
  {83} (\bibinfo {year} {1987})}\BibitemShut {NoStop}%
\bibitem [{\citenamefont {Moortgat-Pick}\ and\ \citenamefont
  {Steiner}(2001)}]{Moortgat-Pick:2001vdh}%
  \BibitemOpen
  \bibfield  {author} {\bibinfo {author} {\bibfnamefont {G.~A.}\ \bibnamefont
  {Moortgat-Pick}}\ and\ \bibinfo {author} {\bibfnamefont {H.~M.}\ \bibnamefont
  {Steiner}},\ }\bibfield  {title} {\bibinfo {title} {{Physics opportunities
  with polarized e- and e+ beams at TESLA}},\ }\href
  {https://doi.org/10.1007/s1010501c0006} {\bibfield  {journal} {\bibinfo
  {journal} {Eur. Phys. J. direct}\ }\textbf {\bibinfo {volume} {3}},\ \bibinfo
  {pages} {6} (\bibinfo {year} {2001})},\ \Eprint
  {https://arxiv.org/abs/hep-ph/0106155} {arXiv:hep-ph/0106155} \BibitemShut
  {NoStop}%
\bibitem [{\citenamefont {Fujii}\ \emph {et~al.}(2018)\citenamefont {Fujii}
  \emph {et~al.}}]{Fujii:2018mli}%
  \BibitemOpen
  \bibfield  {author} {\bibinfo {author} {\bibfnamefont {K.}~\bibnamefont
  {Fujii}} \emph {et~al.},\ }\bibfield  {title} {\bibinfo {title} {{The role of
  positron polarization for the inital $250$ GeV stage of the International
  Linear Collider}},\ }\href@noop {} {\  (\bibinfo {year} {2018})},\ \Eprint
  {https://arxiv.org/abs/1801.02840} {arXiv:1801.02840 [hep-ph]} \BibitemShut
  {NoStop}%
\bibitem [{ILC(2013)}]{ILC:2013jhg}%
  \BibitemOpen
  \bibfield  {title} {\bibinfo {title} {{The International Linear Collider
  Technical Design Report - Volume 2: Physics}},\ }\href@noop {} {\  (\bibinfo
  {year} {2013})},\ \Eprint {https://arxiv.org/abs/1306.6352} {arXiv:1306.6352
  [hep-ph]} \BibitemShut {NoStop}%
\bibitem [{Ado(2013{\natexlab{a}})}]{Adolphsen:2013kya}%
  \BibitemOpen
  \bibfield  {title} {\bibinfo {title} {{The International Linear Collider
  Technical Design Report - Volume 3.II: Accelerator Baseline Design}},\
  }\href@noop {} {\  (\bibinfo {year} {2013}{\natexlab{a}})},\ \Eprint
  {https://arxiv.org/abs/1306.6328} {arXiv:1306.6328 [physics.acc-ph]}
  \BibitemShut {NoStop}%
\bibitem [{\citenamefont {Charles}\ \emph {et~al.}(2018)\citenamefont {Charles}
  \emph {et~al.}}]{CLICdp:2018cto}%
  \BibitemOpen
  \bibfield  {author} {\bibinfo {author} {\bibfnamefont {T.~K.}\ \bibnamefont
  {Charles}} \emph {et~al.} (\bibinfo {collaboration} {CLICdp, CLIC}),\
  }\bibfield  {title} {\bibinfo {title} {{The Compact Linear Collider (CLIC) -
  2018 Summary Report}}\ }\textbf {\bibinfo {volume} {2/2018}},\ \href
  {https://doi.org/10.23731/CYRM-2018-002} {10.23731/CYRM-2018-002} (\bibinfo
  {year} {2018}),\ \Eprint {https://arxiv.org/abs/1812.06018} {arXiv:1812.06018
  [physics.acc-ph]} \BibitemShut {NoStop}%
\bibitem [{\citenamefont {Abada}\ \emph {et~al.}(2019)\citenamefont {Abada}
  \emph {et~al.}}]{FCC:2018evy}%
  \BibitemOpen
  \bibfield  {author} {\bibinfo {author} {\bibfnamefont {A.}~\bibnamefont
  {Abada}} \emph {et~al.} (\bibinfo {collaboration} {FCC}),\ }\bibfield
  {title} {\bibinfo {title} {{FCC-ee: The Lepton Collider}: {Future Circular
  Collider Conceptual Design Report Volume 2}},\ }\href
  {https://doi.org/10.1140/epjst/e2019-900045-4} {\bibfield  {journal}
  {\bibinfo  {journal} {Eur. Phys. J. ST}\ }\textbf {\bibinfo {volume} {228}},\
  \bibinfo {pages} {261} (\bibinfo {year} {2019})}\BibitemShut {NoStop}%
\bibitem [{\citenamefont {Lonnblad}\ \emph {et~al.}(1990)\citenamefont
  {Lonnblad}, \citenamefont {Peterson},\ and\ \citenamefont
  {Rognvaldsson}}]{Lonnblad:1990bi}%
  \BibitemOpen
  \bibfield  {author} {\bibinfo {author} {\bibfnamefont {L.}~\bibnamefont
  {Lonnblad}}, \bibinfo {author} {\bibfnamefont {C.}~\bibnamefont {Peterson}},\
  and\ \bibinfo {author} {\bibfnamefont {T.}~\bibnamefont {Rognvaldsson}},\
  }\bibfield  {title} {\bibinfo {title} {{Finding Gluon Jets With a Neural
  Trigger}},\ }\href {https://doi.org/10.1103/PhysRevLett.65.1321} {\bibfield
  {journal} {\bibinfo  {journal} {Phys. Rev. Lett.}\ }\textbf {\bibinfo
  {volume} {65}},\ \bibinfo {pages} {1321} (\bibinfo {year}
  {1990})}\BibitemShut {NoStop}%
\bibitem [{\citenamefont {Pumplin}(1991)}]{Pumplin:1991kc}%
  \BibitemOpen
  \bibfield  {author} {\bibinfo {author} {\bibfnamefont {J.}~\bibnamefont
  {Pumplin}},\ }\bibfield  {title} {\bibinfo {title} {{How to tell quark jets
  from gluon jets}},\ }\href {https://doi.org/10.1103/PhysRevD.44.2025}
  {\bibfield  {journal} {\bibinfo  {journal} {Phys. Rev. D}\ }\textbf {\bibinfo
  {volume} {44}},\ \bibinfo {pages} {2025} (\bibinfo {year}
  {1991})}\BibitemShut {NoStop}%
\bibitem [{ATL(2017)}]{ATLAS:2017dfg}%
  \BibitemOpen
  \bibfield  {title} {\bibinfo {title} {{Quark versus Gluon Jet Tagging Using
  Jet Images with the ATLAS Detector}},\ }\href@noop {} {\  (\bibinfo {year}
  {2017})}\BibitemShut {NoStop}%
\bibitem [{\citenamefont {Komiske}\ \emph {et~al.}(2017)\citenamefont
  {Komiske}, \citenamefont {Metodiev},\ and\ \citenamefont
  {Schwartz}}]{Komiske:2016rsd}%
  \BibitemOpen
  \bibfield  {author} {\bibinfo {author} {\bibfnamefont {P.~T.}\ \bibnamefont
  {Komiske}}, \bibinfo {author} {\bibfnamefont {E.~M.}\ \bibnamefont
  {Metodiev}},\ and\ \bibinfo {author} {\bibfnamefont {M.~D.}\ \bibnamefont
  {Schwartz}},\ }\bibfield  {title} {\bibinfo {title} {{Deep learning in color:
  towards automated quark/gluon jet discrimination}},\ }\href
  {https://doi.org/10.1007/JHEP01(2017)110} {\bibfield  {journal} {\bibinfo
  {journal} {JHEP}\ }\textbf {\bibinfo {volume} {01}},\ \bibinfo {pages}
  {110}},\ \Eprint {https://arxiv.org/abs/1612.01551} {arXiv:1612.01551
  [hep-ph]} \BibitemShut {NoStop}%
\bibitem [{\citenamefont {Cheng}(2018)}]{Cheng:2017rdo}%
  \BibitemOpen
  \bibfield  {author} {\bibinfo {author} {\bibfnamefont {T.}~\bibnamefont
  {Cheng}},\ }\bibfield  {title} {\bibinfo {title} {{Recursive Neural Networks
  in Quark/Gluon Tagging}},\ }\href {https://doi.org/10.1007/s41781-018-0007-y}
  {\bibfield  {journal} {\bibinfo  {journal} {Comput. Softw. Big Sci.}\
  }\textbf {\bibinfo {volume} {2}},\ \bibinfo {pages} {3} (\bibinfo {year}
  {2018})},\ \Eprint {https://arxiv.org/abs/1711.02633} {arXiv:1711.02633
  [hep-ph]} \BibitemShut {NoStop}%
\bibitem [{\citenamefont {Kasieczka}\ \emph {et~al.}(2019)\citenamefont
  {Kasieczka}, \citenamefont {Kiefer}, \citenamefont {Plehn},\ and\
  \citenamefont {Thompson}}]{Kasieczka:2018lwf}%
  \BibitemOpen
  \bibfield  {author} {\bibinfo {author} {\bibfnamefont {G.}~\bibnamefont
  {Kasieczka}}, \bibinfo {author} {\bibfnamefont {N.}~\bibnamefont {Kiefer}},
  \bibinfo {author} {\bibfnamefont {T.}~\bibnamefont {Plehn}},\ and\ \bibinfo
  {author} {\bibfnamefont {J.~M.}\ \bibnamefont {Thompson}},\ }\bibfield
  {title} {\bibinfo {title} {{Quark-Gluon Tagging: Machine Learning vs
  Detector}},\ }\href {https://doi.org/10.21468/SciPostPhys.6.6.069} {\bibfield
   {journal} {\bibinfo  {journal} {SciPost Phys.}\ }\textbf {\bibinfo {volume}
  {6}},\ \bibinfo {pages} {069} (\bibinfo {year} {2019})},\ \Eprint
  {https://arxiv.org/abs/1812.09223} {arXiv:1812.09223 [hep-ph]} \BibitemShut
  {NoStop}%
\bibitem [{\citenamefont {Kasieczka}\ \emph {et~al.}(2020)\citenamefont
  {Kasieczka}, \citenamefont {Marzani}, \citenamefont {Soyez},\ and\
  \citenamefont {Stagnitto}}]{Kasieczka:2020nyd}%
  \BibitemOpen
  \bibfield  {author} {\bibinfo {author} {\bibfnamefont {G.}~\bibnamefont
  {Kasieczka}}, \bibinfo {author} {\bibfnamefont {S.}~\bibnamefont {Marzani}},
  \bibinfo {author} {\bibfnamefont {G.}~\bibnamefont {Soyez}},\ and\ \bibinfo
  {author} {\bibfnamefont {G.}~\bibnamefont {Stagnitto}},\ }\bibfield  {title}
  {\bibinfo {title} {{Towards Machine Learning Analytics for Jet
  Substructure}},\ }\href {https://doi.org/10.1007/JHEP09(2020)195} {\bibfield
  {journal} {\bibinfo  {journal} {JHEP}\ }\textbf {\bibinfo {volume} {09}},\
  \bibinfo {pages} {195}},\ \Eprint {https://arxiv.org/abs/2007.04319}
  {arXiv:2007.04319 [hep-ph]} \BibitemShut {NoStop}%
\bibitem [{\citenamefont {Guest}\ \emph {et~al.}(2016)\citenamefont {Guest},
  \citenamefont {Collado}, \citenamefont {Baldi}, \citenamefont {Hsu},
  \citenamefont {Urban},\ and\ \citenamefont {Whiteson}}]{Guest:2016iqz}%
  \BibitemOpen
  \bibfield  {author} {\bibinfo {author} {\bibfnamefont {D.}~\bibnamefont
  {Guest}}, \bibinfo {author} {\bibfnamefont {J.}~\bibnamefont {Collado}},
  \bibinfo {author} {\bibfnamefont {P.}~\bibnamefont {Baldi}}, \bibinfo
  {author} {\bibfnamefont {S.-C.}\ \bibnamefont {Hsu}}, \bibinfo {author}
  {\bibfnamefont {G.}~\bibnamefont {Urban}},\ and\ \bibinfo {author}
  {\bibfnamefont {D.}~\bibnamefont {Whiteson}},\ }\bibfield  {title} {\bibinfo
  {title} {{Jet Flavor Classification in High-Energy Physics with Deep Neural
  Networks}},\ }\href {https://doi.org/10.1103/PhysRevD.94.112002} {\bibfield
  {journal} {\bibinfo  {journal} {Phys. Rev. D}\ }\textbf {\bibinfo {volume}
  {94}},\ \bibinfo {pages} {112002} (\bibinfo {year} {2016})},\ \Eprint
  {https://arxiv.org/abs/1607.08633} {arXiv:1607.08633 [hep-ex]} \BibitemShut
  {NoStop}%
\bibitem [{\citenamefont {Sirunyan}\ \emph
  {et~al.}(2018{\natexlab{b}})\citenamefont {Sirunyan} \emph
  {et~al.}}]{CMS:2017wtu}%
  \BibitemOpen
  \bibfield  {author} {\bibinfo {author} {\bibfnamefont {A.~M.}\ \bibnamefont
  {Sirunyan}} \emph {et~al.} (\bibinfo {collaboration} {CMS}),\ }\bibfield
  {title} {\bibinfo {title} {{Identification of heavy-flavour jets with the CMS
  detector in pp collisions at 13 TeV}},\ }\href
  {https://doi.org/10.1088/1748-0221/13/05/P05011} {\bibfield  {journal}
  {\bibinfo  {journal} {JINST}\ }\textbf {\bibinfo {volume} {13}}\bibfield
  {number} {\bibinfo  {number} { (05)},\ \bibinfo {pages} {P05011}},\ }\Eprint
  {https://arxiv.org/abs/1712.07158} {arXiv:1712.07158 [physics.ins-det]}
  \BibitemShut {NoStop}%
\bibitem [{\citenamefont {Erdmann}\ \emph {et~al.}(2021)\citenamefont
  {Erdmann}, \citenamefont {Nackenhorst},\ and\ \citenamefont
  {Zei\ss{}ner}}]{Erdmann:2020ovh}%
  \BibitemOpen
  \bibfield  {author} {\bibinfo {author} {\bibfnamefont {J.}~\bibnamefont
  {Erdmann}}, \bibinfo {author} {\bibfnamefont {O.}~\bibnamefont
  {Nackenhorst}},\ and\ \bibinfo {author} {\bibfnamefont {S.~V.}\ \bibnamefont
  {Zei\ss{}ner}},\ }\bibfield  {title} {\bibinfo {title} {{Maximum performance
  of strange-jet tagging at hadron colliders}},\ }\href
  {https://doi.org/10.1088/1748-0221/16/08/P08039} {\bibfield  {journal}
  {\bibinfo  {journal} {JINST}\ }\textbf {\bibinfo {volume} {16}}\bibfield
  {number} {\bibinfo  {number} { (08)},\ \bibinfo {pages} {P08039}},\ }\Eprint
  {https://arxiv.org/abs/2011.10736} {arXiv:2011.10736 [hep-ex]} \BibitemShut
  {NoStop}%
\bibitem [{\citenamefont {Bedeschi}\ \emph {et~al.}(2022)\citenamefont
  {Bedeschi}, \citenamefont {Gouskos},\ and\ \citenamefont
  {Selvaggi}}]{Bedeschi:2022rnj}%
  \BibitemOpen
  \bibfield  {author} {\bibinfo {author} {\bibfnamefont {F.}~\bibnamefont
  {Bedeschi}}, \bibinfo {author} {\bibfnamefont {L.}~\bibnamefont {Gouskos}},\
  and\ \bibinfo {author} {\bibfnamefont {M.}~\bibnamefont {Selvaggi}},\
  }\bibfield  {title} {\bibinfo {title} {{Jet flavour tagging for future
  colliders with fast simulation}},\ }\href
  {https://doi.org/10.1140/epjc/s10052-022-10609-1} {\bibfield  {journal}
  {\bibinfo  {journal} {Eur. Phys. J. C}\ }\textbf {\bibinfo {volume} {82}},\
  \bibinfo {pages} {646} (\bibinfo {year} {2022})},\ \Eprint
  {https://arxiv.org/abs/2202.03285} {arXiv:2202.03285 [hep-ex]} \BibitemShut
  {NoStop}%
\bibitem [{\citenamefont {Nakai}\ \emph {et~al.}(2020)\citenamefont {Nakai},
  \citenamefont {Shih},\ and\ \citenamefont {Thomas}}]{Nakai:2020kuu}%
  \BibitemOpen
  \bibfield  {author} {\bibinfo {author} {\bibfnamefont {Y.}~\bibnamefont
  {Nakai}}, \bibinfo {author} {\bibfnamefont {D.}~\bibnamefont {Shih}},\ and\
  \bibinfo {author} {\bibfnamefont {S.}~\bibnamefont {Thomas}},\ }\bibfield
  {title} {\bibinfo {title} {{Strange Jet Tagging}},\ }\href@noop {} {\
  (\bibinfo {year} {2020})},\ \Eprint {https://arxiv.org/abs/2003.09517}
  {arXiv:2003.09517 [hep-ph]} \BibitemShut {NoStop}%
\bibitem [{\citenamefont {Erdmann}(2020)}]{Erdmann:2019blf}%
  \BibitemOpen
  \bibfield  {author} {\bibinfo {author} {\bibfnamefont {J.}~\bibnamefont
  {Erdmann}},\ }\bibfield  {title} {\bibinfo {title} {{A tagger for strange
  jets based on tracking information using long short-term memory}},\ }\href
  {https://doi.org/10.1088/1748-0221/15/01/P01021} {\bibfield  {journal}
  {\bibinfo  {journal} {JINST}\ }\textbf {\bibinfo {volume} {15}}\bibfield
  {number} {\bibinfo  {number} { (01)},\ \bibinfo {pages} {P01021}},\ }\Eprint
  {https://arxiv.org/abs/1907.07505} {arXiv:1907.07505 [physics.ins-det]}
  \BibitemShut {NoStop}%
\bibitem [{\citenamefont {Boudjema}\ and\ \citenamefont
  {Singh}(2009)}]{Boudjema:2009fz}%
  \BibitemOpen
  \bibfield  {author} {\bibinfo {author} {\bibfnamefont {F.}~\bibnamefont
  {Boudjema}}\ and\ \bibinfo {author} {\bibfnamefont {R.~K.}\ \bibnamefont
  {Singh}},\ }\bibfield  {title} {\bibinfo {title} {{A Model independent spin
  analysis of fundamental particles using azimuthal asymmetries}},\ }\href
  {https://doi.org/10.1088/1126-6708/2009/07/028} {\bibfield  {journal}
  {\bibinfo  {journal} {JHEP}\ }\textbf {\bibinfo {volume} {07}},\ \bibinfo
  {pages} {028}},\ \Eprint {https://arxiv.org/abs/0903.4705} {arXiv:0903.4705
  [hep-ph]} \BibitemShut {NoStop}%
\bibitem [{\citenamefont {Smillie}(2007)}]{Smillie:2006cd}%
  \BibitemOpen
  \bibfield  {author} {\bibinfo {author} {\bibfnamefont {J.~M.}\ \bibnamefont
  {Smillie}},\ }\bibfield  {title} {\bibinfo {title} {{Spin correlations in
  decay chains involving W bosons}},\ }\href
  {https://doi.org/10.1140/epjc/s10052-007-0330-7} {\bibfield  {journal}
  {\bibinfo  {journal} {Eur. Phys. J. C}\ }\textbf {\bibinfo {volume} {51}},\
  \bibinfo {pages} {933} (\bibinfo {year} {2007})},\ \Eprint
  {https://arxiv.org/abs/hep-ph/0609296} {arXiv:hep-ph/0609296} \BibitemShut
  {NoStop}%
\bibitem [{Ado(2013{\natexlab{b}})}]{Adolphsen:2013jya}%
  \BibitemOpen
  \bibfield  {title} {\bibinfo {title} {{The International Linear Collider
  Technical Design Report - Volume 3.I: Accelerator \textbackslash{}\& in the
  Technical Design Phase}},\ }\href@noop {} {\  (\bibinfo {year}
  {2013}{\natexlab{b}})},\ \Eprint {https://arxiv.org/abs/1306.6353}
  {arXiv:1306.6353 [physics.acc-ph]} \BibitemShut {NoStop}%
\bibitem [{\citenamefont {Moortgat-Pick}\ \emph {et~al.}(2008)\citenamefont
  {Moortgat-Pick} \emph {et~al.}}]{Moortgat-Pick:2005jsx}%
  \BibitemOpen
  \bibfield  {author} {\bibinfo {author} {\bibfnamefont {G.}~\bibnamefont
  {Moortgat-Pick}} \emph {et~al.},\ }\bibfield  {title} {\bibinfo {title} {{The
  Role of polarized positrons and electrons in revealing fundamental
  interactions at the linear collider}},\ }\href
  {https://doi.org/10.1016/j.physrep.2007.12.003} {\bibfield  {journal}
  {\bibinfo  {journal} {Phys. Rept.}\ }\textbf {\bibinfo {volume} {460}},\
  \bibinfo {pages} {131} (\bibinfo {year} {2008})},\ \Eprint
  {https://arxiv.org/abs/hep-ph/0507011} {arXiv:hep-ph/0507011} \BibitemShut
  {NoStop}%
\bibitem [{\citenamefont {Alwall}\ \emph {et~al.}(2014)\citenamefont {Alwall},
  \citenamefont {Frederix}, \citenamefont {Frixione}, \citenamefont {Hirschi},
  \citenamefont {Maltoni}, \citenamefont {Mattelaer}, \citenamefont {Shao},
  \citenamefont {Stelzer}, \citenamefont {Torrielli},\ and\ \citenamefont
  {Zaro}}]{Alwall:2014hca}%
  \BibitemOpen
  \bibfield  {author} {\bibinfo {author} {\bibfnamefont {J.}~\bibnamefont
  {Alwall}}, \bibinfo {author} {\bibfnamefont {R.}~\bibnamefont {Frederix}},
  \bibinfo {author} {\bibfnamefont {S.}~\bibnamefont {Frixione}}, \bibinfo
  {author} {\bibfnamefont {V.}~\bibnamefont {Hirschi}}, \bibinfo {author}
  {\bibfnamefont {F.}~\bibnamefont {Maltoni}}, \bibinfo {author} {\bibfnamefont
  {O.}~\bibnamefont {Mattelaer}}, \bibinfo {author} {\bibfnamefont {H.~S.}\
  \bibnamefont {Shao}}, \bibinfo {author} {\bibfnamefont {T.}~\bibnamefont
  {Stelzer}}, \bibinfo {author} {\bibfnamefont {P.}~\bibnamefont {Torrielli}},\
  and\ \bibinfo {author} {\bibfnamefont {M.}~\bibnamefont {Zaro}},\ }\bibfield
  {title} {\bibinfo {title} {{The automated computation of tree-level and
  next-to-leading order differential cross sections, and their matching to
  parton shower simulations}},\ }\href
  {https://doi.org/10.1007/JHEP07(2014)079} {\bibfield  {journal} {\bibinfo
  {journal} {JHEP}\ }\textbf {\bibinfo {volume} {07}},\ \bibinfo {pages}
  {079}},\ \Eprint {https://arxiv.org/abs/1405.0301} {arXiv:1405.0301 [hep-ph]}
  \BibitemShut {NoStop}%
\bibitem [{\citenamefont {Frederix}\ \emph {et~al.}(2018)\citenamefont
  {Frederix}, \citenamefont {Frixione}, \citenamefont {Hirschi}, \citenamefont
  {Pagani}, \citenamefont {Shao},\ and\ \citenamefont
  {Zaro}}]{Frederix:2018nkq}%
  \BibitemOpen
  \bibfield  {author} {\bibinfo {author} {\bibfnamefont {R.}~\bibnamefont
  {Frederix}}, \bibinfo {author} {\bibfnamefont {S.}~\bibnamefont {Frixione}},
  \bibinfo {author} {\bibfnamefont {V.}~\bibnamefont {Hirschi}}, \bibinfo
  {author} {\bibfnamefont {D.}~\bibnamefont {Pagani}}, \bibinfo {author}
  {\bibfnamefont {H.~S.}\ \bibnamefont {Shao}},\ and\ \bibinfo {author}
  {\bibfnamefont {M.}~\bibnamefont {Zaro}},\ }\bibfield  {title} {\bibinfo
  {title} {{The automation of next-to-leading order electroweak
  calculations}},\ }\href {https://doi.org/10.1007/JHEP11(2021)085} {\bibfield
  {journal} {\bibinfo  {journal} {JHEP}\ }\textbf {\bibinfo {volume} {07}},\
  \bibinfo {pages} {185}},\ \bibinfo {note} {[Erratum: JHEP 11, 085 (2021)]},\
  \Eprint {https://arxiv.org/abs/1804.10017} {arXiv:1804.10017 [hep-ph]}
  \BibitemShut {NoStop}%
\bibitem [{\citenamefont {Bierlich}\ \emph {et~al.}(2022)\citenamefont
  {Bierlich} \emph {et~al.}}]{Bierlich:2022pfr}%
  \BibitemOpen
  \bibfield  {author} {\bibinfo {author} {\bibfnamefont {C.}~\bibnamefont
  {Bierlich}} \emph {et~al.},\ }\bibfield  {title} {\bibinfo {title} {{A
  comprehensive guide to the physics and usage of PYTHIA 8.3}}\ }\href
  {https://doi.org/10.21468/SciPostPhysCodeb.8} {10.21468/SciPostPhysCodeb.8}
  (\bibinfo {year} {2022}),\ \Eprint {https://arxiv.org/abs/2203.11601}
  {arXiv:2203.11601 [hep-ph]} \BibitemShut {NoStop}%
\bibitem [{\citenamefont {Cacciari}\ \emph {et~al.}(2012)\citenamefont
  {Cacciari}, \citenamefont {Salam},\ and\ \citenamefont
  {Soyez}}]{Cacciari:2011ma}%
  \BibitemOpen
  \bibfield  {author} {\bibinfo {author} {\bibfnamefont {M.}~\bibnamefont
  {Cacciari}}, \bibinfo {author} {\bibfnamefont {G.~P.}\ \bibnamefont
  {Salam}},\ and\ \bibinfo {author} {\bibfnamefont {G.}~\bibnamefont {Soyez}},\
  }\bibfield  {title} {\bibinfo {title} {{FastJet User Manual}},\ }\href
  {https://doi.org/10.1140/epjc/s10052-012-1896-2} {\bibfield  {journal}
  {\bibinfo  {journal} {Eur. Phys. J. C}\ }\textbf {\bibinfo {volume} {72}},\
  \bibinfo {pages} {1896} (\bibinfo {year} {2012})},\ \Eprint
  {https://arxiv.org/abs/1111.6097} {arXiv:1111.6097 [hep-ph]} \BibitemShut
  {NoStop}%
\bibitem [{\citenamefont {Cacciari}\ \emph {et~al.}(2008)\citenamefont
  {Cacciari}, \citenamefont {Salam},\ and\ \citenamefont
  {Soyez}}]{Cacciari:2008gp}%
  \BibitemOpen
  \bibfield  {author} {\bibinfo {author} {\bibfnamefont {M.}~\bibnamefont
  {Cacciari}}, \bibinfo {author} {\bibfnamefont {G.~P.}\ \bibnamefont
  {Salam}},\ and\ \bibinfo {author} {\bibfnamefont {G.}~\bibnamefont {Soyez}},\
  }\bibfield  {title} {\bibinfo {title} {{The anti-$k_t$ jet clustering
  algorithm}},\ }\href {https://doi.org/10.1088/1126-6708/2008/04/063}
  {\bibfield  {journal} {\bibinfo  {journal} {JHEP}\ }\textbf {\bibinfo
  {volume} {04}},\ \bibinfo {pages} {063}},\ \Eprint
  {https://arxiv.org/abs/0802.1189} {arXiv:0802.1189 [hep-ph]} \BibitemShut
  {NoStop}%
\bibitem [{\citenamefont {Catani}\ \emph {et~al.}(1993)\citenamefont {Catani},
  \citenamefont {Dokshitzer}, \citenamefont {Seymour},\ and\ \citenamefont
  {Webber}}]{Catani:1993hr}%
  \BibitemOpen
  \bibfield  {author} {\bibinfo {author} {\bibfnamefont {S.}~\bibnamefont
  {Catani}}, \bibinfo {author} {\bibfnamefont {Y.~L.}\ \bibnamefont
  {Dokshitzer}}, \bibinfo {author} {\bibfnamefont {M.~H.}\ \bibnamefont
  {Seymour}},\ and\ \bibinfo {author} {\bibfnamefont {B.~R.}\ \bibnamefont
  {Webber}},\ }\bibfield  {title} {\bibinfo {title} {{Longitudinally invariant
  $K_t$ clustering algorithms for hadron hadron collisions}},\ }\href
  {https://doi.org/10.1016/0550-3213(93)90166-M} {\bibfield  {journal}
  {\bibinfo  {journal} {Nucl. Phys. B}\ }\textbf {\bibinfo {volume} {406}},\
  \bibinfo {pages} {187} (\bibinfo {year} {1993})}\BibitemShut {NoStop}%
\bibitem [{\citenamefont {Quertenmont}(2011)}]{QUERTENMONT201195}%
  \BibitemOpen
  \bibfield  {author} {\bibinfo {author} {\bibfnamefont {L.}~\bibnamefont
  {Quertenmont}},\ }\bibfield  {title} {\bibinfo {title} {Particle
  identification with ionization energy loss in the cms silicon strip
  tracker},\ }\href
  {https://doi.org/https://doi.org/10.1016/j.nuclphysbps.2011.03.145}
  {\bibfield  {journal} {\bibinfo  {journal} {Nuclear Physics B - Proceedings
  Supplements}\ }\textbf {\bibinfo {volume} {215}},\ \bibinfo {pages} {95}
  (\bibinfo {year} {2011})}\BibitemShut {NoStop}%
\bibitem [{\citenamefont {Lippmann}(2012)}]{LIPPMANN2012148}%
  \BibitemOpen
  \bibfield  {author} {\bibinfo {author} {\bibfnamefont {C.}~\bibnamefont
  {Lippmann}},\ }\bibfield  {title} {\bibinfo {title} {Particle
  identification},\ }\href
  {https://doi.org/https://doi.org/10.1016/j.nima.2011.03.009} {\bibfield
  {journal} {\bibinfo  {journal} {Nuclear Instruments and Methods in Physics
  Research Section A: Accelerators, Spectrometers, Detectors and Associated
  Equipment}\ }\textbf {\bibinfo {volume} {666}},\ \bibinfo {pages} {148}
  (\bibinfo {year} {2012})}\BibitemShut {NoStop}%
\bibitem [{\citenamefont {Boyarski}\ \emph {et~al.}(1989)\citenamefont
  {Boyarski}, \citenamefont {Coupal}, \citenamefont {Feldman}, \citenamefont
  {Hanson}, \citenamefont {Nash}, \citenamefont {O'Shaughnessy}, \citenamefont
  {Rankin},\ and\ \citenamefont {Van~Kooten}}]{Boyarski:1989gw}%
  \BibitemOpen
  \bibfield  {author} {\bibinfo {author} {\bibfnamefont {A.}~\bibnamefont
  {Boyarski}}, \bibinfo {author} {\bibfnamefont {D.}~\bibnamefont {Coupal}},
  \bibinfo {author} {\bibfnamefont {G.~J.}\ \bibnamefont {Feldman}}, \bibinfo
  {author} {\bibfnamefont {G.}~\bibnamefont {Hanson}}, \bibinfo {author}
  {\bibfnamefont {J.}~\bibnamefont {Nash}}, \bibinfo {author} {\bibfnamefont
  {K.~F.}\ \bibnamefont {O'Shaughnessy}}, \bibinfo {author} {\bibfnamefont
  {P.}~\bibnamefont {Rankin}},\ and\ \bibinfo {author} {\bibfnamefont {R.~J.}\
  \bibnamefont {Van~Kooten}},\ }\bibfield  {title} {\bibinfo {title} {{Particle
  Identification Using De / Dx in the Mark-{II} Detector at the {SLC}}},\
  }\href {https://doi.org/10.1016/0168-9002(89)91427-7} {\bibfield  {journal}
  {\bibinfo  {journal} {Nucl. Instrum. Meth. A}\ }\textbf {\bibinfo {volume}
  {283}},\ \bibinfo {pages} {617} (\bibinfo {year} {1989})}\BibitemShut
  {NoStop}%
\bibitem [{\citenamefont {Va'vra}(2000)}]{Vavra:2000vag}%
  \BibitemOpen
  \bibfield  {author} {\bibinfo {author} {\bibfnamefont {J.}~\bibnamefont
  {Va'vra}},\ }\bibfield  {title} {\bibinfo {title} {{Particle identification
  methods in high-energy physics}},\ }\href
  {https://doi.org/10.1016/S0168-9002(00)00644-6} {\bibfield  {journal}
  {\bibinfo  {journal} {Nucl. Instrum. Meth. A}\ }\textbf {\bibinfo {volume}
  {453}},\ \bibinfo {pages} {262} (\bibinfo {year} {2000})}\BibitemShut
  {NoStop}%
\bibitem [{\citenamefont {Hauschild}(1996)}]{Hauschild:1996np}%
  \BibitemOpen
  \bibfield  {author} {\bibinfo {author} {\bibfnamefont {M.}~\bibnamefont
  {Hauschild}},\ }\bibfield  {title} {\bibinfo {title} {{Progress in dE/dx
  techniques used for particle identification}},\ }\href
  {https://doi.org/10.1016/0168-9002(96)00607-9} {\bibfield  {journal}
  {\bibinfo  {journal} {Nucl. Instrum. Meth. A}\ }\textbf {\bibinfo {volume}
  {379}},\ \bibinfo {pages} {436} (\bibinfo {year} {1996})}\BibitemShut
  {NoStop}%
\bibitem [{\citenamefont {Einhaus}\ \emph {et~al.}(2019)\citenamefont
  {Einhaus}, \citenamefont {Kr\"amer},\ and\ \citenamefont
  {Malek}}]{Einhaus:2019jvg}%
  \BibitemOpen
  \bibfield  {author} {\bibinfo {author} {\bibfnamefont {U.}~\bibnamefont
  {Einhaus}}, \bibinfo {author} {\bibfnamefont {U.}~\bibnamefont {Kr\"amer}},\
  and\ \bibinfo {author} {\bibfnamefont {P.}~\bibnamefont {Malek}},\ }\bibfield
   {title} {\bibinfo {title} {{Studies on Particle Identification with dE/d$x$
  for the ILD TPC}},\ }in\ \href@noop {} {\emph {\bibinfo {booktitle}
  {{International Workshop on Future Linear Colliders}}}}\ (\bibinfo {year}
  {2019})\ \Eprint {https://arxiv.org/abs/1902.05519} {arXiv:1902.05519
  [physics.ins-det]} \BibitemShut {NoStop}%
\bibitem [{\citenamefont {Lehraus}\ \emph {et~al.}(1982)\citenamefont
  {Lehraus}, \citenamefont {Matthewson},\ and\ \citenamefont
  {Tejessy}}]{Lehraus:1982em}%
  \BibitemOpen
  \bibfield  {author} {\bibinfo {author} {\bibfnamefont {I.}~\bibnamefont
  {Lehraus}}, \bibinfo {author} {\bibfnamefont {R.}~\bibnamefont
  {Matthewson}},\ and\ \bibinfo {author} {\bibfnamefont {W.}~\bibnamefont
  {Tejessy}},\ }\bibfield  {title} {\bibinfo {title} {{PARTICLE IDENTIFICATION
  BY DE/DX SAMPLING IN HIGH PRESSURE DRIFT DETECTORS}},\ }\href
  {https://doi.org/10.1016/0029-554X(82)90100-8} {\bibfield  {journal}
  {\bibinfo  {journal} {Nucl. Instrum. Meth.}\ }\textbf {\bibinfo {volume}
  {196}},\ \bibinfo {pages} {361} (\bibinfo {year} {1982})}\BibitemShut
  {NoStop}%
\bibitem [{\citenamefont {Allison}\ and\ \citenamefont
  {Cobb}(1980)}]{Allison:1980vw}%
  \BibitemOpen
  \bibfield  {author} {\bibinfo {author} {\bibfnamefont {W.~W.~M.}\
  \bibnamefont {Allison}}\ and\ \bibinfo {author} {\bibfnamefont {J.~H.}\
  \bibnamefont {Cobb}},\ }\bibfield  {title} {\bibinfo {title} {{Relativistic
  Charged Particle Identification by Energy Loss}},\ }\href
  {https://doi.org/10.1146/annurev.ns.30.120180.001345} {\bibfield  {journal}
  {\bibinfo  {journal} {Ann. Rev. Nucl. Part. Sci.}\ }\textbf {\bibinfo
  {volume} {30}},\ \bibinfo {pages} {253} (\bibinfo {year} {1980})}\BibitemShut
  {NoStop}%
\bibitem [{\citenamefont {Adeva}\ \emph {et~al.}(1990)\citenamefont {Adeva}
  \emph {et~al.}}]{Adeva:1990kd}%
  \BibitemOpen
  \bibfield  {author} {\bibinfo {author} {\bibfnamefont {B.}~\bibnamefont
  {Adeva}} \emph {et~al.},\ }\bibfield  {title} {\bibinfo {title} {{Study of
  Theta inclined tracks in L3 muon chambers}},\ }\href
  {https://doi.org/10.1016/0168-9002(90)90349-B} {\bibfield  {journal}
  {\bibinfo  {journal} {Nucl. Instrum. Meth. A}\ }\textbf {\bibinfo {volume}
  {290}},\ \bibinfo {pages} {115} (\bibinfo {year} {1990})}\BibitemShut
  {NoStop}%
\bibitem [{\citenamefont {Sheaff}(2000)}]{Sheaff:1999iv}%
  \BibitemOpen
  \bibfield  {author} {\bibinfo {author} {\bibfnamefont {M.}~\bibnamefont
  {Sheaff}},\ }\bibfield  {title} {\bibinfo {title} {{Detectors for Particle
  Identification: Time-of-Flight, dE/dx, and Transition Radiation}},\ }\href
  {https://doi.org/10.1063/1.1361759} {\bibfield  {journal} {\bibinfo
  {journal} {AIP Conf. Proc.}\ }\textbf {\bibinfo {volume} {536}},\ \bibinfo
  {pages} {87} (\bibinfo {year} {2000})}\BibitemShut {NoStop}%
\bibitem [{\citenamefont {D'Agostini}\ \emph {et~al.}(1981)\citenamefont
  {D'Agostini}, \citenamefont {Marini}, \citenamefont {Martellotti},
  \citenamefont {Massa},\ and\ \citenamefont {Sciubba}}]{DAgostini:1980jhh}%
  \BibitemOpen
  \bibfield  {author} {\bibinfo {author} {\bibfnamefont {G.}~\bibnamefont
  {D'Agostini}}, \bibinfo {author} {\bibfnamefont {G.}~\bibnamefont {Marini}},
  \bibinfo {author} {\bibfnamefont {G.}~\bibnamefont {Martellotti}}, \bibinfo
  {author} {\bibfnamefont {F.}~\bibnamefont {Massa}},\ and\ \bibinfo {author}
  {\bibfnamefont {A.}~\bibnamefont {Sciubba}},\ }\bibfield  {title} {\bibinfo
  {title} {{High Resolution Time-of-flight Measurements in Small and Large
  Scintillation Counters}},\ }\href
  {https://doi.org/10.1016/0029-554X(81)91193-9} {\bibfield  {journal}
  {\bibinfo  {journal} {Nucl. Instrum. Meth.}\ }\textbf {\bibinfo {volume}
  {185}},\ \bibinfo {pages} {49} (\bibinfo {year} {1981})}\BibitemShut
  {NoStop}%
\bibitem [{\citenamefont {Chen}\ and\ \citenamefont
  {Guestrin}(2016)}]{Chen:2016:XST:2939672.2939785}%
  \BibitemOpen
  \bibfield  {author} {\bibinfo {author} {\bibfnamefont {T.}~\bibnamefont
  {Chen}}\ and\ \bibinfo {author} {\bibfnamefont {C.}~\bibnamefont
  {Guestrin}},\ }\bibfield  {title} {\bibinfo {title} {{XGBoost}: A scalable
  tree boosting system},\ }in\ \href {https://doi.org/10.1145/2939672.2939785}
  {\emph {\bibinfo {booktitle} {Proceedings of the 22nd ACM SIGKDD
  International Conference on Knowledge Discovery and Data Mining}}},\ \bibinfo
  {series and number} {KDD '16}\ (\bibinfo  {publisher} {ACM},\ \bibinfo
  {address} {New York, NY, USA},\ \bibinfo {year} {2016})\ pp.\ \bibinfo
  {pages} {785--794}\BibitemShut {NoStop}%
\bibitem [{\citenamefont {Lewis}(2019)}]{Lewis:2019xzd}%
  \BibitemOpen
  \bibfield  {author} {\bibinfo {author} {\bibfnamefont {A.}~\bibnamefont
  {Lewis}},\ }\bibfield  {title} {\bibinfo {title} {{GetDist: a Python package
  for analysing Monte Carlo samples}},\ }\href@noop {} {\  (\bibinfo {year}
  {2019})},\ \Eprint {https://arxiv.org/abs/1910.13970} {arXiv:1910.13970
  [astro-ph.IM]} \BibitemShut {NoStop}%
\end{thebibliography}%
\end{document}